\documentclass{aa}

\usepackage{times}
\usepackage{graphicx}
\usepackage{xspace}
\usepackage{epsfig}
\usepackage{natbib}
\usepackage{rotating}
\usepackage{dcolumn}
\usepackage{lscape}

\newcommand{\lori}{\ensuremath{\lambda\,Ori}\xspace}

\begin{document}

\title{XMM-{\it Newton} investigations of the Lambda  Orionis star-forming region (XILO)}
  \subtitle{I. The young cluster Collinder\,69}

\author{D. Barrado\inst{1,2}
          \and
        B. Stelzer\inst{3}
          \and
          M. Morales-Calder\'on
          \inst{4}
          \and
	  A. Bayo
          \inst{5}
          \and
	  N. Hu\'elamo
          \inst{2}
          \and
	  J.R. Stauffer
          \inst{4}
	  \and
	  S. Hodgkin
          \inst{6}
          \and
	  F. Galindo
          \inst{2}
	  \and
	  E. Verdugo
          \inst{7}
}

\offprints{D. Barrado, barrado@cab.inta-csic.es}

\institute{
   Calar Alto Observatory, Centro Astron\'omico Hispano Alem\'an, C/ Jes\'us Durb\'an Rem\'on,
   E-04004 Almer\'{\i}a, Spain 
         \and
  Depto. Astrof\'{\i}sica, Centro de Astrobiolog\'{\i}a (INTA-CSIC),  ESAC campus,
  P.O. Box 78,
  E-28691 Villanueva de la Ca\~nada, Spain 
         \and
  INAF - Osservatorio Astronomico di Palermo,
  Piazza del Parlamento 1,
  I-90134 Palermo, Italy 
         \and
             Spitzer Science Center, California Institute of Technology, Pasadena, CA 91125, USA.
         \and
             European Southern Observatory, Alonso de C\'ordova 3107, Vitacura, Santiago, Chile.
         \and
         Institute of Astronomy, Madingley Road, Cambridge CB3 0HA, UK.
         \and
             European Space Agency (ESAC), P.O.Box  78, E-28691 Villanueva de la Ca\~nada (Madrid), Spain.
}

\titlerunning{XMM-{\it Newton} investigation in Collinder\,69}

\date{Received $<$date$>$ / Accepted $<$date$>$}

\abstract
{
This is the first paper of a series devoted to the Lambda Orionis star-forming region, Orion's Head, from
the X-ray perspective.
Our final aim is to provide a comprehensive view of this complex region, which includes several distinct 
associations and dark clouds.
}
{
We aim to uncover the population of the central, young star cluster Collinder\,69, and in particular
those diskless Class III objects not identified by previous surveys based on near- and mid-infrared 
searches, and to establish the X-ray luminosity function for the association.
}
{We have combined two exposures taken with the XMM-{\it Newton} satellite  with an exhaustive data set of optical, near- and mid-infrared photometry  to assess the membership of the X-ray sources  based on different color-color and color-magnitude diagrams, as well as other properties, such as effective temperatures, masses and bolometric luminosities derived from spectral energy distribution fitting  and comparison with theoretical isochrones.  The presence of circumstellar disks is discussed using mid-infrared photometry  from the  {\em Spitzer} {\rm space telescope.}}
{\rm
With an X-ray flux limit of $\sim 5 \cdot 10^{-15}\,{\rm erg/cm^2/s}$
we detected a total of 164 X-ray sources, of which 66 are probable and possible 
cluster members. A total of 16 are newly identified probable members plus another three possible new members.
The two XMM-{\it Newton} pointings east and west of the cluster center have allowed us to verify the 
heterogeneous spatial distribution of young stars, 
which is probably related to the large scale structure of the region.
The disk fraction of the X-ray detected cluster sample (complete down to $\sim 0.3\,M_\odot$) 
is very low, close to 10\%, in remarkable
contrast to the low-mass stellar and substellar population (mostly undetected in X-rays) 
where the disk fraction reaches about 50\%. 
The X-ray luminosity function of Collinder\,69 in different mass bins 
 provides support for an age of several Myr
when  compared  with other well known young associations.
}
{
The X-ray properties of the young stars in Collinder\,69 resemble those found
in other young stellar associations, with saturation at $\log{(L_{\rm x}/L_{\rm bol})} \sim -3$
and low fractional X-ray luminosities for stars with $M \geq 2\,M_\odot$. 
With our improved cluster census we confirm previous reports on the untypically low disk
 fraction compared to other clusters of several Myr age. 
The different disk fractions of X-ray detected (essentially solar-like) and undetected 
(mostly low-mass stars and brown dwarfs) members 
can be understood as a consequence of a mass-dependence of the time-scale for disk evolution.  
}


\maketitle

\section{Introduction}\label{sect:intro}

This is the first paper of a series devoted to the Lambda Orionis star-forming region (hereafter
LOSFR), a complex located at about 400 pc, as seen in X-ray.
(\cite{Murdin77.1}).
This region gives shape to the head of Orion, the legendary
Hellenic hero, and includes several distinct zones at different evolutionary stages: 
from dark clouds with no
obvious star formation (or at the very initial steps) toward  the quasi circular boundary,
to the full fledged \lori cluster, located  at the center  and about 5 Myr old
(also called Collinder\,69, the name we will use throughout this paper), and other structures in between,
such as Barnard 30 and Barnard 35.

We carried out several surveys at different wavelengths in the
past, from the optical to the submillimeter. The main goal of these
studies was to characterize the population of low-mass stars and
brown dwarfs, obtain a stellar and substellar census, specially
identifying members with disks using infrared (IR) excesses, and derive the
initial mass function (IMF). Part of these results have been published by
\cite{Barrado04.1, Barrado07.1}; \cite{Morales08.1} and
\cite{Bayo09.1} --their PhD dissertations, which include the multi-wavelength
search and spectroscopic characterization, respectively; and \cite{Bouy09.1}.  
Additional analyses are forthcoming.

When searching for members of star-forming regions, 
optical surveys are usually not very efficient, 
because the internal reddening caused by the
inhomogeneous dust (and gas) absorbs a significant amount of light. On
the  one hand, near- and mid-infrared  searches, such as the one we have
conducted with the {\em Spitzer} space telescope, avoid this problem to a
large extent. However, they are biased and identify primarily objects with
IR excess emission over the stellar photosphere, 
either Class 0/I, characterized by the presence of an
envelope and increasing fluxes toward longer wavelengths or Class II
(Classical TTauri stars), with an actively accreting circumstellar
disk and a spectral energy distribution  (SED) with a significant
departure from the black--body shape, following the evolutionary classification scheme 
(\cite{Lada87.1,Adams87.1}).

With the main goal of unveiling the Class III population in the LOSFR (i.e., weak-line TTauri stars,
diskless young stars 
or those with transition disks), we have conducted  an X-ray survey with the 
XMM-{\it Newton} observatory. 
X-rays are as efficient as near-IR photometry piercing a dark cloud, and they allow to
detect coronal activity, which is 
known to be very strong in young late-type stars regardless of the presence or absence of
circumstellar material (Feigelson \& Montmerle 1999).
Our project, the 
{\em XMM-Newton Investigations of the Lambda Orionis Star-Forming Region (XILO)},
 concentrates on crucial star formation sites within the LOSFR and so far comprises 
eight XMM-{\em Newton} pointings. 
We present here the results related to the  central cluster, Collinder\,69.
We combine the XMM-{\em Newton} data with the existing multi-wavelength database. The analysis 
of the XMM-{\em Newton} data is described in Sect.~\ref{sect:obs_and_data}. The master catalog
of Collinder\,69 obtained from combining X-ray, optical, and IR data is introduced 
in Sect.~\ref{sect:catalog}.
In Sect.~\ref{sect:select_cand} we explain how we select cluster candidates from this list. 
Section~\ref{sect:properties} deals with the properties of the selected cluster members, 
and Sect.~\ref{sect:discussion}  summarizes the findings of this paper.

   \begin{figure*}
   \centering
   \includegraphics[width=\textwidth]{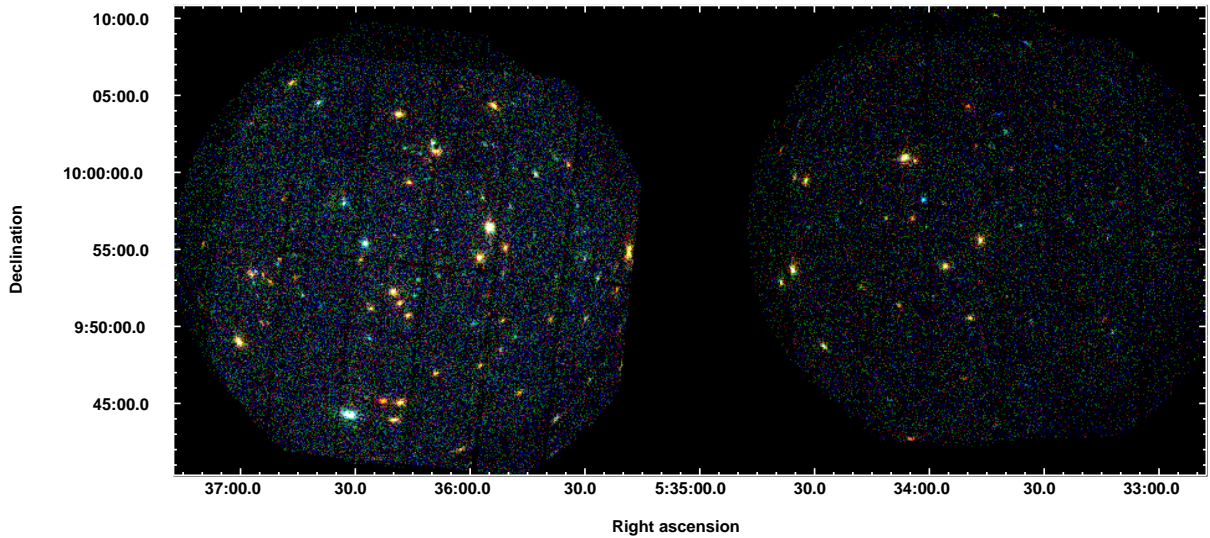}
      \caption{Merged images of XMM-{\em Newton}'s EPIC pn, MOS\,1 and MOS\,2 for Collinder\,69 eastern and western 
fields (Col\,69\,E and Col\,69\,W).
}
         \label{fig:xmm}
   \end{figure*}


\section{Observations and data analysis}\label{sect:obs_and_data}

We obtained two XMM-{\em Newton} observations of
Collinder\,69. The observations were centered east and west of
 \lori, a massive bright O8\,III  star (see Fig.~\ref{fig:xmm}).
Henceforth we refer to these two XMM-{\em Newton} exposures 
as Col\,69\,E and Col\,69\,W. 

The prime instrument EPIC ({\it European Photon Imaging Camera};
see \cite{Jansen01.1}, \cite{Turner01.1}, and \cite{Strueder01.1})
was operated in full-frame mode.
The Optical Monitor (OM; \cite{Mason01.1})
was scheduled for several consecutive imaging exposures in the $V$ and $B$ bands.
The observing log for both pointings is given in Tab.~\ref{tab:obslog}.

\subsection{EPIC}\label{subsect:data_epic}

Both observations were analyzed with the XMM-{\em Newton} Science Analysis System (SAS)
pipeline, version 7.0.0. We created a photon events list with the metatasks {\sc epchain}
and {\sc emchain} for EPIC/pn and EPIC/MOS respectively.
There was some loss of observing time because of high background. We selected good time intervals
(GTIs) using a filter that cuts out times of high background, maximizing
the overall signal-to-noise. The C\,69\,E field is almost unaffected, but 
for C\,69\,W the useful time was reduced to about half  the nominal exposure time
given in Tab.~\ref{tab:obslog}.
The data were further filtered for pixel patterns, and events at the
boundary of individual CCD chips or near bad pixels or those outside the field-of-view (FOV)
were removed.

Images with a pixel size of $5^{\prime\prime}$ were binned from the cleaned events list.
We performed source detection on the images in four energy bands using standard SAS tools.
The most suitable boundaries of the energy bands
depend on the spectral shape of the X-ray sources, which is not known a priori.
We have tested various bounds for the energy bands, and after inspection of the results
from source detection made the following choice for the soft ($S$), medium ($M$),
hard ($H$) and broad ($B$) bands: $S=0.5-1.0$\,keV, $M=1.0-2.0$\,keV,
$H=2.0-7.3$\,keV, and $B = S+M+H$.

For the source detection process we proceeded in two steps.
First, all EPIC detectors were analyzed separately as described above.
To evaluate the relative sensitivity of the detectors we computed
the ratio of count rates measured by EPIC/pn and EPIC/MOS,
${(C_{\rm pn}/C_{\rm MOS})}_{\rm i}$ with $i=S, M, H, B$,
separately for each energy band,
for bright sources detected with all three instruments.
The median of these values were used to scale the exposure maps of EPIC/pn for the joint analysis
of all three instruments. The merged observation is then equivalent to an effective MOS exposure
of duration ${(C_{\rm pn}/C_{\rm MOS})}_{\rm i} \cdot t_{\rm pn} + t_{\rm M1} + t_{\rm M2}$,
where $t_{\rm pn}, t_{\rm M1}, t_{\rm M2}$ are the individual exposure times of pn, MOS\,1, and MOS\,2.

Our final X-ray catalog contains all sources detected in the merged data (`EPIC sources'),
plus all sources detected in the individual instruments that are not
within $10^{\prime\prime}$ of an EPIC source. Detection in one detector but not in the
combined data may occur, e.g. if the source is
located outside the FOV, near a chip gap, or in a region of low exposure (such as a bad column)
in one or more of the instruments.
We selected a detection threshold of $ML \geq 15$.
This yields $112$ sources in the merged EPIC data of C\,69\,E and $52$ sources for C\,69\,W.
The X-ray coordinates of the final source list have been cross-correlated with
the 2\,MASS catalog \citep{Cutri03.1},
and a boresight correction was computed as the median of the astrometric
offsets in RA and DEC.
Tabs.~\ref{tab:XsourcesE} and ~\ref{tab:XsourcesW} summarize the X-ray detections from both fields and
their X-ray parameters (corrected X-ray position and statistical positional error,
off-axis angle, maximum likelihood of
source detection, broadband count rate, and hardness ratios).
We define hardness ratios as $(B_1-B_2)/(B_1+B_2)$ where $B_1$ and $B_2$ can be 
different combinations of $S$, $M$, and $H$. The hardness ratios listed in Tabs.~\ref{tab:XsourcesE}
and ~\ref{tab:XsourcesW} are $HR1 = (M-S)/(M+S)$ and $HR2 = (H+M-S)/(H+M+S)$.
The merged EPIC/pn+MOS images are shown in Fig.~\ref{fig:xmm}.

To examine the conditions in the X-ray emitting plasma,
we performed a spectral analysis within XSPEC version 12.4.
The spectra were grouped depending on the number of counts,
with a minimum of $5$ counts per bin for sources with less than $100$ counts.
For each X-ray source we used the instrument listed in Col.~2
of Tabs.~\ref{tab:XsourcesE} and ~\ref{tab:XsourcesW}.
 Both one- and two-temperature thermal model fits ({\sc APEC})
with photo-absorption ({\sc wabs}) were tested,
starting the fitting process from a range of initial parameters and selecting for each model
as best fit the result with the minimum $\chi^2$. X-ray temperature(s), absorbing column $N_{\rm H}$,
and normalization(s) were free fit parameters, while the global abundances were fixed at $0.3$\,times
the solar value.
We chose as final representation of the
X-ray source between the $1$-T and the $2$-T model based on the null probability
$P(\chi^2 > \chi^2_0)$. Our first choice
is the $1$-T model. Whenever $P < 0.1$ for the $1$-T model, we resorted to the $2$-T fit.
The results and the calculation of X-ray luminosities for possible cluster members is deferred to
Sect.~\ref{subsect:lx} after the Collinder\,69 bona fide sample has been established.

The cross-correlation with cataloged Collinder\,69 membership lists
and previous photometric surveys
 is described in
Sect~\ref{subsect:mastercat}. 
We anticipate here that there is a total of $68$ confirmed cluster
members within our two XMM-{\em Newton} pointings
 that are not detected with any EPIC instrument. 
These members were previously identified  in optical surveys and
 were published by \cite{Dolan99.1}, \cite{Barrado04.1} and  \cite{Morales08.1}. 
We have derived upper limits to their count rates making use of the sensitivity map generated with
the SAS tool {\sc esensmap}.
 The detection limit derived from the sensitivity map is $0.00043$\,cts/s (C69\,E) and
$0.00071$\,cts/s (C69\,W) in the center of the image, and roughly a factor of three lower toward
the edge of the overlapping field of MOS and pn. The sensitivity is further decreased in the outermost
areas that are covered only by one of the detectors.

\subsection{OM}\label{subsect:data_om}

The OM was operated in full frame imaging mode. 
For the C\,69\,W field one $V$ band exposure is available, and for the
C\,69\,E field there are five consecutive exposures with the $V$ and $B$ filters
(see Tab.~\ref{tab:astrometry_om}).

The OM data were reduced with the SAS metatask {\sc omichain} with
default parameters. This task performs all basic data reduction steps, including
flat-fielding, identification of bad pixels, modulo-8 fixed pattern correction,
and source detection with aperture photometry.
The final output of the pipeline is a combined OM source list that contains 
 the Johnson magnitudes. In the case of several exposures in a given filter 
the magnitude is the average of all exposures in that filter, i.e. for C\,69\,E 
the average $B$ magnitude of exposures $007$, $008$, and $010$ and the average $V$ magnitude
of exposures $006$ and $009$.
We corrected the absolute positions of the detections by cross-correlating the
OM source list with the 2\,MASS catalog \citep{Cutri03.1}. A search radius of
$3^{\prime\prime}$ was used and only OM sources with $\geq 15\,\sigma$ detection significance
were considered for the astrometry. Subsequently, the OM coordinates of all detected
sources were shifted by the median position offset
(given in Tab.~\ref{tab:astrometry_om}).

Each exposure yielded
approximately $1000$ detections. In this paper we present only the photometry for those
OM objects that are counterparts to EPIC X-ray sources.

Standard (Johnson) magnitudes are automatically extracted by the SAS tools if observations
are performed in more than one band, allowing us  to calculate the color. Since we obtained
only $V$ band data for C\,69\,W, the {\sc omichain} pipeline does not provide standard magnitudes.
However, we can make use of the C\,69\,E data to calibrate the $V$ band photometry for the
OM detections in C\,69\,W. For the OM sources in the C\,69\,E field we verified that
a linear transformation with the $B-V$ color reproduces the standard magnitudes derived
 by the SAS tools,
$V_{\rm ins} = V_{\rm std} + \kappa_{\rm V} \cdot (B-V)_{\rm std} = V_{\rm std} + CF_{\rm V}$, and
we determined the average coefficient $\langle \kappa_{\rm V} \rangle$.
In order to compute standard magnitudes $V_{\rm std}$ from the instrumental magnitudes $V_{\rm ins}$
measured in the C\,69\,W field, we assume for all objects the mean color observed in C69\,E,
$\langle B-V \rangle = 1.04 \pm 0.35$.
The conversion factor derived in this way is $CF_{\rm V} = 0.03 \pm 0.02$, where the uncertainty
reflects the statistical errors of $\langle \kappa \rangle$ and $\langle B-V \rangle$
of the C\,69\,E field.
Note that the transformation factor $CF_{\rm V}$ is almost negligibly small.

   \begin{figure}
   \centering
   \includegraphics[width=8cm]{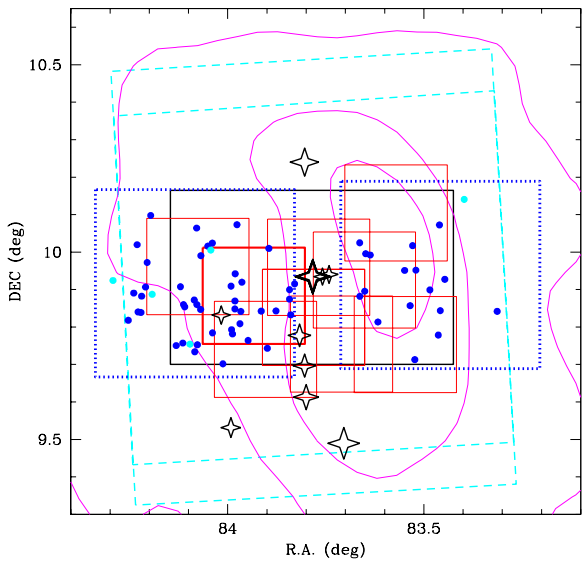}
      \caption{
The young cluster Collinder\,69, with the surveys we have conducted so far at different wavelengths.
The black rectangle represents the CFHT1999 $RI_c$ survey,  the red squares the
near-IR data ($JHKs$) from Omega2000 (thick lines for the deep image), whereas
the IRAC mapping is represented with dashed, cyan lines. Finally, our two XMM-{\em Newton} FOVs
are located with blue, dotted squares. O and B stars are represented by four-point stars, with
increasing size related to increasing brightness.
Our final cluster members (see Section\,\ref{subsect:members}) are represented as solid circles
(blue for probable members, cyan for  possible members).
The magenta contours corresponds to IRAS data at 100 micron.}
         \label{fig_surveys}
   \end{figure}

   \begin{figure}
   \centering
   \includegraphics[width=9cm]{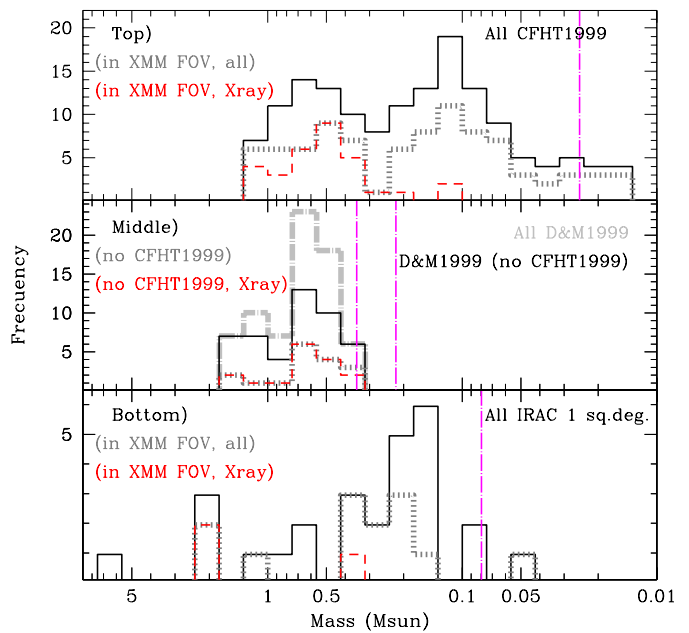}
    \caption{
       Histograms for different samples (and subsamples) described in Sect.\,\ref{sect:catalog}
       and identified with different techniques: optical, mid-infrared, and X-rays.
       The vertical dotted and long-dashed --magenta-- segments locate the detection limit of each survey for
       cluster members.
       In these  three subpanels --a top, middle and bottom, 
       solid --black-- lines correspond to the whole sample analyzed in each subpanel;
       dotted --gray-- lines are used for subsamples spatially included in the XMM-{\em Newton} field-of-view,
       whereas dashed --red-- lines describe the data with X-ray detections.
       {\bf Subpanel at the  top.-} 
       Probable and possible members from \cite{Barrado04.1}  and \cite{Barrado07.1}, with
       an initial selection based on optical photometry. The less
       massive objects detected in the XMM-{\em Newton} data are LOri-CFHT-090 and
       LOri-CFHT-098, with $\sim$0.12 $M_\odot$.
       {\bf Subpanel in the  middle.-} 
       Members listed by \cite{Dolan99.1}, originally selected from optical photometry 
       (including  narrow-band H$\alpha$ -i.e., active stars or with accretion),
       and confirmed based on high-resolution spectroscopy. The
       thick dotted long-dashed --silver-- histogram includes members in common with \cite{Barrado04.1}.
       The completeness limits correspond to the whole photometric survey described in
       \cite{Dolan99.1} and \cite{Dolan02.1} --0.35 and 0.22 $M_\odot$, respectively.
       {\bf Subpanel at the  bottom.-} 
       New members from \cite{Morales08.1}, composed by Class II sources 
       (Classical TTauri stars and few  members with transition disks).
       Initial selection based on mid-IR data from {\em Spitzer}.
}
    \label{fig:histomassA}
   \end{figure}

   \setcounter{figure}{3}
   \begin{figure}
   \centering
   \includegraphics[width=9cm]{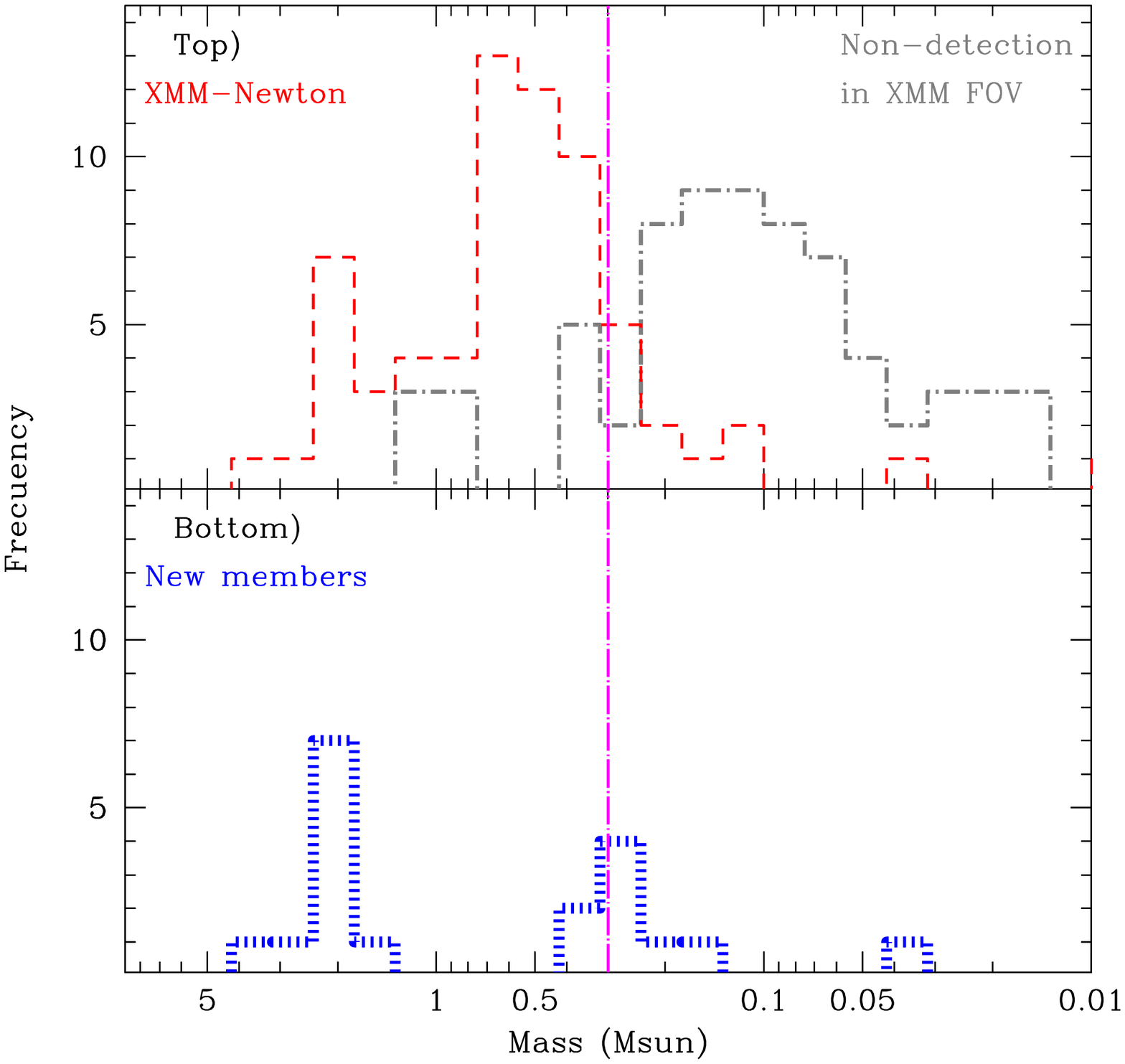}
    \caption{
       {\bf Subpanel at the top.-} The X-ray sample of C69 members --dashed, red line--
       (i.e., possible and probable Collinder 69 members with X-ray detections)
       based on XMM-{\em Newton} data.
       The dotted and long-dashed --grey-- 
       represents previously known Collinder 69 members --probable and possible--
       with no X-ray (only upper limits).
       {\bf Subpanel at the bottom.-} 
       We highlight in blue the new probable and possible members identified with detections by 
       XMM-{\em Newton}  (16 probable and 6 possible new members).
}
    \label{fig:histomassB}
   \end{figure}

\section{Compilation of the master catalog}\label{sect:catalog}

For the search for optical/IR counterparts to the X-ray sources we used the published data
from \cite{Dolan99.1, Dolan01.1, Dolan02.1} and \cite{Barrado04.1, Barrado07.1}.
These tables include
all presently known solar-type and low-mass members of Collinder\,69,
 and a large catalog of photometric measurements ($VR_cI_c$) carried out in the field around the cluster,
without any membership selection (Tab.~4 of \cite{Dolan02.1}). 
 For the high-mass end (spectral types earlier than F5) , see \cite{Hernandez09.1}.
In addition we cross-correlated the
X-ray source list (Tabs.~\ref{tab:XsourcesE} and ~\ref{tab:XsourcesW}) with the {\em Spitzer} data
(from the Guaranteed Time Observation program  PID \#37).
We also made use of the $V$ and $B$ band data obtained with the OM. 
Near-IR data  were obtained from 2MASS and Calar Alto using the Omega2000 camera.

Next we describe the individual catalogs and their cross-correlation with the X-ray source list.
 A graphical overview for the spatial coverage of the multi-wavelength database in
Collinder\,69 is shown in Fig.~\ref{fig_surveys}.
In Figs.~\ref{fig:histomassA} and ~\ref{fig:histomassB} we compare the mass ranges (for cluster members)
 covered by the samples described in the various datasets. Details about how masses were
derived can be found in Sections\,\ref{sect:vosa} and ~\ref{subsect:masses}.

\subsection{Optical data from the CFHT survey}\label{subsect:cat_cfht}

\cite{Barrado04.1}  present deep optical photometry in an area of
$42^\prime \times 28^\prime$ centered on the star \lori,
 taken with the CFHT telescope and the cousins $R_c, I_c$  filters (hereafter CFHT1999).
Owing to the field-of-view of the CFHT1999 survey, these optical data only
cover part of the two XMM-{\em Newton} fields.
Since this survey includes  shallow ($10$\,s) and deep
($600$\,s and $900$\,s for the $I_c$ and $R_c$ filter, respectively) exposures,
 providing photometry
for bright and faint objects,    we used both datasets.
The deep images are considered only for X-ray sources that have no counterpart
in the shallow images. We removed from the final catalog all optical counterparts
with photometry outside the detector linear regime. For cluster members, this survey is complete
down to 0.025 $M_\odot$ (additional information can be found in 
\cite{Barrado04.1}). A histogram with a distribution of masses is displayed in  Fig.~\ref{fig:histomassA}.

\subsection{Optical data from Dolan \& Mathieu 1999, 2001, and 2002}\label{subsect:cat_dol}

One of the most comprehensive studies of the region around the LOSFR  has
been published by  \cite{Dolan99.1, Dolan01.1, Dolan02.1}. They include optical photometry for
cluster candidate members selected by H$\alpha$ narrow-band photometry and
subsequent optical spectroscopy.
They have also published the complete photometry for those stars not selected
 as cluster members (i.e., not active in H$\alpha$ at the time of the observation).
We have cross-correlated their data --both the identified members and the whole photometric database--
with our X-ray
detections using the same strategy as in the case of the CFHT1999 survey,
and detected  all cluster members within the XMM-{\em Newton} pointings except one (DM048).
Note that these optical surveys include completely
our XMM-{\em Newton} pointings.
In total, out of the 205 optical and IR counterparts located near
our 164  XMM-{\em Newton} X-ray sources, 53 have photometry coming from these
catalogs (35 in \cite{Dolan99.1} and another 18 in \cite{Dolan02.1}).

For the cluster members described in \cite{Dolan99.1}, and 
using a detection limit --for cluster members- 
$I_{lim}$$\sim$15 mag, this value corresponds  to a mass of 0.35 $M_\odot$, for
 a NextGen 5Myr isochrone (\cite{Baraffe98.1}) and a distance of 400 pc.
Figure~\ref{fig:histomassA} includes a histogram for the mass distribution for cluster members 
listed by \cite{Dolan99.1}, although masses were derived using bolometric luminosities
(see Section~\ref{subsect:masses}).
Of the 36 cluster members located within the XMM-{\em Newton} FOVs, 35 have been detected in X-rays.
Only  DM048, a Class II member, appears to have  a low activity and we were only able to derive an upper
limit for its X-ray flux.

In the case of the comprehensive photometric survey contained in \cite{Dolan02.1},
 with a  $R_{compl}$$\sim$17.5 mag, this value
 translates for cluster members to $\sim$0.22 $M_\odot$.

\subsection{Optical data from the XMM-{\em Newton} Optical Monitor}\label{subsect:cat_om}

%
%
\begin{figure}
\begin{center}
   \includegraphics[width=9cm]{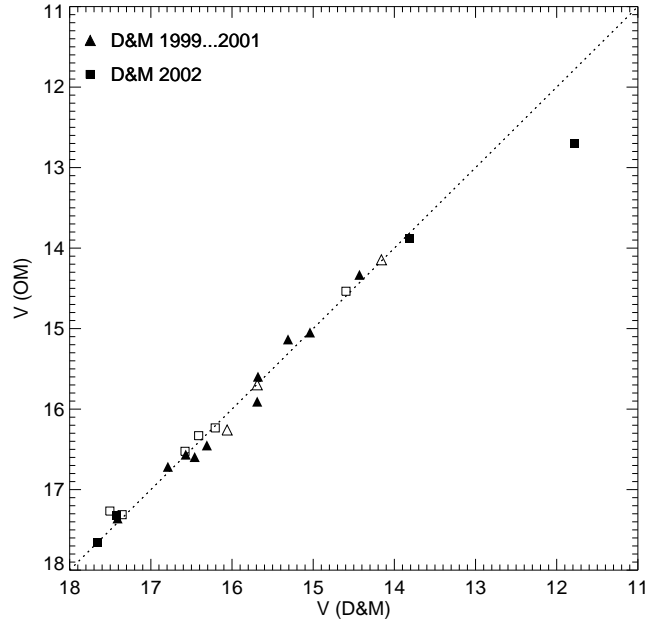}
\caption{Comparison of $V$ magnitudes measured with the OM and from the literature
for optical counterparts to EPIC sources in Collinder\,69: {\em filled symbols} - C\,69\,E,
{\em open symbols} - C\,69\,W. The good agreement shows that
OM photometry is reliable and that the errors associated with the transformation of
instrumental into standard magnitudes for the C\,69\,W field are negligibly small.}
\label{fig:v_om_cat}
\end{center}
\end{figure}

About a third of the X-ray sources with OM counterpart have  $V$ band data
obtained by  \cite{Dolan99.1, Dolan02.1}.
This allows us to check the quality of the OM photometry
by comparing the magnitudes measured with the OM to those from the \cite{Dolan99.1, Dolan01.1, Dolan02.1} surveys.
In Fig.~\ref{fig:v_om_cat} filled symbols represent data from the C\,69\,E field
with pipeline computed
standard magnitudes for the OM. Open symbols are stars from the C\,69\,W field, for which
we used the transformation into standard magnitudes described in Sect.~\ref{subsect:data_om}.

Owing to the high sensitivity of its detectors, the OM is very well suited for observing
faint stars. Indeed, we find good agreement with published photometry from
\cite{Dolan99.1, Dolan01.1, Dolan02.1} down to $V \sim 18$\,mag.
The exception are bright stars where the
OM photometry is seen to deviate strongly from the \cite{Dolan99.1, Dolan01.1, Dolan02.1} catalogs.
Indeed the OM cannot be used for observations of optically bright
sources (Ehle et al. 2003, XMM-{\em Newton} Users' Handbook), 
and we do not use OM data for stars with $V<12$\,mag in the
master catalog.

\subsection{Near-IR data from the 2MASS All Sky Survey}\label{subsect:cat_tm}

We have correlated the positions of the X-ray detections with near-IR photometry in
the 2MASS catalog (Cutri et al. 2003). This All-Sky survey has limiting magnitudes of
$J_{lim}$$\sim$16.8 mag, $H_{lim}$$\sim$16.1 mag, $Ks_{lim}$$\sim$15.3 mag, which does
not suffice to provide a complete coverage of our sources detected in the X-ray images.
Using a completeness limit of $H_{compl}$$\sim$15.1 mag, this value translates to a mass of 0.05 $M_\odot$
for a NextGen 5Myr isochrone and 400 pc.
 Therefore, cluster members less massive than this value, as well as other 
sources (background stars and galaxies), might not be detected with 2MASS.
In any event, as we will show in Sect.5.3, the completeness limit of the 
XMM-{\em Newton} data for
Collinder 69 is  $\sim$$0.3M_\odot$ --although a higher mass might be possible,
 such that all cluster members,  except C69-X-w012 because it is too bright,
 have a 2MASS counterpart.

\subsection{Near-IR data from CAHA and Omega2000}\label{subsect:cat_omega}

We have obtained deep near-IR photometry
in October 2005 with the Calar Alto 3.5m (Almeria, Spain) and the
Omega2000 camera under a Director Discretionary Time (DDT) program. Omega2000 has a 15.36$\times$15.36 arcmin
FOV and we observed several pointings in the $J$, $H$, and $Ks$ totaling 5 minutes per
filter and field.  The seeing was about 1.2 arcsec and the photometric calibration
was obtained with stars in the 2MASS catalog.  Details about the photometry extraction and the
calibration can be found in \cite{Barrado07.1}. The data are complete
down to $J_{compl}$=20.00 mag, $H_{compl}$=19.00 mag and $Ks_{compl}$=18.00 mag, well below the deuterium burning limit
at 13 $M$(Jupiter) for cluster members.  The detection limits are
$J_{lim}$=20.50 mag, $H_{lim}$=19.75 and $Ks_{lim}$=18.75 mag.

In addition, we later requested  on 
another DDT observation with the same setup
in one specific field located south-east of the O8III star
 \lori, where a significant number of cluster members are located
 (see \cite{Barrado04.1, Barrado07.1}). This observation took place in November 2007. Each image
is composed of 30 individual
60 seconds exposures, with a dithering of 15 arcsec. Therefore, the total exposure time for each
 filter is 30 minutes. The reduction and the calibration have been performed in the same way as 
for the data collected in 2005.  This dataset is complete
down to $J_{compl}$=21.00 mag, $H_{compl}$=20.00 mag and $Ks_{compl}$=19.50 mag.
The detection limits are $J_{lim}$=21.50 mag, $H_{lim}$=20.50 and $Ks_{lim}$=20.00 mag.

\subsection{The {\em Spitzer} data}\label{subsect:cat_spi}

Our {\em Spitzer} data were collected during March 15, with the Multiband Imaging
Photometer for {\em Spitzer} (MIPS, \cite{Rieke04.1}) and on October 11, with the
InfraRed Array Camera (IRAC, \cite{Fazio04.1}), 2004, as part of a
Guaranteed Time Observation program (GTO, PID:37). The layout of
observations is explained in detail in \cite{Barrado07.1}.

IRAC imaging was performed in mapping mode with individual exposures
of 12\,sec œôòüframetimeœôòý and three dithers at each map step. The
Collinder\,69 map was broken into two segments, one offset west of the
star \lori --a very bright O8 III--
and the other offset to the east, with the combined
image covering an area of 57'$\times$61.5', leaving the star
\lori approximately at the center. Each of the IRAC images
from the {\em Spitzer} Science Center pipeline were corrected for
instrumental artifacts using the IDL code provided by the {\em Spitzer}
Science Center and then combined into mosaics at each of the four
bandpasses using the MOPEX package (\cite{Makovoz05.1}). Note that
the IRAC images do not cover exactly the same FOV in all bands,
providing a slice north of the star with data at 3.6 and 5.8 $\mu$m, and
another slice south of it with photometry at 4.5 and 8.0
$\mu$m. Figure~\ref{fig_surveys}  shows the IRAC pointings as cyan dashed lines.
However, our X-ray sources are not affected by the lack of spatial coincidence between
 the IRAC channels. Note, however, that one X-ray source (C69-X-e041=DM71) is located outside of the 
IRAC FOV and lack mid-IR photometry (although we do have MIPS data at 24 micron).

The MIPS instrument was used to map the cluster with a medium rate
scan mode and 12 legs separated by 302'' in the cross scan
direction. The total effective integration time per point on the sky at
24 $\mu$m for most points in the map was 40 seconds, and the mosaic
covered an area of 60.5'$\times$98.75' centered around the star \lori. 
Since there were no visible artifacts in the pipeline mosaics
for MIPS 24 $\mu$m we used them as our starting point to extract the
photometry.

The analysis of the data was done with IRAF. We  performed
aperture photometry to derive fluxes for all objects in our field. For
the IRAC mosaics we used an aperture of 3 pixels radius (about 3.6 arcsec),
 and the sky was computed using a circular annulus 3 pixels wide, starting at a
radius 3 pixels away from the center. It is necessary to apply an
aperture correction to our 3-pixel aperture photometry in order to
estimate the flux for a 10-pixel aperture, because the latter is the
aperture size used to determine the IRAC flux calibration. For the MIPS
photometry at 24 $\mu$m, we used a 5.31 pixels (13 arcsec) aperture
and a sky annulus from 8.16 pixels (20 arcsec) to 13.06 pixels (32
arcsec).

The {\em Spitzer} survey was used to identify new cluster members with mid-IR excesses.
These additional members, spread on an area about 1 sq. deg., are presented in   \cite{Morales08.1}.
Additional information can be found in  Morales-Calder\'on et al. (2010, in prep).
The  mass distribution of this sample is displayed in Fig.~\ref{fig:histomassA}.
 The completeness
limit for {\em Spitzer} new members, imposed by the IRAC band at 8 $\mu$m, is $\sim$0.08 $M_\odot$.

\subsection{Cross-correlation with the X-ray catalog}\label{subsect:mastercat}

Each of the catalogs described from 
Sect.\,\ref{subsect:cat_cfht} to 
Sect.\,\ref{subsect:cat_spi} was cross-correlated with the X-ray source list. 
We   searched for
optical and IR counterparts for our X-ray detections, using a radius of $5.1^{\prime\prime}$. 
This search radius is motivated by the astrometry of XMM-{\em Newton}
($\sim 1-2^{\prime\prime}$) and the statistical errors of the X-ray sources
($\leq 4^{\prime\prime}$).

We  found multiple counterparts for several X-ray sources within our search radius. 
The visual inspection of  all optical and IR  images
indicated that in a  few cases there are additional possible counterparts 
even slightly beyond this search radius. 
In order to be as comprehensive as possible, we have also retained them.
We compiled a master catalog with all  sources
 that are present in at least one of the mappings (optical, near-IR or mid-IR)
and extracted the photometry from these
surveys. The photometry of all possible counterparts to X-ray sources 
is listed in Tabs.~\ref{tab:photometryE} and \ref{tab:photometryW} 
(XMM-{\em Newton} eastern and western pointings, respectively).
The additional columns at the end of  these tables reference
the catalog/instrument from which the data was extracted as described
in the following paragraphs.

When there is ambiguity in the optical or IR  identification.
these possibilities were identified, after visual inspection,
by a letter denoting whether the optical or IR source is close to the 
 center of the X-ray error box
--c, or south --s,  north --n, west --w-- or east --e.
Although we are confident that the closest optical/IR counterpart provides the correct
identification of the X-ray source, we have  considered the secondary counterparts and
assessed 
whether all these optical/infrared sources might be cluster members, and also the likelihood 
of them  being the actual  source of the X-ray emission (i.e., the correct identification between the 
X-ray and the optical/IR coordinates).

For the selection of counterparts that were detected in a given
passband in more than one of the catalogs and/or surveys we proceed as follows:
For the blue side of the optical range  ($B$ and $V$), we listed first the data coming  from 
the OM;
otherwise, the $V$ magnitude from \cite{Dolan99.1, Dolan01.1, Dolan02.1} is listed.
For the reddest portion of the optical range  ($R_c$ and $I_c$), we have chosen first the data from
\cite{Barrado07.1}
(i.e., candidate members detected in \cite{Barrado04.1}).
Then, we selected the photometry from \cite{Dolan99.1, Dolan01.1}
(i.e., members based on photometry and spectroscopy). If the X-ray source was
 not present in any of those
catalogs, we first looked for an optical counterpart in the CFHT1999 survey
 (only for objects with $I_c$$>$13.2
 mag, and if it did not appear there, 
in the database from \cite{Dolan02.1}, which is shallower but
covers a much larger area.
For the near-IR photometry ($JHKs$),
 we  prioritized the data coming from the deep images acquired with Omega2000
in November 2007 ($J$$>$17.2 mag). If no data were recorded, we selected the photometry
 coming from  Omega2000 taken in  October 2005, shallower  ($J$$>$13.0 mag)
 but covering a much larger area. Otherwise, data from
 the 2MASS All Sky Survey are listed.

{\em Spitzer} data have been published  for the known members of Collinder\,69 (\cite{Barrado07.1}) 
of which 31 are
found to be X-ray sources with XMM-{\em Newton}.  Another 16 X-ray sources were previously identified
by \cite{Dolan99.1}  as cluster members, and
we have obtained the mid-IR fluxes with {\em Spitzer} for them
(see \cite{Morales08.1} and Morales-Calder\'on et al. 2010, in prep).

We present here for the first time the {\em Spitzer} data for additional 95 X-ray objects
that were not  identified as young stars before (two of them only with data at 24 micron from MIPS).

In total, we have 164 X-ray sources and identified 205 possible counterparts either in
the optical or in the  infrared.
A summary table, with the identification in different surveys, can be found in 
Tab.~\ref{tab:summaryTAB}.

\section{Selection of  Collinder\,69 candidate members}\label{sect:select_cand}

The X-ray, optical, and IR data presented above were used to select new candidate
members of Collinder\,69. The result is summarized in Tabs.~\ref{tab:membershipcriteriaE} and \ref{tab:membershipcriteriaW}.
Here we describe our selection criteria.
 In the next two figures (Figures~\ref{fig:qso} and ~\ref{fig:HRD_all}) we will include all optical,
near- and mid-IR sources detected 
 in the search box around the X-ray position, 
as examples of how we have proceeded with different membership diagnostic tools.
 Therefore, these diagrams include all  possible
optical and IR counterparts of the X-ray sources, when the data are available.
However, after finishing the membership analysis, we  reached the
conclusion that when several identifications are possible,
the object responsible of the X-ray emission is the closest
to the X-ray coordinate, and they
are the only ones plotted in the subsequent color-color and color-magnitude diagrams
(such as the panels in Fig.~\ref{fig:ccd_cmd}).
Out of  our $164$ XMM-{\em Newton} detections, there are 61 probable members
and another five possible members.
These are the clusters members we have included in our calculation of Collinder\,69
 X-ray luminosity function  and disk fraction, 
together  with
Collinder\,69 members not  detected in X-rays (Sect.~\ref{subsubsect:membersnotdetected}),
restricted to specific mass intervals.

\subsection{Removing the contamination by quasars}\label{subsect:select_QSO}

   \begin{figure}
   \centering
   \includegraphics[width=9cm]{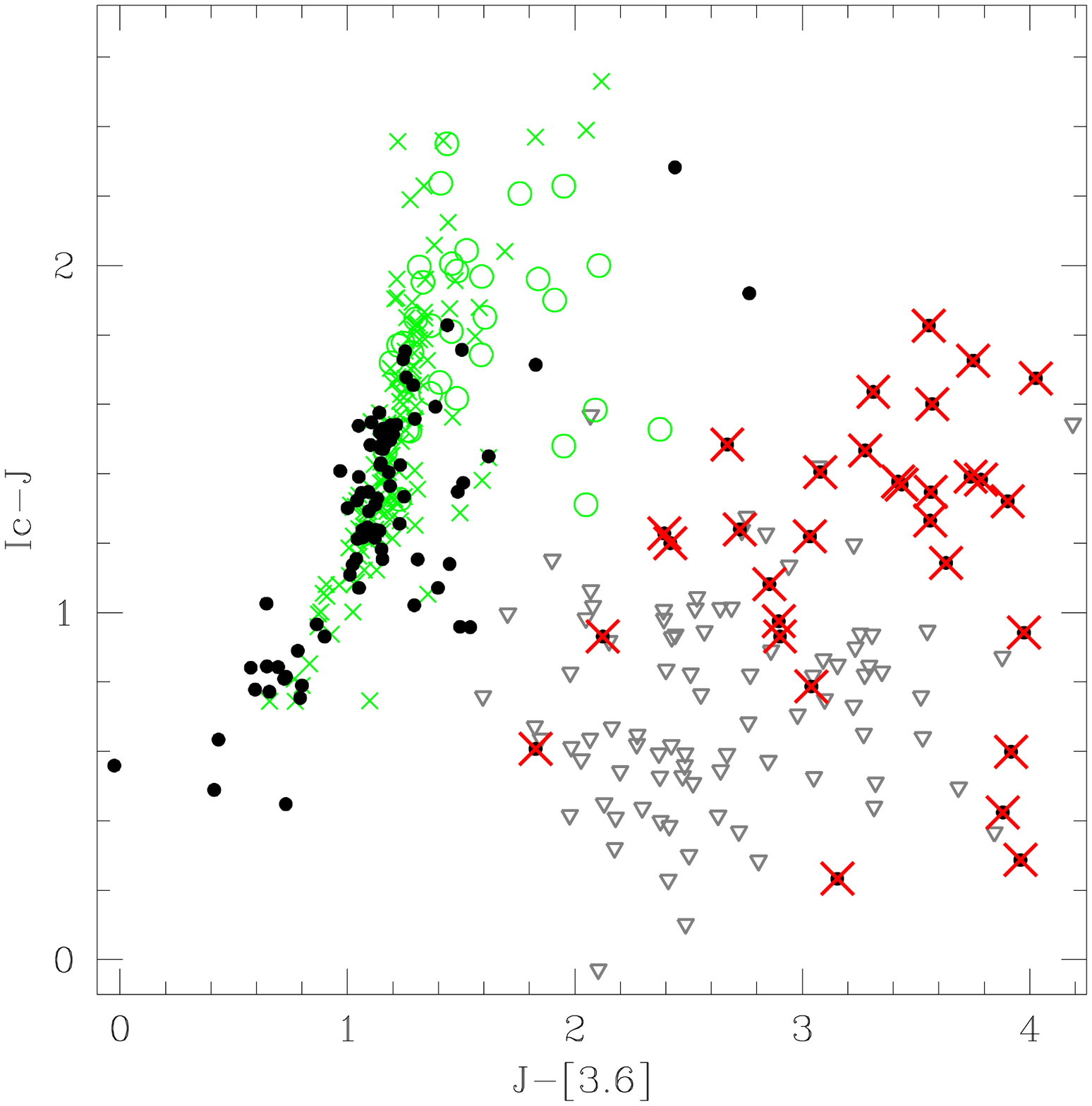}
      \caption{ This graph allows the identification of contamination by quasars: previously
known members of Collinder\,69 from \cite{Dolan99.1} and \cite{Barrado04.1, Barrado07.1}
are displayed as green crosses and open circles (for Class III and II members, respectively).
  Quasars observed by the SWIRE team (\cite{Harvey06.1})
appear as  gray open triangles.  
Counterparts to our XMM-{\em Newton} X-ray sources are shown  as
as black circles
over-imposed by  big red crosses (for quasars, rejected as cluster members based on this diagram)
and as   solid black circles
(other sources, including  possible cluster members  and other contaminants).
After \cite{Bouy09.1}.
}
         \label{fig:qso}
   \end{figure}

As shown by \cite{Bouy09.1}, in a deep search of faint Collinder\,69 members close to the bright star $\lambda$ Orionis,
 a color-color diagram that includes data in the $I$, $J$, and
 $[3.6]$
filters is a powerful tool to identify QSOs. This diagram is shown in Fig.~\ref{fig:qso}, where we
 included several samples corresponding to previously known members of the cluster from
\cite{Dolan99.1}, 
from \cite{Barrado04.1, Barrado07.1}, and from the {\em Spitzer}/SWIRE survey
 (extragalactic sources,  \cite{Harvey06.1}), 
which identified a significant number of quasars. 
The location of our X-ray sources in the figure allows
an easy classification between objects located within the same  
region as the SWIRE QSOs 
 and those compatible with stellar nature and cluster membership.
The first group -QSO- includes 31 from the total sample (23 of the
objects we have, at the end of the process, identified as the origin of the X-ray  emission). 
The second group  contains 80 of the initial sample of possible counterparts (72 of the final sample of X-ray sources).
 The rest of the X-ray sources cannot be classified because 
they lack  IR data in
one or more bands used
for this analysis (94 of the initial sample  of counterparts, 
69 of the final list of counterparts).

 Based on this information, we classified the 31 counterparts
that lie in the QSO region  as non-members of the Collinder\,69 cluster.  
As explained above, when there are several identifications,  this includes
 objects detected in our multi-wavelength surveys that are not the closest to the X-ray source.
In any case, we used additional diagnostics (see next subsections) and verify, 
when the data are available, that they fulfill several other non-membership criteria, 
such as luminosities too low for their effective
temperatures, for example. This information is listed in Col. \#4 of Tabs.~\ref{tab:membershipcriteriaE} and 
\ref{tab:membershipcriteriaW}.

\subsection{The Hertzsprung-Russell diagram}\label{subsect:HRD}

   \begin{figure}
   \centering
   \includegraphics[width=9cm]{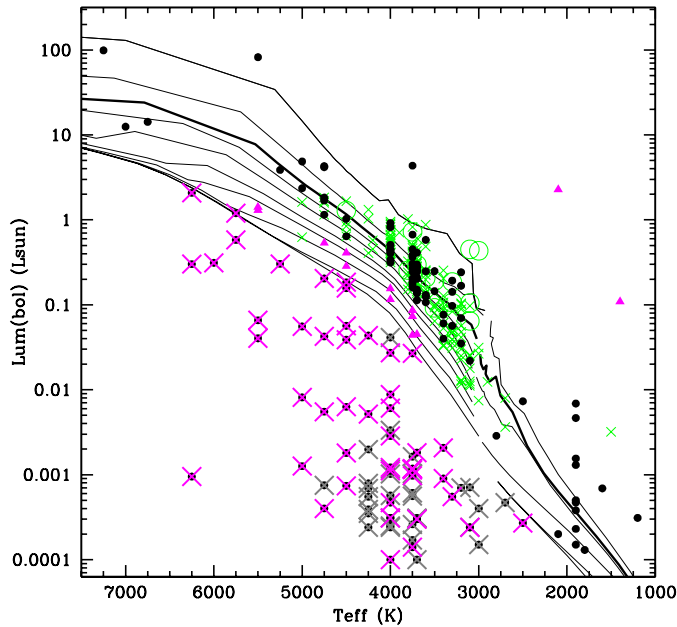}
      \caption{
HR diagram with our rejection of non-members based on their location
 (solid circles with over-imposed big crosses --magenta-- for non-members below the Main-Sequence).
Other possible non-members, inmediately above and below the cluster sequence are displayed as magenta triangles.
Previously rejected  quasars are displayed as solid   circles with over-imposed big crosses --in  gray.
Other  candidate members, not rejected by this test, are shown as solid black circles. Some will be later removed
from the membership list using other color-color and color-magnitude diagrams.
Previously known members for the cluster appear as green crosses (Class III) or
open green circles (Class II), and  were selected from \cite{Dolan99.1},
\cite{Barrado07.1}
 and \cite{Morales08.1}.
Isochrones correspond to \cite{Siess00.1}  --1, 3, 5, 7, 10, 15, 20, 30, 50 and 100 Myr--
  and COND models --1, 5, 10, 100, 1000 and 10,000 Myr-- from the Lyon group
(\cite{Baraffe02.1}  or \cite{Chabrier00.1} .
}
         \label{fig:HRD_all}
   \end{figure}

We  applied Virtual Observatory techniques to derive the properties of the  SEDs for 
the  counterparts to the X-ray emitters. In particular,
we  used the VOSA tool (``Virtual Observatory SED Analyzer'', \cite{Bayo08.1}) 
to fit theoretical models and to estimate effective
temperatures and bolometric luminosities, assuming  a gravity value of $\log{g}=4.0$ 
and a distance of $400$\,pc.
We used models by Kurucz \citep{castelli97} for fits with temperatures $T_{eff} \ge$3750 K,
NextGen models for 3700 K $\ge T_{\rm eff} \ge$ 2500K,
Dusty models  for 2400 K $\ge T_{\rm eff} \ge$ 1900K,  and
COND models for  $T_{\rm eff} \ge$ 1800K (\cite{Baraffe98.1}, \cite{Chabrier00.1}, \cite{Baraffe02.1}).
Individual masses and ages were also estimated with this tool. In addition, using the bolometric luminosities and
5 Myr isochrones (\cite{Siess00.1}, \cite{Chabrier00.1}; where the link point is at 0.1 $M_\odot$ or 3000 K),
 we also derived masses and effective temperatures.
We additionally computed the fraction between the measured part of the observed flux 
(from the photometry) and the total bolometric luminosity --the flux fraction.
All this information is listed in  Tab.~\ref{tab:finalmembers} for our possible and probable members
 of the cluster.

An Hertzsprung-Russell (HR)
 diagram is presented in Fig.~\ref{fig:HRD_all}. As we  forced the fits with specific
 values of gravity and distance,  we expect that field stars and extragalactic sources will not
 appear close to the cluster isochrone at 5 Myr. Therefore, we classified all X-ray detections
well below the Main Sequence (MS) 
as not belonging to the Collinder\,69  cluster (probable non-members --NM--
 if they also fulfill other non-membership criteria, or possible non-members --NM?-- if we 
were unable to reject membership using other diagrams). 
Other objects located between the MS and the 15 Myr isochrone or well above the 1 Myr isochrone have
been flagged as possible  non-member (NM?) and classified later on depending on their location in several 
CMDs and CCD.
(Col. \#5 in Tabs.~\ref{tab:membershipcriteriaE} and \ref{tab:membershipcriteriaW}).

Based on our experience (\cite{Bayo08.1}), this methodology to estimate bolometric luminosities based on 
photospheric SED fitting can provide only lower limits for objects such as edge-on disks. This way we 
might be misclassifying some good candidates. In any case  this type of object would have been easy 
to recover according to their IRAC photometry, and indeed  we  did not detected any of them
(see \cite{Morales08.1} or Morales-Calder\'on et al. (2010)).

\subsection{Additional pollutants based on CM and CC diagrams}\label{subsect:select_CCD_CMD}

   \begin{figure*}
   \centering
   \includegraphics[width=7.5cm]{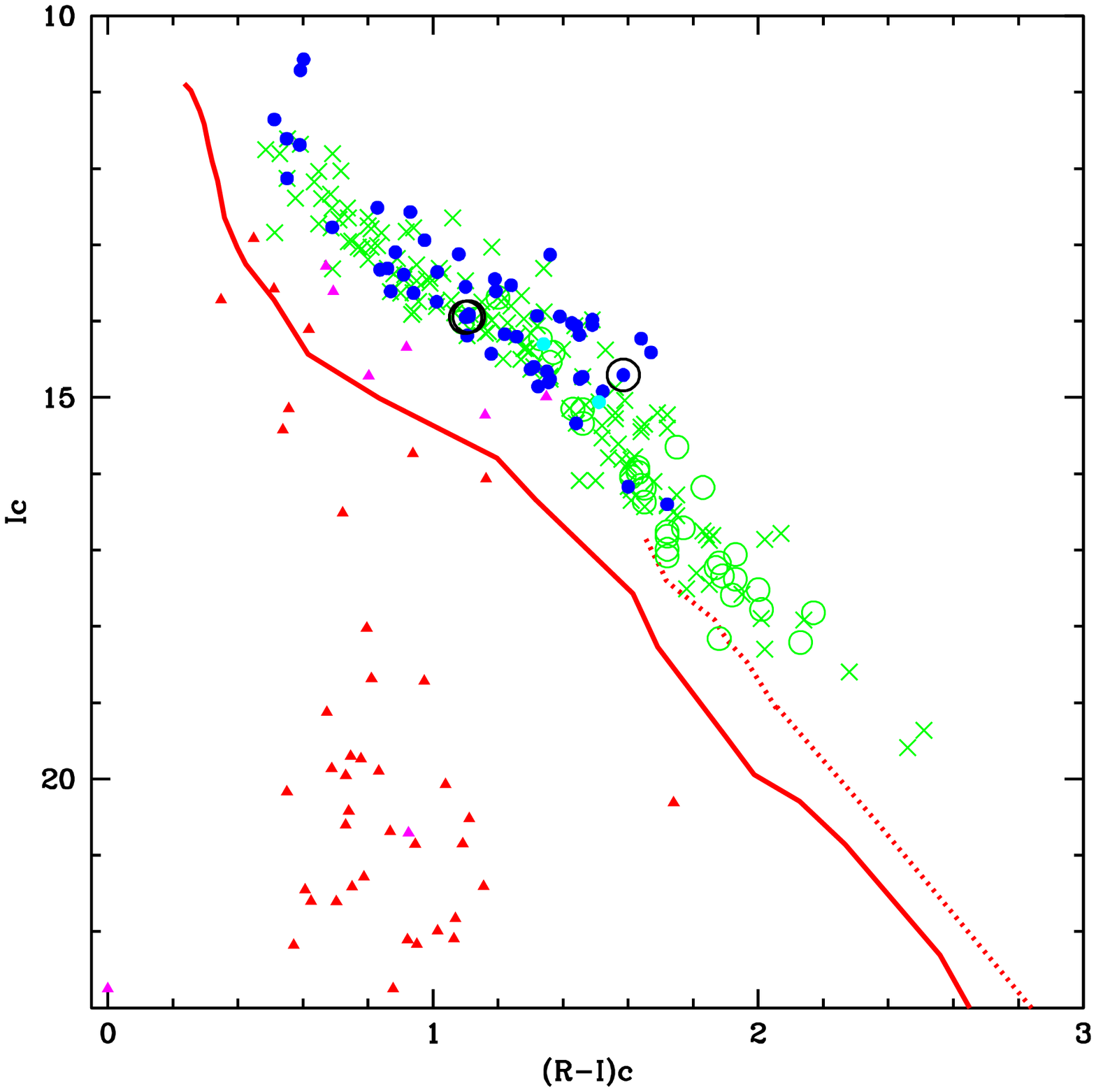}
   \includegraphics[width=7.5cm]{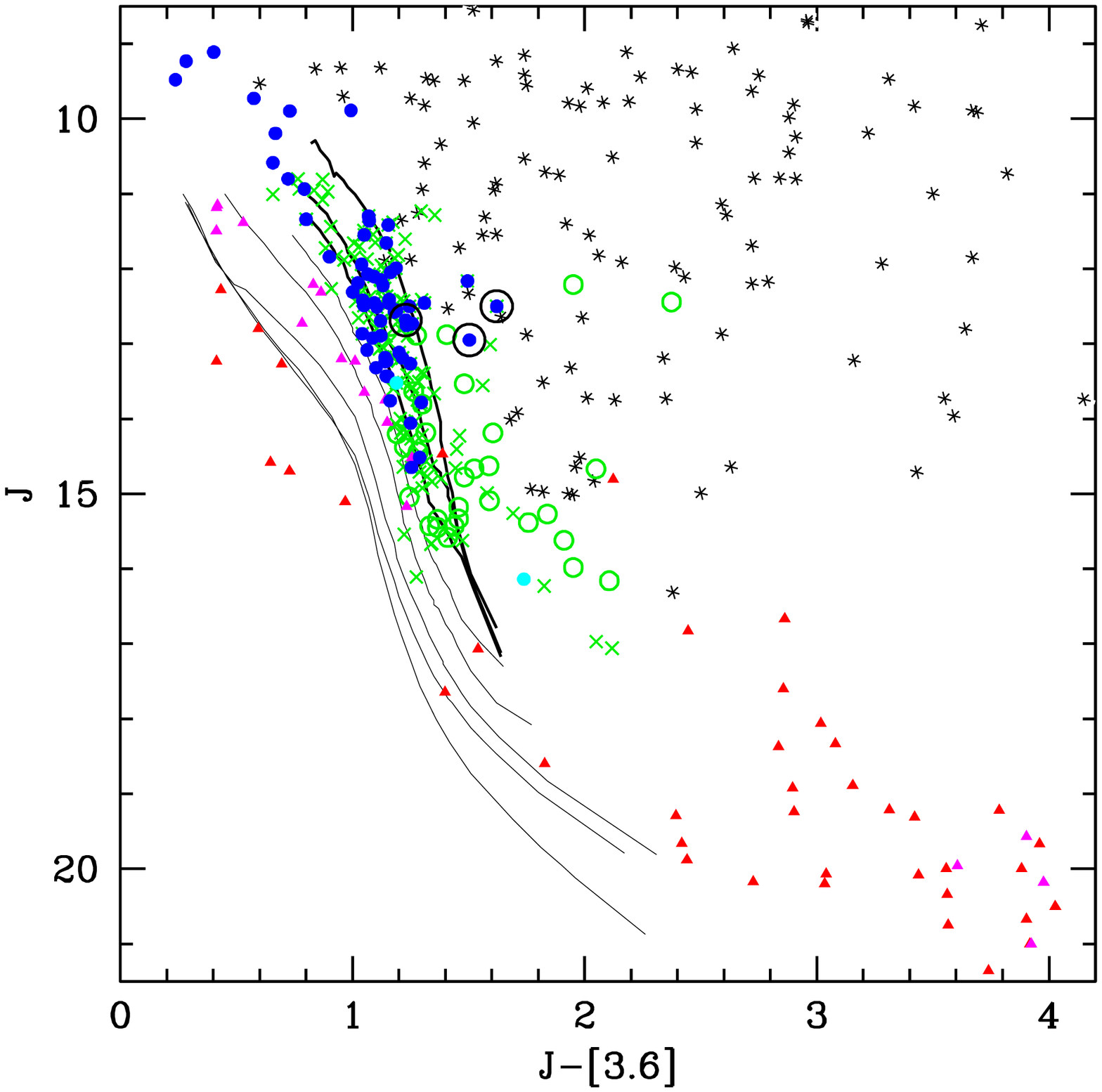}
   \includegraphics[width=7.5cm]{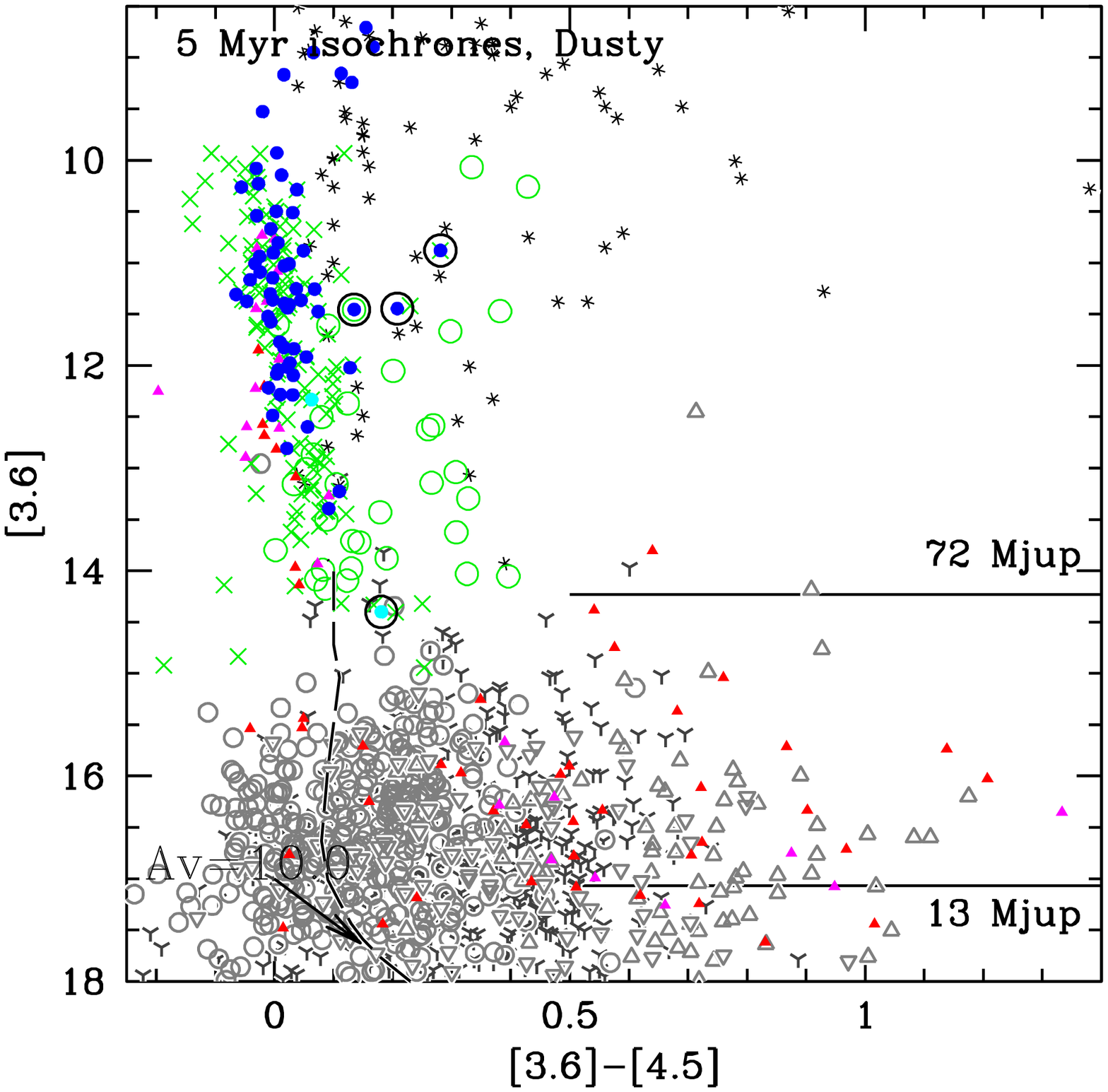}
   \includegraphics[width=7.5cm]{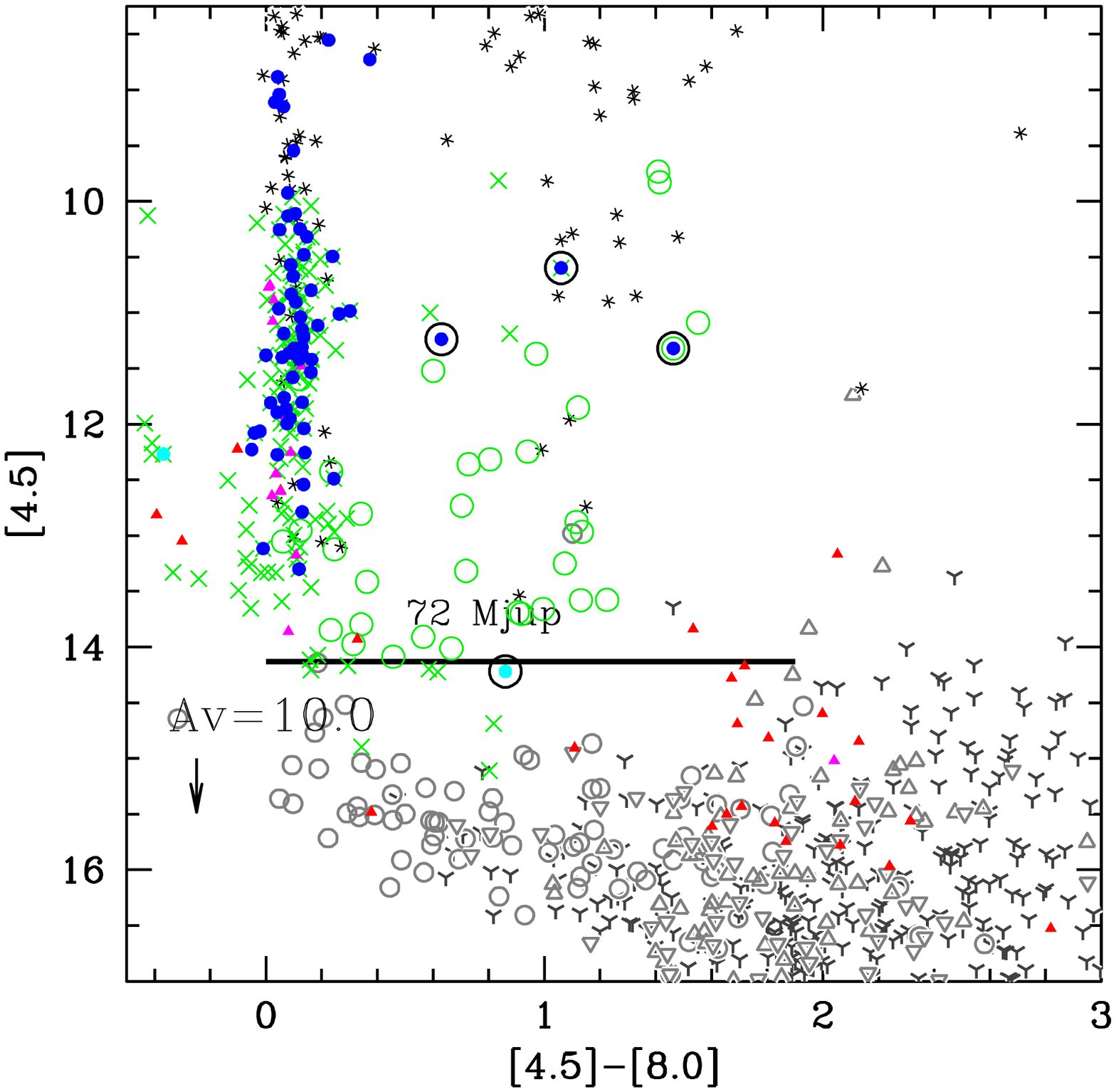}
   \includegraphics[width=7.5cm]{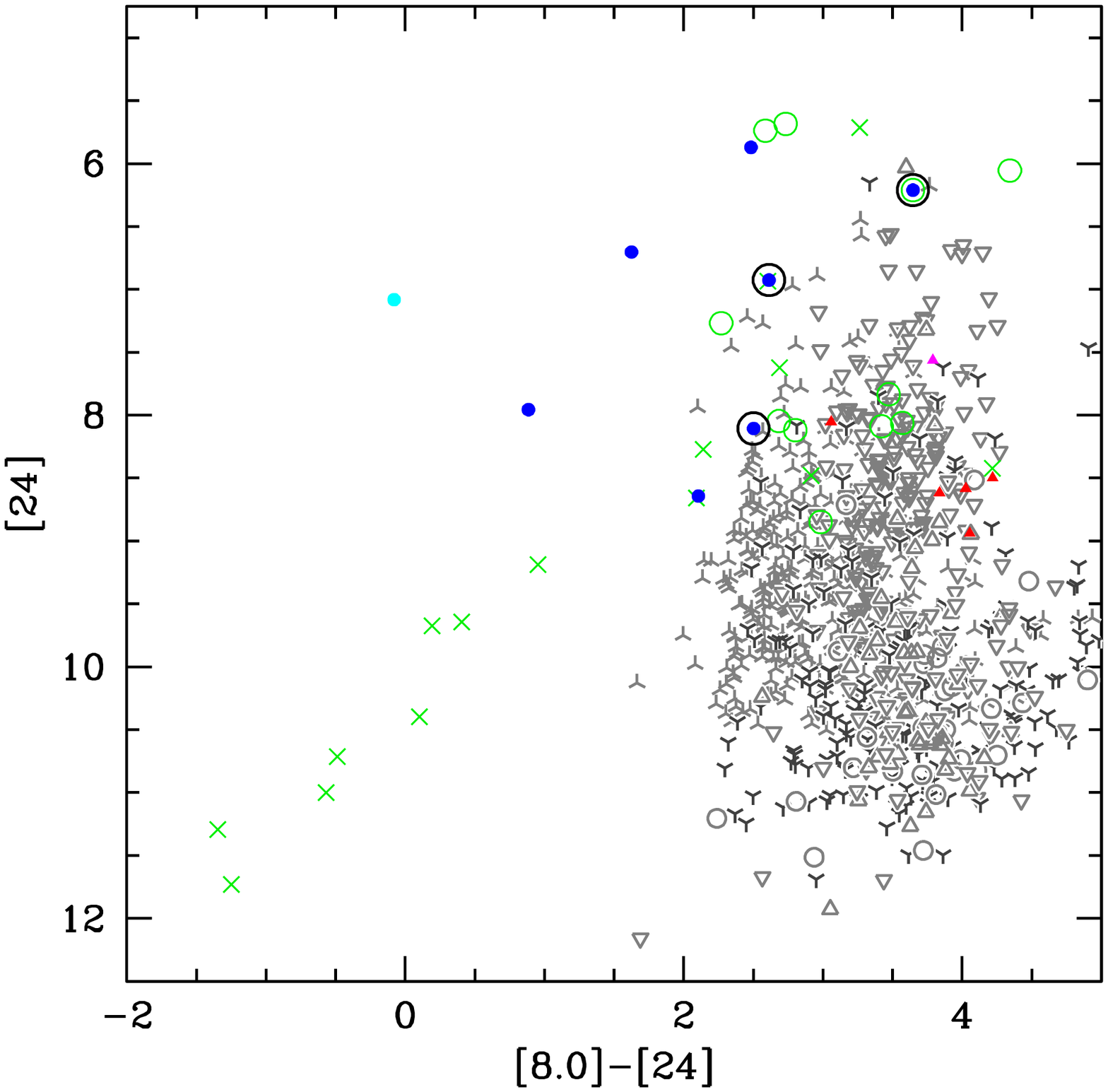}
   \includegraphics[width=7.5cm]{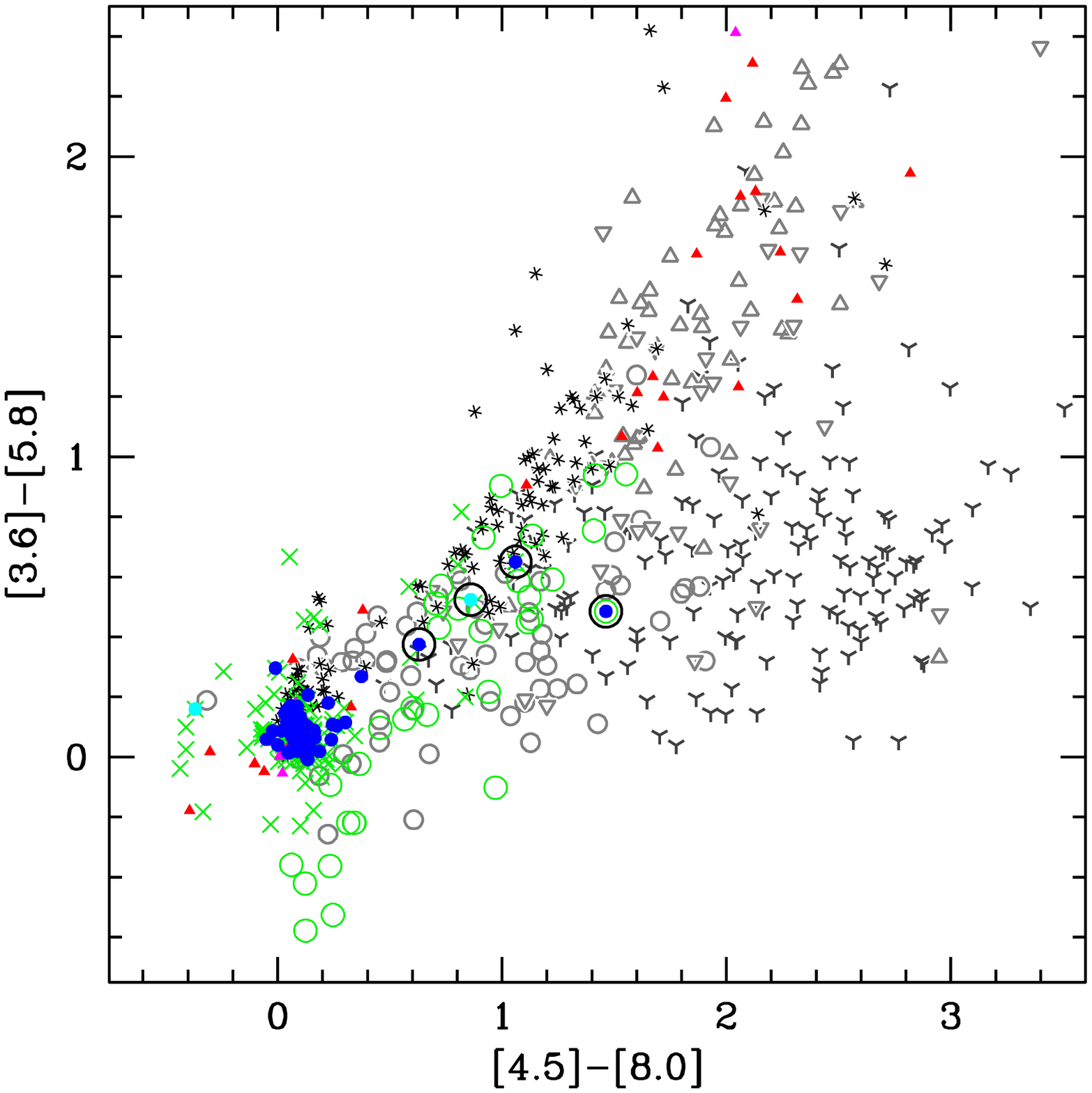}
      \caption{
        Optical, near- and mid-IR Color-Color and Color-Magnitude diagrams of 
        counterparts to X-ray sources in
        the Collinder\,69 cluster  and comparison with other  samples.
        Small solid circles -blue and cyan- represent probable and possible (Y?)
        members of the cluster, whereas solid triangles -red and magenta- correspond to the
        location of probable (NM) and possible (NM?) non-members.
        Class II stellar objects are identified as big open, black circles.
        Previously known members for the cluster appear as green crosses (Class III) or
        open green circles (Class II), and were selected from \cite{Dolan99.1},
        \cite{Barrado07.1} and \cite{Morales08.1}.
        The asterisks represent the location of Taurus known
        members, with photometry coming from \cite{Luhman06.1}. 
        The extragalactic samples were selected from \cite{Sacchi09.1},
        and correspond to the SWIRE/XMM-{\em Newton}/ELAIS-S1 Field with spectroscopic data.
        Up- and down-ward open triangles --light gray-- correspond to the
        AGN1 --QSO1, broad lines-- and AGN2 --obscured QSO, narrow lines, respectively,
        whereas open circles --light gray-- and three-point stars --dark gray--
        represent resolved galaxies and Emission Line Galaxies (which include AGN and star-forming galaxies),
        respectively.
        The first
        color-magnitude diagram includes a 5 Myr isochrone --Dusty--
        from the Lyon group (\cite{Baraffe02.1}). A
        reddening vector also has been included.  
      }
         \label{fig:ccd_cmd}
   \end{figure*}
%

Because our spatial coverage is not complete or because of the completeness limits of our 
optical and mid-IR surveys, we do not have a complete photometric coverage for all 
X-ray sources and we cannot
discuss their nature based on the previous diagrams (Figs.~\ref{fig:qso} and~\ref{fig:HRD_all}).
 Therefore, we now discuss their location in optical and
near-IR color-magnitude and color-color diagrams.

Figure~\ref{fig:ccd_cmd} 
contains six panels with a very complete dataset and, for the sake of clarity, the final classification
for the $164$ sources identified as the origin of the X-ray emission.
We also included several samples for comparison:\\
i) Previously known members of the cluster appear as green crosses (Class III) or
        open green circles (Class II), and were selected from \cite{Dolan99.1},
        \cite{Barrado07.1} and \cite{Morales08.1}.\\
ii)  Known Taurus
        members, with photometry coming from 
        \cite{Luhman06.1}, have been included as asterisks. We included these  stars, 
        younger than Collinder 69 members, in order
        to compare their location in these diagrams with those characteristic 
        for extragalactic sources (see below).
        Bona fide members of the young cluster Collinder\,69 should coincide  with 
        the area where Taurus members are found. \\
iii) Extragalactic sources have also been included, taken from  \cite{Sacchi09.1}.
        Additional samples can be found in \cite{Grazian06.1} (GOODS/MUSICS,
        mainly AGNs), \cite{Hatziminaoglou08.1} (SWIRE/QSO), and \cite{Surace04.1} (SWIRE/ELAIS).
        The extragalactic sources were selected from
        the SWIRE/XMM-Newton/ELAIS-S1 Field and were classified using spectroscopic data.
        Up- and down-ward open triangles --light gray-- correspond to the
        AGN1 --QSO1, broad lines-- and AGN2 --obscured QSO, narrow lines, respectively,
        whereas open circles --light gray-- and three-point stars --dark gray--
        represent resolved galaxies and Emission Line Galaxies (ELG, 
        which include AGN and star-forming galaxies),
        respectively.

We  identified probable and possible cluster non-members based on failures to be located in the area
corresponding to cluster members. Note, though, that in several cases the situation is not clear, 
because  the overlap
of the properties of several of the comparison samples we have used.
Probable and possible non-members are included as red and magenta triangles, whereas possible and probable
cluster members in our final list appear as cyan and blue solid circles, respectively.
Those of the possible and probable members that are Class II are highlighted
by big open, black circles on top of the solid circles.

\subsection{A final list of good cluster candidate members}\label{subsect:members}

After our membership discussion 
we  cataloged 61 objects as probable members (Y) 
 and another five as possible members (Y?),
out of the 164 X-ray sources. They are listed in Tab.~\ref{tab:finalmembers}.

 All probable members have complete IRAC photometry. We classified three of them 
as Class II objects (Classical TTauri stars) based on IRAC data and $58$ of them as 
Class III objects (weak-line TTauri stars). Sixteen  of these probable members are new, 
whereas 30 were listed in  \cite{Barrado04.1} and another 15 were identified by \cite{Dolan99.1}.

Regarding the possible members (five in total), we have one Class II (C69-X-w028),
 another Class III  (C69-X-e096c) and another three
(C69-X-e013, C69-X-e041, C69-X-e042) lack data at  some of the four IRAC bands. 
However, their SED resembles a blackbody (i.e., they should be Class III),
although one of them (C69-X-e041)
 has an obvious excess at 24 microns, which suggests it has a transition disk.
Of these five possible members, C69-X-e096c is LOri058  (\cite{Barrado04.1})
and C69-X-e041  has been identified
previously as DM071  (\cite{Dolan99.1}), which again reinforces the validity of our selection procedure.

Another 86 X-ray  sources were
 classified as non-members based on the different criteria described above,
 and another 12 do not have enough data  to be discussed, 
which indicates that they  are probably
 heavily extincted and that  their nature is probably extragalactic (see next section). 
None of these 98 (86 non-members plus 12 without data) sources have
been listed either by \cite{Dolan99.1} or \cite{Barrado04.1}  as possible members, which provides
additional support to our classification scheme.

\subsubsection{New candidates for  visual binaries}\label{subsubsect:binaries}

Three X-ray sources classified as members have multiple counterparts within the XMM-{\em Newton}
 pointing uncertainty. 
One of them, LOri025 (C69-X-e011c), is a confirmed member of the cluster for which we have now discovered
a possible faint companion (C69-X-e011e, a possible member).
Another probable member (C69-X-e104c) has a possible companion nearby, which fulfills some membership
criteria (C69-X-e104e), whereas the probable member  C69-X-w001c  might have a companion north of it, but
it is  too close  and fainter compared to the first one (the central source)  to be able to say anything more.

Finally,  LOri\,113 (C69-X-w032w), a cluster member from \cite{Barrado04.1, Barrado07.1},
 is relatively close to the X-ray source C69-X-w032c, which has been
classified as a  probable non-member.
However, we do not believe they correspond to the same source and
assumed that this X-ray source does not belong to the Collinder\,69 cluster,
 while LOri\,113 is considered an X-ray undetected Collinder\,69 member.

\subsubsection{Cluster members not detected in X-rays}\label{subsubsect:membersnotdetected}

Within the XMM-{\em Newton} pointings, there are a  number of probable and possible members of 
Collinder\,69 that are not  detected as X-ray sources. There is one from
\cite{Dolan99.1}, another 10 from \cite{Morales08.1},  and 57 objects from \cite{Barrado04.1}.
This is not surprising since most of the undetected candidate members from  \cite{Barrado04.1}
 are near or  within the 
substellar domain (see Figs.~\ref{fig:histomassA} and ~\ref{fig:histomassB})
 and therefore they are expected to be faint in X-rays. 
The mass distribution of this subsample can be found in Fig.~\ref{fig:histomassB},
as thick long dash-point line -- dark gray.
We  derived upper limits for all of them, with the goal of producing an unbiased X-ray luminosity function.
Their X-ray data are listed in Tabs.~\ref{tab:stelpar_ulE} and \ref{tab:stelpar_ulW} 
 together with their properties derived from optical/IR
observations. 

There are few cluster members with masses around 1 $M_\odot$ that have not been detected in X-ray 
(see Figs.~\ref{fig:histomassB} and ~\ref{fig:lglxlbol_mass}). This is not surprising, since
they have Log   $L_{\rm x}/L_{\rm bol}$$\sim$-4 dex (upper limits). As a comparison, \cite{Briggs07.1} have measured 
values for Taurus members (younger and closer) with similar mass at Log   $L_{\rm x}/L_{\rm bol}$$\sim$-4.5 dex.
Therefore, the explanation is that our XMM-{\em Newton}
 observations are not deep enough to detect them, not
because they lack X-ray emission.

 For the calculation of the upper limits to the X-ray luminosities we  assumed an isothermal
absorbed spectrum with column density and temperature corresponding to the mean values measured from
the spectral fits of detected sources (see Sect.~\ref{subsect:lx}).
Under these assumptions the limiting flux of our observations is 
$f_{\rm lim} \sim 5 \cdot 10^{-15}\,{\rm erg/cm^2/s}$ or $\log{L_{\rm x}}\,{\rm [erg/s]} \sim 29$. 
The upper limits we derive for $L_{\rm x} > 10^{29}$\,erg/s (see Tabs.~\ref{tab:stelpar_ulE} and \ref{tab:stelpar_ulW})
are consistent with this sensitivity limit.

\subsection{The nature of the X-ray sources not selected as cluster members}\label{subsect:nonmembers}

   \begin{figure}
   \centering
   \includegraphics[width=9cm]{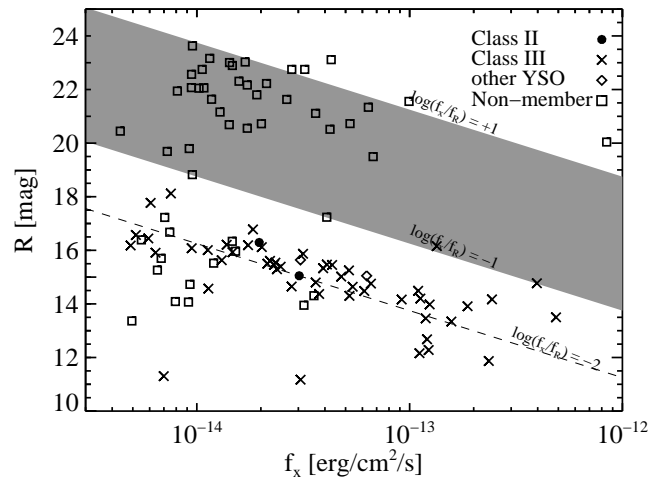}
      \caption{
$R$ magnitude vs. X-ray flux for all XMM-{\em Newton} sources with $R$-band counterpart.
 The plotting symbols are explained in the top right of the figure.  
The gray area is the typical location of AGNs, and the dashed line marks the empirical
upper envelope for quiescent galaxies;  see e.g. \cite{Barger03.1}.
}
         \label{fig:fxfr}
   \end{figure}

According to our selection, the fraction of cluster members among all detected X-ray sources
is $\sim 40$\,\% (66/164). In particular, we identified $48$ of $112$ sources in C69\,E and $18$ of $52$
sources in C69\,W as Collinder\,69 member. As a rough estimate for the extragalactic contamination
of the XMM-{\em Newton}
source lists we consider the $\log{N} - \log{S}$ distributions shown \cite[e.g. by][]{Baldi02.1}.
For our flux limit of $f_{\rm x} \sim 5 \cdot 10^{-15}\,{\rm erg/cm^2/s}$ about $40$ extragalactic
sources are expected within each XMM-{\em Newton} pointing.
 This compares well to the $64$ and $34$
 X-ray sources classified as non-cluster members 
in C69\,E and~W, respectively.

The nature of the X-ray sources that are not Collinder\,69 members is examined in Fig.~\ref{fig:fxfr}.
 The location of extragalactic sources
in this diagram was studied e.g. by \cite{Barger03.1}. 
Among the $164$ X-ray sources, $102$ have measured $R$ magnitude. 
The typical area for AGN  is highlighted and coincides with the bulk of the optically faint X-ray
emitters. None of them has been classified as a cluster member by our selection procedure. 
Quiescent galaxies are typically located below the dashed line in an area that overlaps with that
of the cluster members and contains the majority of the remaining non-members.

\section{Properties of Collinder\,69  members}\label{sect:properties}

\subsection{Spectral energy distributions}\label{sect:seds}

   \begin{figure}
   \centering
   \includegraphics[width=9cm]{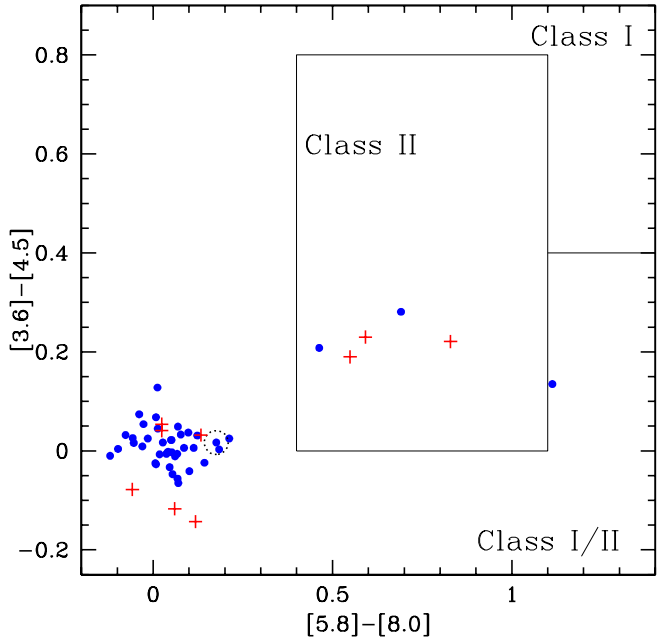}
      \caption{
        {\em Spitzer}/IRAC color-color diagram for Collinder\,69 for all X-ray sources with masses in the
        mass range 0.3--1.2 $M_\odot$.
        Solid circles -blue- represent probable members identified in the XMM-{\em Newton} survey.
        The large, dotted circle (in black) represent one of the two objects with excess at 24 micron
        (the second one has no IRAC data).
        Previously known members from  \cite{Dolan99.1},
        \cite{Barrado07.1} and \cite{Morales08.1} within the
        XMM-{\em Newton} field-of-view and with no X-ray detections are included with 
        plus --red-- symbols.
}
         \label{fig:CCD_IRAC}
   \end{figure}

Our multi-wavelength  photometry, which covers a range  from the $B$ ($V$) band at 0.3 (0.5 micron)
up to 24 micron, allows us to construct  SEDs for our $164$ 
 X-ray sources (more precisely the $205$ original possible counterparts, although in this
section we will restrict the discussion to the selected probable and possible members). 
A visual inspection  allows us an initial classification as Class I --increasing slope in the SED,
II --an obvious IR excess- and III --black-body SED-- for each X-ray source
 (or for the several identifications, when there are several optical or IR possible counterparts),
keeping in mind that this is tentative and that some among them could be extragalactic. 
If galactic, a Class I or II object is a good candidate cluster member.
This initial classification based on the visual inspection
 has been confirmed by the classification based on
an IRAC color-color diagram. 
We list a tag  in Col. \#10 of Tab.~\ref{tab:finalmembers} 
indicating whether each object (probable or possible member)
 has an SED corresponding to a blackbody
(BB or Class III), Class II or Class I.

For those sources with at least three {\em Spitzer}/IRAC datapoints,
 we also computed the slope and used the criterion
 by \cite{Lada06.1}  to establish whether they have a disk, and whether it is optically
 thin or thick. Both the disk type and slope appear listed in Tab.~\ref{tab:finalmembers}
(Cols. \#11 and \#12). They can be compared with the results based on
 the IRAC CCD listed in Col. \#10. 
We emphasize that the presence of a disk is a good membership indicator for a stellar source.

\subsection{Bolometric luminosity, effective temperature, and mass for cluster members}\label{sect:vosa}

As already mentioned in Sect.\,\ref{subsect:HRD}, we made 
of the  VOSA tool (\cite{Bayo08.1})
to derive effective temperatures and bolometric luminosities for the XMM-{\em Newton} sample of X-ray
sources. For those among them classified as probable and possible members (Tab.~\ref{tab:finalmembers}) we 
also
 individual masses and effective temperatures using a 5 Myr isochrone constructed using
data from  \cite{Siess00.1} --masses above 0.1 $M_\odot$-- and \cite{Chabrier00.1} --masses below 
this value, by  interpolating the bolometric luminosity. 

The same procedure was applied to several samples used throughout the paper (\cite{Dolan99.1},
\cite{Dolan02.1}, \cite{Barrado04.1}, \cite{Morales08.1}). A comparison  of the mass distributions
can be found in Figs.~\ref{fig:histomassA} and  ~\ref{fig:histomassB}.

\subsection{Mass distribution and comparison with other  subsamples}\label{subsect:masses}

Figure ~\ref{fig:histomassB} shows the mass distribution of the cluster members in the X-ray sample
described in Sect.~\ref{subsect:members}
 (both the detected --dashed, red line-- and the non-detected --thick dashed-dotted, gray line-- Collinder\,69 members).
This graph suggests that the X-ray survey is complete down to $\sim$0.3 $M_\odot$ 
(or even as high as 0.5 $M_\odot$), 
although a couple of previously known members with masses around 0.15 $M_\odot$ have been detected.
Note that our X-ray flux limit of $\log{L_{\rm x}} \sim 29$ erg/s 
corresponds to the lower $1$\,sigma boundary at $0.3\,M_\odot$ for the $L_{\rm x} - M$ relation 
derived for pre-MS stars in the Orion Nebula Cluster \citep{Preibisch05.1}. Therefore, if the X-ray
emission of LOSFR stars is similar to those in the ONC, most but not all 
$0.3\,M_\odot$ stars should have been detected confirming 
qualitatively the above estimate for the XILO mass limit. 
This is not in contradiction with the detection of a few less massive stars given the various
uncertainties of the X-ray flux limit and the large spread of X-ray luminosities for a given mass.  
Only the very lowest
mass cluster member detected in XILO (C69-X-e028 with $M=0.035\,M_\odot$) may not
be plausible (see Sect.~\ref{subsect:lx}).

The mass distribution of cluster members with X-ray emission (dashed, red line in Fig.~\ref{fig:histomassB}) 
 also has a bimodal distribution, with a very prominent
peak around 0.6 $M_\odot$ and a secondary peak for higher masses (1.75-2.35 $M_\odot$), with a low  total number of 
objects, including two possible members with masses around 3  $M_\odot$. This peak actually resembles what we 
found with a different technique (mid-IR excesses due to a circumstellar disk) with our {\em Spitzer}/IRAC 
mapping (panel 3c). 

A word of caution is required for the mass distribution for high-mass end in the sample (solar-type stars
and larger masses).
 Among the 16 probable and possible members with X-ray emission and mass
larger than 1  $M_\odot$, there are seven whose photometric data are incomplete, 
and who lack any optical information
(C69-X-e009, C69-X-e013, C69-X-e023, C69-X-e042, C69-X-w001, C69-X-w014, C69-X-w020).
Therefore, the fits we carried out with VOSA used only the Rayleigh-Jeans part of the SED and 
not the portion with the highest fluxes.
This means that the bolometric luminosities might be biased and that the masses obtained from 
them are shifted (lower values)
from the real values by an amount enough to produce the distribution we derived for high masses. 
We note, however, that five among them are probable members and that the offset of the mass could not 
be larger than few tenths of a solar mass, i.e., moving the histogram one or two bins at most for the 
high mass end. Indeed, we performed a 
search with VOSA for additional photometry in VO-compliant databases for those objects, fitting 
again bolometric luminosities and effective temperatures, and deriving masses using a 5 Myr isochrone. 
As expected, the final individual mass does not change
 by a significant amount, because it is almost identical for four of them. For the other three, 
the mass estimate moves from 
1.23 to  1.02    $M_\odot$ (C69-X-e023) , from 3.17  to 2.92   $M_\odot$  (C69-X-e042),
 and from  3.04 to 2.94   $M_\odot$  (C69-X-e013).

Masses around $\sim 2-3\,M_\odot$ are critical for dynamo action because the 
convection zone becomes very thin and low chromospheric and coronal activity is expected. 
The comparison of the most massive stars in the sample
($M \geq 1.8\,M_\odot$) with  the evolutionary models of \cite{Siess00.1}
shows  that only two of them,  those with $M \sim 3\,M_\odot$
(C69-X-e042 and C69-X-e013), have very shallow convection zones. 
These two stars indeed have by far the lowest $L_{\rm x}/L_{\rm bol}$ level of all cluster members
(see Sect.~\ref{subsect:lx}). 

The  combination of the histograms  (Figs.~\ref{fig:histomassA} and ~\ref{fig:histomassB}) 
corresponding to X-ray detection (dashed, red line)
 and non-detections (dashed-dotted line, in gray) may suggest 
that some  solar-type cluster members, are missing, and that   the existing surveys 
leave a gap around 1  $M_\odot$.
This
could be owing
 to the lack of completeness for the bright and faints ends of overlapping
surveys.
 Again, the small number of objects makes it difficult to establish this
  fact with a high degree of certainty.
 However, it is clear that previous  photometric surveys in the optical missed some cluster
members with masses around 0.3 $M_\odot$ (specifically \cite{Dolan99.1,Barrado04.1}).  
\cite{Dolan99.1} 
did not reach this mass and since they were looking for strong emission in 
 H$\alpha$ (i.e., stars with accretion and/or stellar activity), the X-ray search is able to add new, 
fainter objects.

\subsection{Spatial distribution for members with X-ray emission}\label{subsect:spatial}

We note that there is a dichotomy in the distribution of X-ray sources (already obvious in 
Fig.~\ref{fig:xmm}).
Forty-eight 
 members are located east of the central star \lori,
whereas there are 20 west of this O8III star.
This 
was already discussed in \cite{Barrado07.1} in the
context of the distribution of members with circumstellar disks.
In any event, our final cluster members have been displayed in Fig.~\ref{fig:spatial}  as solid circles
(blue for probable members, and  cyan for possible members). For the sake of homogeneity and clarity,
we 
only represented cluster members in the mass range 0.3--1.2 $M_\odot$. Non-detected cluster members
in the same mass range  
were also included as solid triangles --dark gray. 

One possible explanation would be the different effective  exposure times
 for both XMM-{\em Newton} pointings (37 and 28 kilosec for east and west, respectively). 
Yet, this explanation seems unlikely or  to be at least only partial, because 
the  average for actual measurements and detection limits
of $L_{\rm x}$ and    $L_{\rm x}/L_{\rm bol}$ are very similar in both pointings. 
Averages for the masses are slightly different (being  0.17  $M_\odot$ higher for the western pointing),
 so it seems that the reason
 should lie on the actual distribution of cluster members.

Indeed, this dichotomy might be related to the original structure
of the parental cloud. 
The large-scale structure of the 
 Lambda Orionis Complex (the Head of Orion and related structures
--see for instance a IRAS image\footnote{http://irsa.ipac.caltech.edu/data/IRIS/} of the region),
and as discussed in \cite{Dolan02.1}, suggests that some sort of large filament-like structure 
might be present, which  was affected  by a supernova explosion or, more likely, 
the strong winds from the O and early B stars located in the region.

   \setcounter{figure}{10}
   \begin{figure}
   \centering
   \includegraphics[width=9cm]{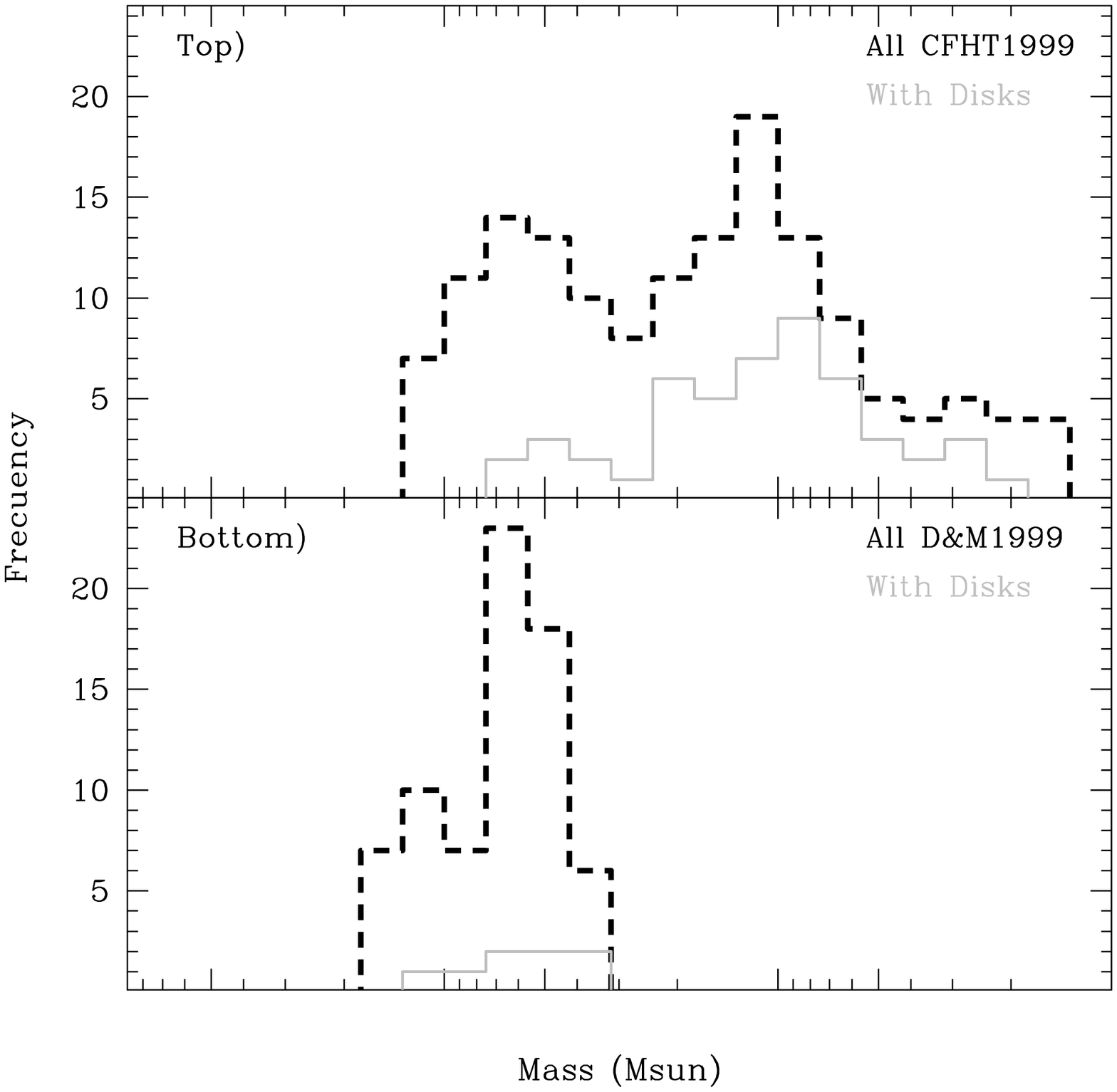}
    \caption{
       Histograms for different samples --dashed, black lines with and without disks-- and subsamples only with disks
        --solid, gray lines--
       as described in Sect.\,\ref{subsect:masses}
       {\bf Subpanel at the top.-} 
       Probable and possible members from \cite{Barrado04.1}  and \cite{Barrado07.1}, with
       an initial selection based on optical photometry. 
       {\bf Subpanel at the  bottom.-} 
       Members listed by \cite{Dolan99.1}, originally selected from optical photometry (including 
       narrow-band H$\alpha$ -i.e., active stars or with accretion),
       and confirmed based on high-resolution spectroscopy. 
}
    \label{fig:histomassdiskA}
   \end{figure}

   \setcounter{figure}{11}
   \begin{figure}
   \centering
   \includegraphics[width=9cm]{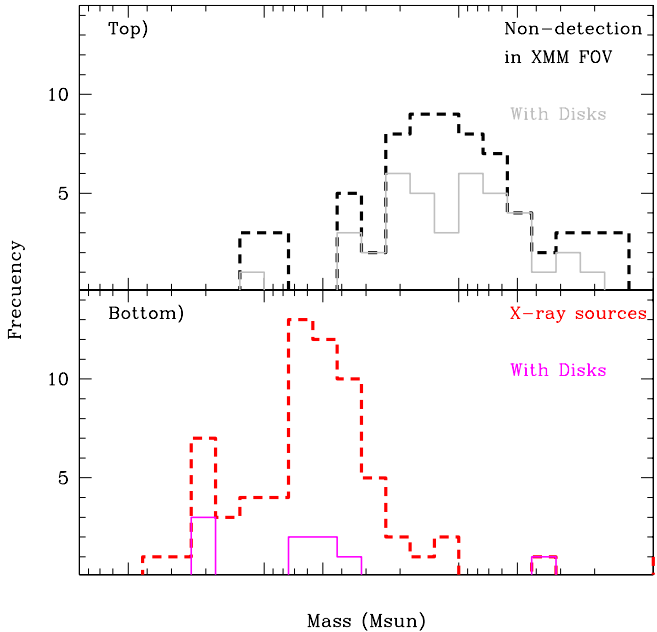}
    \caption{
       Histograms for different samples --dashed, black lines with and with disks-- and subsamples only with disks
        --solid, gray lines--
       as described in Sect.\,\ref{subsect:masses}. Data in the XMM-{\em Newton} field-of-view. 
       {\bf Subpanel at the top).-} 
       In black-dashed and gray-solid the C69 members with 
       X-ray upper limits, corresponding to the whole subsample and those members with disks.
       {\bf Subpanel at the bottom).-} 
       In red-dashed and magenta-solid the C69 members with 
       X-ray detections, also corresponding to the whole subsample and those members with disks.
}
    \label{fig:histomassdiskB}
   \end{figure}

\subsection{Disks in the X-ray sample and comparison with other subsamples}\label{subsect:disks}

For those X-ray sources with complete IRAC data (four bands),
there are only four X-ray sources (three probable and one possible member)
 that can be classified as Class II objects (or Classical TTauri stars),
 whereas there are 59 classified as Class III (for instance, a subsample with 0.3--1.2 $M_\odot$
 in Fig.~\ref{fig:CCD_IRAC}, see next subsection).

 Among the Class III subsample,  four have thin disks, according to the slope of
the IRAC data (C69-X-e009 and C69-X-e012)
or as revealed by the presence of the flux at 24 micron 
(more precisely, transition disks, as in 
the case of C69-X-e030 and C69-X-w014).
We do not have enough information for
another three cluster members regarding the presence or absence of a circumstellar disk.
However, the SED of C69-X-e041 (DM071), which is located outside of the IRAC mapping, suggests
the presence of  a circumstellar disk because of its excess at 24 micron.

Considering the whole sample of $63$ probable and possible
cluster members with four-band IRAC photometry 
in the XMM-{\em Newton} fields, 
the minimum disk fraction is 6.1\% (Class II versus total) and 10.6\% (disk versus total), respectively.
 Our sample is incomplete at low and high masses, and in any case we want to verify whether 
there is a dependence with mass. Therefore, we present in Tab.~\ref{tab:diskfraction} 
the disk fractions for different mass ranges.  
Three of the four Class\,II objects have $M < 0.5\,M_\odot$. 
The lower fraction of disks among higher mass stars is consistent with results in 
other SFRs (see, for instance, \cite{Bayo09.1}; or \cite{Lada06.1} for IC348 or 
\cite{Scholz07.1} in the case of Upper Sco).

\subsubsection{The mass range 0.3--1.2 $M_\odot$:  comparison between homogeneous  subsamples}\label{subsubsect:diskmass1}

We  compared the disk fraction of different samples using the mass range
 0.3--1.2 $M_\odot$, common to all
of them and without the presence of any obvious bias (for instance, owing to completeness limits).

In Fig.~\ref{fig:CCD_IRAC} we show Collinder\,69 cluster members with masses
 in this range (0.3--1.2 $M_\odot$)
in the {\em Spitzer}/IRAC color-color diagram.
The diagram includes the X-ray sources (solid circles, in blue) 
and non-detections within the XMM-{\em Newton} 
pointings (plus symbols, in red).

In the case of the  \cite{Dolan99.1} sample (originally selected based on H$\alpha$ photometry),
 which includes $14$ members that are not listed in \cite{Barrado04.1}, 
there are two Class\,II objects plus one  
transition disk. This makes a disk fraction of $21.4$\,\%. 
If the only object without X-ray detection is removed (DM048, a Class II),
 the fraction goes down to
 15.4\%. Note, however, the low numbers, involving few stars. In an analogous way, 
for the CFHT1999 sample 
(\cite{Barrado04.1}, \cite{Barrado07.1}), which contains 33 probable and possible members, 
the disk fraction 
is 3.0\% for the undifferentiated sample, 3.7\% for the members (27) 
with X-ray emission, and 0\% for the upper
 limits  (six members).

Finally, for the sample extracted with the XMM-{\em Newton} data (43 sources), 
three are Class II and another two have thin/transition disks (disk fraction of
 7.0--11.6\%). 
Among the nine  members with X-ray upper limits, there are three with
 disks  (all of them Class II, disk fraction of  33\%). 
In total, regardless the X-ray detection, the disk fraction in the 0.3--1.2 $M_\odot$ is 16.7\%.

\subsubsection{Disk fractions and the  mass distribution}\label{subsubsect:diskmassdistr}

In order to clarify the issue of disk fraction in each subsample, we  represented the
star number versus the mass in another set of histograms 
(see Figs.~\ref{fig:histomassdiskA} and ~\ref{fig:histomassdiskB}). 
The top panel  compares the distribution of cluster members extracted using optical photometry,
 from \cite{Barrado04.1},
with a subsample that has mid-infrared excesses, from \cite{Barrado07.1}. A significant fraction do have 
circumstellar disks, but there is a strong dependency on mass, and the disk frequency is higher for low-mass 
members, peaking at 0.1 $M_\odot$. This  seems to reflect a larger time 
scale for the disk dissipation at the end of the
 Main Sequence, as already suggested by different authors (see, for instance, \cite{BarradoMartin03.1}).
 Figure~\ref{fig:histomassdiskB} displays more massive members,
 as selected by \cite{Dolan99.1}, based on optical photometry,
 including H$\alpha$ narrow-band photometry, and confirmed with 
high-resolution optical spectroscopy and properties such as 
radial velocity and the detection of the lithium doublet, a signpost of youth, at 6708  \AA. In this particular case,
the method seems to be very efficient for  picking up Class III cluster members.
 That is, they are  diskless, with high activity level
probably due to lack of disk-locking. This might be because disks have been already dissipated for 
solar-type stars or because of  an initial different distribution in the angular momentum 
(higher rotation rates). As shown by \cite{Bouvier97.1}, 
rotation increases notably during the first tens of million years. In particular, from
 $\sim$1 to 5 Myr (Taurus and Collinder 69 ages, respectively). 

Finally,   Fig.~\ref{fig:histomassdiskB} contains four different distributions in two panels:
  two corresponding to cluster  members with X-ray detections
 (dashed-red for the whole sample and solid-magenta for those  with  mid-IR excesses),
 and another two for cluster members within the XMM-{\em Newton} field-of-view with only upper limits 
 (dashed-black for the whole sample and solid-gray for those with mid-IR excesses).
Clearly, both sets are very different, peaking at $\sim$0.6 and $\sim$0.15 $M_\odot$, respectively. This is probably
because of  the sensitivity limit of our XMM-{\em Newton} imaging, but it may also be 
related to intrinsic properties of very low-mass stars and brown dwarfs,
 in the sense that different
empirical studies have found a correlation between X-ray activity and mass (the lower the mass, 
the lower the  coronal emission). 
And, again, the disk fractions are very different:  almost negligible for the X-ray 
sample, close to half of them for the X-ray non-detected members. 
Naively, one may think that
this is a consequence of the technique we are using, 
since X-ray data are very efficient for  selecting Class\,III stars  known to have strong  
coronal activity, and the low-mass sample is dominated by  less active 
Class II objects (Classical TTauri). However, the average difference between the X-ray luminosities
of Class\,II and Class\,III objects is a factor two at most \citep[e.g.][]{Preibisch05.1, Telleschi07.1}, 
and cannot explain a strong bias of the X-ray selection against diskless stars.
We note that we do not have rotational velocities or rotational periods, but 
the low-disk fraction in the X-ray  detected sample (mostly Class\,III) 
is likely not directly connected to rotation  because
\cite{Preibisch05.1} and \cite{Briggs07.1} have found that 
at least for $\sim$1 Myr stellar associations such as 
 Orion and Taurus there is no rotation/activity connection in pre-MS stars.

Most members from \cite{Dolan99.1} have been detected in X-ray (not surprising, since the were 
selected based on their are chromospheric activity or accretion rate),  and about a quarter of them have disks.
The disk fraction of the X-ray sample is much lower than the \cite{Dolan99.1} (a fourth), and
almost a third of the probable and possible Collinder\,69 members with no X-ray detections have a
 circumstellar disk.

\subsubsection{Disks for higher masses}\label{subsubsect:diskmass2}

Our sample of X-ray sources does not contain any Class II object with mass $M>1.2 M_\odot$ 
(16 probable and possible cluster members). 
However, two probable members
 have thin disks and another one presents excess at 24 micron, which is a characteristic
of a transition disk. Nine are diskless and another two  do not have enough information. Therefore,
 the disk fraction
seems to be around 25\%. However, these disks are in general more evolved
 than those found for the mass range 0.3--1.2  $M_\odot$, suggesting again a different time scale 
for disk dissipation, which is faster for more massive objects.

\subsection{The completeness of the Collinder 69 census}\label{subsect:completecensus}

As discussed above (see also Figs.~\ref{fig:histomassdiskA} and  ~\ref{fig:histomassdiskB}), 
most Class\,II sources --whether detected  in X-rays or not-- seem to be of very low-mass.
The disk fraction for the higher-mass stars ($M$$>$0.3-0.5 $M_\odot$) is very low.
 However, the disk fraction for masses below 0.3  $M_\odot$ is much higher. This dichotomy
might  mean that a
very large number of very low-mass stars and brown dwarfs 
without disks is still undiscovered and that our X-ray observations were not
sensitive enough for them. As a consequence, the derived cluster 
mass function would be very incomplete.

According to \cite{Morales08.1} and \cite{Bayo09.1}, the disk fraction among very low-mass stars and 
brown dwarfs in Collinder 69 is higher than for more massive objects. This 
has been observed in other clusters (like Upper Sco, \cite{Scholz07.1} and \cite{Bouy07};
 or see the general comparison in  \cite{Barrado03.1}).
This can be interpreted  as proof that low-mass objects have  disks
 that last longer. 

Two questions have to be considered: Most of our low-mass members come from the CFHT1999 optical survey,
which should be  independent of the presence of disks and the level of X-ray emission. 
In principle, only few underluminous members, such as edge-on disks 
(see the case of LS-CrA 1 in \cite{Fernandez01}, \cite{Barrado04.2}; or Par-Lup-3-4, 
\cite{Huelamo10.1}) may have been overlooked.
 On the other hand, the  initial mass function 
has been 
derived by several authors (\cite{Barrado04.1}; \cite{Bayo09.1}),  even covering a very large mass range
(from 20 to 0.02 $M_\odot$, \cite{Barrado05}) and it seems canonical. 
In recent comparisons with a large number of associations,
Bouvier (2009, private communication) has shown that the log-form is very similar for all of them. Therefore,
the completeness of our census of Collinder 69 members should be located at 
very low masses, 
well below the substellar limit.

In any event, this discussion about the completeness of the Collinder 69  census does not
affect the present analysis of the X-ray properties because  the mass range covered does not reach
below 0.3 $M_\odot$.
A  discussion about the Collinder 69 census will be presented 
in Morales-Calder\'on et al. (2010)   and Bayo et al. (2010).


\subsection{X-ray hardness ratios}\label{subsect:select_hrs}

   \begin{figure}
   \centering
   \includegraphics[width=9cm]{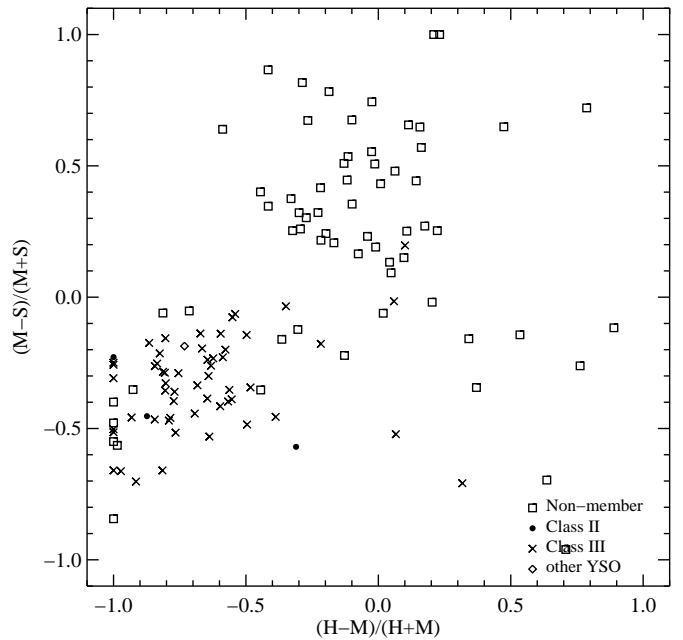}
      \caption{X-ray hardness ratios for sources detected with EPIC/pn. Collinder\,69
members, possible members, and non-members are distinguished by circles of different colors.
}
         \label{fig:hardnessratios}
   \end{figure}

   \begin{figure}
   \centering
   \includegraphics[width=9cm]{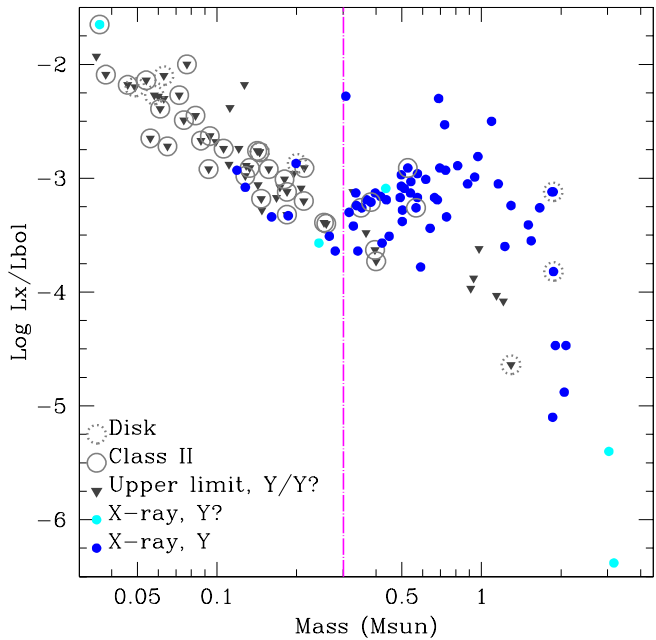}
      \caption{
       Fractional X-ray luminosity as a function of mass for Collinder\,69 members
       within the field-of-view of the two XMM-{\em Newton} observations.
       X-ray detections are represented as solid circles (blue and cyan for probable 
      and possible members,
       respectively). Cluster members without X-ray detection appear as solid --dark gray-- triangles. 
       Class II and class III sources with thick or thin/transition disks, as defined by IRAC photometry,
       are highlighted with large solid and dotted open circles, respectively. 
       The vertical dotted and long-dashed --magenta-- line locates the completeness limit (in mass) of 
       the X-ray observations.
}
         \label{fig:lglxlbol_mass}
   \end{figure}

We examined the X-ray hardness ratios for all $124$ X-ray sources detected on EPIC/pn.
 In principle, an analogous analysis can be done for EPIC/MOS hardness ratios. However, 
the EPIC/pn detections comprise  more than $75$\,\% of the X-ray sources  and not much
information would be added by considering the remaining objects. 
In Fig.~\ref{fig:hardnessratios}
we show one of various combinations of EPIC/pn hardness diagrams computed from the counts in the
$S$, $M$, and $H$ band. We distinguish Collinder\,69 members  belonging to different 
young stellar object classes with different
plotting symbols, and non-members are represented by squares.

From Fig.~\ref{fig:hardnessratios} it is evident that the cluster stars
are characterized by relatively soft X-ray emission,
 while most of the objects classified as non-members
on basis of optical and IR data have a much harder emission.  This is expected for objects
in the distant background whose soft photons are removed by interstellar extinction.
This separation of cluster members and non-members in the hardness plot lends additional support
to our candidate selection criteria described in  the previous section.

   \begin{figure}
   \centering
   \includegraphics[width=9cm]{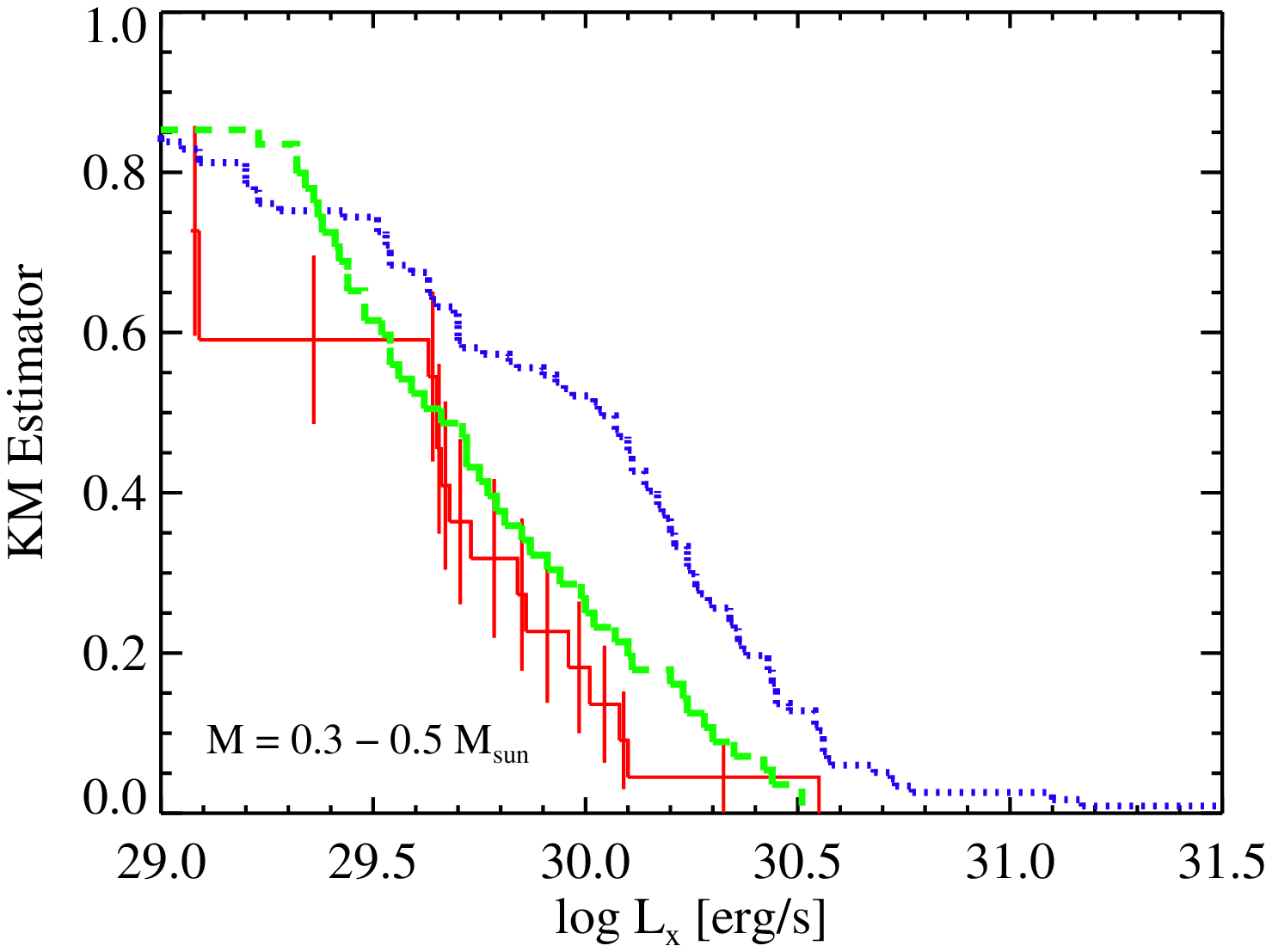}
   \includegraphics[width=9cm]{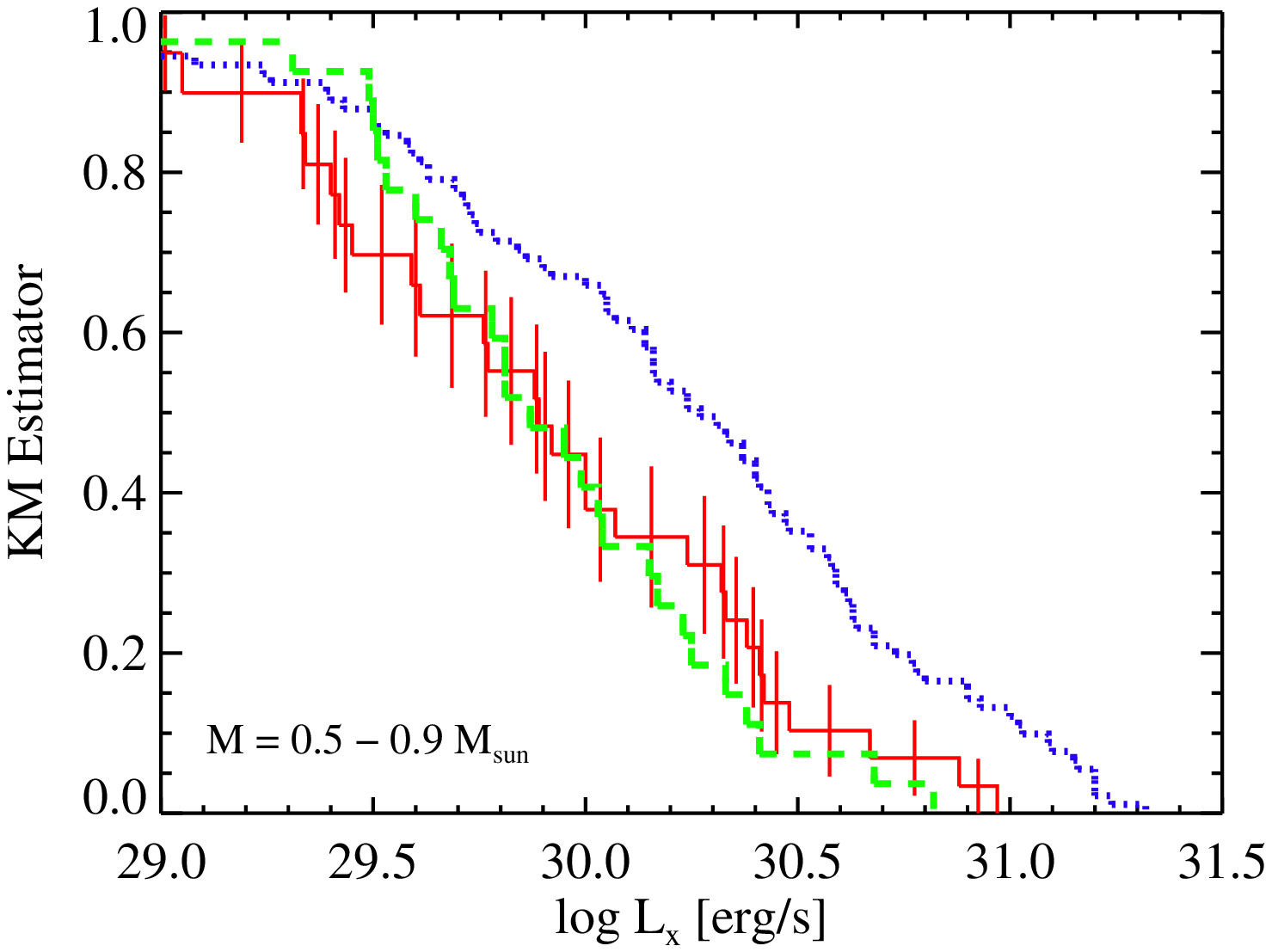}
      \caption{X-ray luminosity functions for Collinder\,69 (red, thin line) compared to those of
NGC\,2264 (green, thick line) and the ONC (blue, dotted line). From top to bottom the mass ranges
$0.3-0.5\,M_\odot$ and  $0.5-0.9\,M_\odot$ are shown. See text in
Sect.~\ref{subsect:lx} for more details.}
         \label{fig:xlf}
   \end{figure}


\subsection{X-ray spectra}\label{subsect:spectra}

 Spectra were analyzed for all X-ray sources detected on at least one of the individual
EPIC detectors. Among the $124$ X-ray sources seen on EPIC/pn there are $59$ Collinder\,69 members,
and among the additional $23$ X-ray sources present on EPIC/MOS there are $6$ Collinder\,69 members. 
One Collinder\,69 member (C69-X-e105) is detected only in the merged pn+MOS dataset and has no spectral
fit. 

Below,
we evaluate only the results from EPIC/pn. This is our `prime' choice for most sources according
to Col. \#2 of Tabs.~\ref{tab:XsourcesE} and \ref{tab:XsourcesW}
 because of its higher sensitivity with respect to EPIC/MOS.
 According to the criteria described in Sect.~\ref{subsect:data_epic}, the spectra of $40$ Collinder\,69 
members could be adequately fitted with a $1$-T model and $19$ required a $2$-T approach.
There are two EPIC/pn sources for which even a $2$-T model does not provide a
statistically acceptable fit ($P(\chi^2 > \chi_0) < 0.05$). 
Most of the sources have spectra of only moderate quality and we abstain from a detailed
discussion of individual source spectra. We are especially interested to use 
 the information from
the spectral fitting for deriving X-ray luminosities.
 
The median gas absorption column of the $59$  Collinder\,69 members with EPIC/pn
spectrum is ${(N_{\rm H})}_{\rm med} \sim 5 \cdot 10^{20}\,{\rm cm^{-2}}$. This low absorption 
agrees well
 with the average dust extinction 
 $A_{\rm V} = 0.38$\,mag of Collinder\,69 (\cite{Diplas94.1}), which corresponds
to $N_{\rm H} \sim 7 \cdot 10^{20}\,{\rm cm^{-2}}$
for standard gas/dust extinction conversions \citep{Ryter96.1}.
A small number of stars, exhibit much higher $N_{\rm H}$ according to the spectral analysis.
However, all of these have either very poor statistics or the fit yields an untypically low temperature
combined with extremely high emission measure to compensate for the strong absorption. In these latter
cases, the fitting procedure likely ended up in an unphysical $\chi^2$ minimum.
The median X-ray temperature derived from the spectral models is 
${(\log{T})}_{\rm med}\,{\rm [K]} \sim 7.0$ with a standard deviation of $0.31$\,logarithmic dex.


\subsection{X-ray luminosities}\label{subsect:lx}

%
%
   \begin{figure}
   \centering
   \includegraphics[width=9cm]{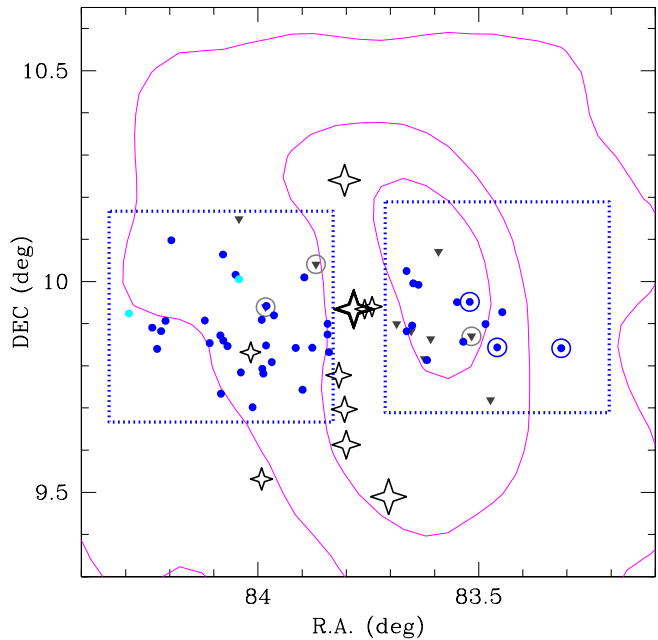}
    \caption{
       Spatial distribution of the stellar population in Collinder\,69 in the mass range  0.3--1.2 $M_\odot$.
       X-ray detections are represented as solid circles (blue for probable and possible members).
        Cluster members without X-ray detection appear as solid --dark gray-- triangles. 
       Class II, as defined by IRAC photometry,
       are highlighted with large solid open circles, respectively.      
}
    \label{fig:spatial}
   \end{figure}

The most accurate way of computing X-ray luminosities is to infer them from spectral
fitting on an individual basis. However,
most X-ray sources in Collinder\,69 are fairly faint and occasionally the fitting procedure results in
unphysical parameters (see above). 
Therefore, we prefer the use of a global 
 count-rate to flux conversion (CF) based on an average
spectrum for Collinder\,69. 
We base the computation of X-ray luminosities on a 1-T thermal model with plasma temperature and absorbing
column corresponding to the median values derived above from the EPIC/pn spectra,
i.e. ${(\log{T})}_{\rm med}\,{\rm [K]} \sim 7.0$ and $N_{\rm H} \sim 5 \cdot 10^{20}\,{\rm cm^{-2}}$.
Among the X-ray sources with statistically acceptable spectral fits
    the 50\,\% quartile range for $N_{\rm H}$ is $0.03-0.12 \cdot 10^{22}\,{\rm cm^{-2}}$
    and for $\log{T}$ it is $6.93-7.16$.
    For the median values of temperature and column density,
PIMMS\footnote{PIMMS is accessible at http://asc.harvard.edu/toolkit/pimms.jsp}
 yields $CF = 8.6 \cdot 10^{-12}\,{\rm cts/erg/cm^2}$
for the conversion of the $0.5-7.3$\,keV MOS count rate into a $0.5-8$\,keV  absorption-corrected flux.
The $CF$ is not strongly dependent on small variation of the temperature (it changes by $< 20$\,\% for the
observed spread of X-ray temperatures). 

In Tab.~\ref{tab:finalmembers}, we present the X-ray luminosities and  $L_{\rm x}/L_{\rm bol}$ ratios 
for all confirmed and suspected
Collinder\,69 members,  and Tabs.~\ref{tab:stelpar_ulE} and \ref{tab:stelpar_ulW} 
include the upper limits of the X-ray luminosities
and   $L_{\rm x}/L_{\rm bol}$ ratios
of undetected cluster members (for the XMM-{\em Newton} fields C69\,E and C69\,W, respectively.).

The fractional X-ray luminosity, $\log{(L_{\rm x}/L_{\rm bol})}$, 
as a function of stellar mass is shown in Fig.~\ref{fig:lglxlbol_mass}. 
The mean level for stars with $M \leq 1\,M_\odot$ is
$\log{(L_{\rm x}/L_{\rm bol})} = -2.85$ with a spread of $\sim 1.5$\,dex.
For higher-mass stars (starting at $M$$\sim$2\,$M_\odot$) the 
typical decline of the $L_{\rm x}/L_{\rm bol}$ ratio is seen.
The two highest-mass stars in our sample (C69-X-e013 and C69-X-e042, both newly identified 
as Collinder\,69 members) with $M \sim 3\,M_\odot$ have by far the lowest 
$L_{\rm x}/L_{\rm bol}$ levels. Indeed, according to  \cite{Siess00.1}, 
they are the only two stars with very shallow convective envelopes 
($<1$\,\% of the stellar radius). Their fractional X-ray luminosities are similar 
to those of Herbig Ae/Be stars (see \cite{Stelzer09.1}). Note, however,
that C69-X-e013 and C69-X-e042 are Class\,III objects (based on the SED and the lack of excess at 24 micron), 
i.e. they represent a more evolved  phase than Herbig stars.
In  Fig.~\ref{fig:lglxlbol_mass},
X-ray sources and upper limits within the XMM-{\em Newton} field-of-view are displayed as solid circles 
(blue and cyan for probable and possible members) and inverted triangles (dark gray).
Class II objects and transition/thin disks,
 as defined by the use of IRAC photometry (Fig.~\ref{fig:CCD_IRAC} and the IRAC slope) are
highlighted by  large circles with solid and dotted lines, respectively.

A few facts that can be noted from Fig.~\ref{fig:lglxlbol_mass} are:\\
(i) Only three Class\,II sources  with masses $>$ 0.3 M$_{\odot}$ (probable members) are detected in X-rays,
whereas the very low-mass regime 
($M$$<$0.3M $_{\odot}$ ) is dominated by upper limits of Class\,II sources. 
This  may partly be owing to the limited sensitivity of our X-ray observations 
(as described before, $\sim 0.3\,M_\odot$),
 and the subsequent bias of the cluster sample toward 
stars with IR excess in the very low-mass regime.
That is, we believe we have obtained a nearby complete census
for solar-like stars, because all our surveys and techniques go down to
this mass limit, but some low-mass members without IR excess 
 may  have remained unidentified.
However, the number of missing members
 cannot be significant, as discussed in  Sect.~\ref{subsect:completecensus},
 because the cluster mass function looks comparable to those corresponding to
well known stellar association in its power-law
or lognormal form (\cite{Barrado04.1}; \cite{Barrado05}; \cite{Bayo09.1}; Bouvier 2009, private communication)\\
(ii) The stars with the highest masses have no young stellar object
 classification because of incomplete {\em Spitzer}
photometry related to their brightness  (the cyan solid circles at $M\sim3\,{\rm M_\odot}$). \\
(iii) A few stars are  detected above the average saturation level $\log(L_x/L_{bol}) \sim -3$. 
However,  as we verified from the analysis of their X-ray light-curves none of them has shown a flare.\\
(iv)
One faint object, with a mass estimate about 0.035 $M_\odot$ 
 has been detected in the XMM-{\em Newton} 
survey (C69-X-e028). We classified it as possible member based on our selection criteria
(see Sect.~\ref{sect:select_cand}). 
The high X-ray activity level combined with very low mass suggests 
it might be a non-member. 
If its membership is confirmed, it would be one of the least massive brown dwarfs detected 
in X-rays. Additional data (i.e., spectroscopy) are required to determine its nature and membership to the
cluster.

To take account of the known mass dependence of pre-MS X-ray emission \citep[e.g.][]{Preibisch05.1}
 and mass-dependent incompleteness  of our sample,
we computed the X-ray luminosity functions (XLFs) for Collinder\,69 in different mass bins. 
 We restrict this analysis to mass ranges where the sample is (nearly) complete.
The Kaplan-Meier
Estimator implemented in ASURV\  \citep{Feigelson85.1} takes into account the upper limits.
 In Fig.~\ref{fig:xlf}
we compare the XLF to those of two other star-forming regions, NGC\,2264 ($\sim 3$\,Myr) and the
 ONC ($\sim 1$\,Myr).
The XLF of these latter two  associations were computed for the same mass ranges.
The data for the ONC were extracted from the {\em Chandra Orion Ultradeep Project}  
\citep[COUP; see][]{Getman05.2}.
We eliminated the probable non-members identified by \cite{Getman05.1} from the COUP source list.
The XLF of NGC\,2264 is obtained from data presented by \cite{Flaccomio06.1}.
For  both mass bins the XLF of Collinder\,69 is
in excellent agreement with that of NGC\,2264, while the stars in the ONC are younger and on average more
X-ray luminous.
The median X-ray luminosities and the sample size are given in Tab.~\ref{tab:diskfraction}.

\section{Conclusions and summary}\label{sect:discussion}

We report first results from our XMM-{\em Newton} based multi-wavelength project XILO 
to study the LOSFR. We updated the young stellar object census for the central
cluster of the LOSFR, Collinder\,69. We detected 164 sources in our 
two X-ray images, located east and west of the hot star
$\lambda$\,Ori (actually, a close, massive visual binary).
By  combining with our optical/IR database
and through a careful selection process using
optical/IR CMD and CCDs,
 we evaluated cluster membership for all detected X-ray sources and
provide a new, more complete and unbiased cluster census,
specially adding new members with masses about 2 $M_\odot$.
Out of the 164 X-ray sources, only approximately 
$40$\,\% turned out to be members. 
The remainder are mostly quasars and AGN, as judging from the
expected number of extragalactic objects in our X-ray images and the
X-ray to optical flux ratios, $f_{\rm x}/f_{\rm opt}$. 
 With our multi-wavelength approach, we 
were
 able to identify these pollutants from the true census of cluster members.
We  identified three possible companions to our X-ray members, which fulfill the membership criteria.
Therefore, we  detected in X-rays a total of 61 and 5 probable and possible 
 members, respectively (without including the three possible companions to cluster members with X-ray emission). 
Most of them  have been known from previous surveys, but 
 16  are newly identified probable cluster members, and three are additional
new possible members.

We estimate our completeness limit --for cluster members-- 
of the X-ray data to be $\sim0.3\,M_\odot$, based on the histograms
displayed in Figs.~\ref{fig:histomassA} and ~\ref{fig:histomassB}, and the peaks of the distributions. 
This is in agreement with the
estimated sensitivity limit of XILO, $L_{\rm x} \sim 10^{29}$\,erg/s 
 according to
 the empirical $L_{\rm x} - M$ relation derived for pre-MS stars in Orion. 
Therefore, the non-detection of very low-mass stars and 
substellar objects in  Collinder 69 is not surprising.
We detect only one possible brown dwarf, C69-X-w028, which might actually be a 
non-member. Additional data, specially spectroscopy, will eventually allow us to 
establish whether this object belongs to the stellar association.

In principle, the Collinder\,69 sample is a particularly reliable testbed for measuring X-ray temperatures, 
because the generally low $N_{\rm H}$ reduces biases in the spectral fitting process.
Indeed, an anti-correlation between X-ray
temperature and absorbing column density for the few cluster members with 
$\log{N_{\rm H}}\,{\rm [cm^{-2}]} > 21.5$ can be explained by a known bias of the
fitting procedure. We find no other trends between spectral parameters. 
The photon statistics of most sources are low and 
 and we thought a detailed discussion of the individual X-ray spectra unnecessary" (and inaccurate).
The mean temperature of $\log{T}\,{\rm [K]} \sim 7.0$ is similar to those of the 
coronae of young main-sequence stars \citep{Briggs03.1},
but somewhat lower than the temperatures found in the youngest star-forming regions
\citep[e.g.][]{Preibisch05.1}. Even the NGC\,2264 cluster ($\sim3$\,Myr), 
shows higher average X-ray plasma temperature ($1.3-1.5$\,keV for CTTS and wTTS, 
respectively; \cite{Flaccomio06.1}. This hints at an age older than $3$\,Myr for Collinder\,69.
On the other hand, the XLF of Collinder\,69 agrees very well with that of NGC\,2264, suggesting
a similar age of the two clusters.
Note that Collinder\,69 cannot be older than about 7 Myr, because of the presence of the central, very massive
 star $\lambda$ Ori.

Stellar properties such as effective temperatures, bolometric luminosities, masses
 and presence of circumstellar disk have been derived for cluster  members falling 
within the two  XMM-{\em Newton} pointings, independently of  whether they are
detected or not in X-rays.
We have done so using a Virtual Observatory tool (VOSA, \cite{Bayo08.1}) in an homogeneous way. The same
procedure was carried out with a larger sample (covering  a larger area) in the cluster.
The spatial dichotomy observed  previously in the distribution of members
with more stars  east of $\lambda$\,Orionis than west  
(in particular those diskless versus those with  disks) are  reproduced with our updated
cluster sample.

We derived the disk fraction for  the XILO sample using {\em Spitzer} mid-infrared photometry 
and  compared it with other  samples  for Collinder\,69 extracted with different techniques. 
For Class II members, we have  6.1 \%, whereas when we take into account thin and transition disk, 
it increases up to 10.6 \%.
Most members from \cite{Dolan99.1} --solar-like-- are detected in X-rays (not surprising, 
because they were selected based on their chromospheric activity or high accretion rate),  
and about a quarter of them have disks. 
Therefore, the disk fraction of the  XILO sample is much lower than  that of
the sample of \cite{Dolan99.1}.
This difference between the disk fraction
 is more striking when the XILO sample is compared with cluster members not detected in X-rays. 
 More than half of the probable and possible Collinder\,69 members with no X-ray  detection
--mostly low-mass stars and brown dwarfs-  have a  circumstellar disk. 
Therefore, it seems that there are significant differences, which are mass-dependant,
 despite the low number of stars. 

 The small disk fraction is a characteristic of Collinder\,69 members with masses around $\sim$1 $M_\odot$,
as discussed in Sect.~\ref{subsect:disks}. 
However, the {\it measured} disk fraction for very low-mass  stars and brown dwarfs
--mostly undetected in X-rays-- is significantly
higher than for more massive members --of which most are detected in X-rays.
If the {\it real} disk fraction  were similarly low for both groups,
 the majority of diskless objects with the 
smallest masses in Collinder\,69 would still be undiscovered.
 X-ray observations would be an adequate tool for 
identifying  such objects, but these new observations would have
to be at least one order of magnitude deeper than XILO, our actual survey. 
Yet, because the mass function of the cluster has been shown to be similar to that corresponding
to other associations, we do not believe this is the case, and therefore it seems there
is a significant dichotomy between the disk fraction of solar-like and very low-mass members. 
The obvious,  simplest interpretation is a longer dissipation time-scale 
for disks around low-mass objects.
This result has been already established in young cluster such as IC348 (\cite{Lada06.1}),
or suggested in larger samples (\cite{Barrado03.1}).
 Our results suggest that Collinder 69 seems to be at a critical age, about 5 Myr, 
for disk dissipation, and, therefore, it
can provide an interesting stepping stone to understand  disk evolution. 

Our improved census of cluster members
 would allow us to derive a more accurate  IMF and compare it with 
those of other SFRs. However, since our XMM-{\em Newton} survey covers a limited area overlapping 
our previous optical and infrared surveys \cite{Barrado04.1,Barrado07.1,Morales08.1},
 and  the overall shape of the IMF (for instance, the slopes in the mass spectrum form) is not
affected in significant way by the new cluster members, 
we refer the reader to upcoming papers where we analyze
photometry and spectroscopy (Morales-Calder\'on et al. 2010; and Bayo et al. 2010, both in prep).
 
Finally, we have shown that only an ambitious study involving the use of different techniques 
with multi-wavelength coverage avoid most of the biases present in more limited surveys based 
on few photometric bands, since they left out potential cluster members.  
In addition, the proper interpretation of any  observed phenomenology (disk fraction
and properties, evolution of activity) greatly benefit from a large,  comprehensive database.

 One missing piece of information of the Collinder\,69 population is a systematic assessment
of rotation.
Several previous works (\cite{Stelzer03.1}; \cite{Preibisch05.1}; \cite{Briggs07.1})
 have focused on the  X-ray activity $-$ rotation connection 
 on the pre-MS. Contrary to early findings,
it seems now that T\,Tauri stars occupy the saturated regime of the X-ray/rotation relation,
and therefore $L_{\rm x}$ is not correlated with rotation. Testing this scenario for Collinder\,69
requires. 
High resolution spectroscopy and a systematic monitoring to derive rotational periods.

\begin{acknowledgements}
 We thank the  Calar Alto Observatory, specially J. Alves, for the  allocation of director's discretionary
time to this program.
BS wishes to thank E. Flaccomio for stimulating discussions. 
Discussions about the properties of extragalactic sources with
Almudena Alonso Herrero were also very revealing.
The careful reading by the referee, Manuel Guedel, and his suggestions,  have been very helpful.
 This research has been funded by Spanish grants ESP2007-65475-C02-02,
 CSD2006-00070 and  PRICIT-S2009/ESP-1496.
BS has been supported by ASI/INAF contract I/088/06/0. The support of the ESAC 
faculty is also recognized.
It makes use of VOSA, developed under the Spanish Virtual Observatory project
 supported from the Spanish MICINN through grant AyA2008-02156, and
 of the SIMBAD database, operated at CDS, Strasbourg, France.
\end{acknowledgements}

\bibliographystyle{aa} 
\bibliography{14732_C69}

%
%
\clearpage

\setcounter{table}{0}
%
\begin{table*}
\begin{center}
\caption{Observing log for the XMM-{\em Newton} observations of Collinder\,69.}
\label{tab:obslog}
\begin{tabular}{lcrrrcccrr}\hline
Field & ObsID & \multicolumn{1}{c}{$\alpha_{\rm 2000}^{(a)}$} & \multicolumn{1}{c}{$\delta_{\rm 2000}^{(a)}$} & \multicolumn{1}{c}{PA$^{(a)}$} & \multicolumn{3}{c}{-----------Filter-----------} & \multicolumn{1}{c}{UT$^{(b)}$} & \multicolumn{1}{c}{Exp.$^{(c)}$} \\
      &       & [hh:mm:ss.ss]                                 & [dd:mm:ss.s]                                  & [$^\circ$]                     & PN & MOS\,1 & MOS\,2       &                                                      & [ksec]                             \\
\hline
Col\,69\,W  & 0300100201 & 05 33 46.29 & +09 47 48.5 &  90.72 & Medium & Medium & Medium & 2005-10-01 03:48:05 & 28 \\
Col\,69\,E  & 0405210601 & 05 35 15.42 & +09 56 19.2 &  83.86 & Thin   & Medium & Medium & 2006-08-31 02:24:17 & 37 \\
\hline
\multicolumn{10}{l}{$^{(a)}$ Mean pointing position and spacecraft position angle.} \\
\multicolumn{10}{l}{$^{(b)}$ Start of EPIC/MOS exposure, EPIC/PN observation starts about $30$\,min later.} \\
\multicolumn{10}{l}{$^{(c)}$ Nominal exposure time; the useful exposure time has been reduced because of  high background, especially for C\,69\,W.} \\
\end{tabular}
\end{center}
\end{table*}

\clearpage

\setcounter{table}{1}
%
%
\begin{table}
\begin{center}
\caption{Observing log and astrometric correction of OM data with respect to 2\,MASS.}
\label{tab:astrometry_om}
\begin{tabular}{lcrlcc}\hline
Exp.no. & OM mode$^a$ & Exp.time & Filter & $\Delta_{\rm \alpha}$ & $\Delta_{\rm \delta}$ \\
        &             & [sec]    &        & [${\prime\prime}$] & [${\prime\prime}$] \\ \hline
\multicolumn{6}{c}{ObsID 0300100201 (C\,69\,W)} \\ \hline
006     & FF High-res & $5000$   & $V$    & $+1.68$   & $-1.49$ \\
\hline
\multicolumn{6}{c}{ObsID 0405210601 (C\,69\,E)} \\ \hline
006     & FF Low-res  & $5000$   & $V$    & $+1.37$ & $-0.21$ \\
007     & FF Low-res  & $5000$   & $B$    & $-0.30$ & $+0.25$ \\
008     & FF Low-res  & $5000$   & $V$    & $+0.61$ & $-0.03$ \\
009     & FF Low-res  & $5000$   & $B$    & $-0.46$ & $+0.34$ \\
010     & FF Low-res  & $2630$   & $B$    & $-0.61$ & $+0.47$ \\
\hline
\multicolumn{6}{l}{$^a$ Full Frame low-resolution imaging mode yields a $1024 \times 1024$} \\
\multicolumn{6}{l}{image with $1^{\prime\prime}$ pixel size.} \\
\end{tabular}
\end{center}
\end{table}

\setcounter{table}{2}
%
%
\begin{table*} \begin{center}
\caption{X-ray sources detected with $ML > $15.0 in the EPIC observation of Collinder 69, East field (C69E).}
\label{tab:XsourcesE} \newcolumntype{d}[1]{D{.}{.}{#1}}
\begin{tabular}{lcrrrrrrrr} \\ \hline
Designation  &
Instr.$^1$   &
$\alpha_{\rm x,2000}$ & 
$\delta_{\rm x,2000}$ &
Poser               & 
Offax               & 
$ML$                & 
\multicolumn{1}{c}{Rate} & 
\multicolumn{1}{c}{${\rm HR_1}$} &  
\multicolumn{1}{c}{${\rm HR_2}$}  \\  
C69-X-      &         &                      &                       & [$^{\prime\prime}$] & [$^{\prime}$] &      & \multicolumn{1}{c}{[$\cdot 10^{-3}$\,cts/s]} &    &  \\ 
\hline
e001 & E,PN &  5 35 55.4 &   9 56 30.2 &      0.1 & $     6.2$ & $ 15620.9$ & $   5.67 \pm   0.89$ & $-0.14 \pm  0.03$ & $-0.00 \pm  0.03$ \\   
e002 & E,PN &  5 36 32.0 &   9 44 21.1 &      0.1 & $    11.2$ & $ 15212.9$ & $   9.81 \pm   1.49$ & $ 0.35 \pm  0.03$ & $ 0.58 \pm  0.02$ \\   
e003 & E,PN &  5 37  0.8 &   9 49  6.1 &      0.2 & $    11.8$ & $  5082.0$ & $   2.74 \pm   0.71$ & $-0.26 \pm  0.05$ & $-0.16 \pm  0.06$ \\   
e004 & E,PN &  5 35 58.0 &   9 54 31.6 &      0.2 & $     5.4$ & $  3915.3$ & $   2.18 \pm   0.62$ & $-0.23 \pm  0.04$ & $-0.12 \pm  0.05$ \\   
e005 & E,PN &  5 36 27.8 &   9 55 27.7 &      0.2 & $     2.0$ & $  2566.0$ & $   1.16 \pm   0.36$ & $ 0.45 \pm  0.06$ & $ 0.65 \pm  0.04$ \\   
e006 & E,PN &  5 36 20.5 &   9 52 19.0 &      0.3 & $     2.9$ & $  2736.4$ & $   1.07 \pm   0.34$ & $-0.30 \pm  0.05$ & $-0.21 \pm  0.07$ \\   
e007 & E,PN &  5 36 19.0 &  10  3 49.7 &      0.2 & $     8.6$ & $  2967.0$ & $   1.44 \pm   0.47$ & $-0.20 \pm  0.05$ & $-0.08 \pm  0.06$ \\   
e008 & E,PN &  5 36 18.6 &   9 45  8.6 &      0.3 & $    10.0$ & $  2353.2$ & $   1.40 \pm   0.50$ & $-0.36 \pm  0.06$ & $-0.31 \pm  0.07$ \\   
e009 & E,PN &  5 35 19.0 &   9 54 52.6 &      0.3 & $    15.0$ & $  3880.7$ & $   6.68 \pm   2.14$ & $-0.20 \pm  0.04$ & $-0.11 \pm  0.05$ \\   
e010 & E,PN &  5 35 54.2 &  10  4 22.5 &      0.3 & $    11.2$ & $  2120.2$ & $   1.43 \pm   0.54$ & $-0.39 \pm  0.06$ & $-0.30 \pm  0.07$ \\   
e011 & E,PN &  5 36 20.2 &   9 44  2.6 &      0.4 & $    11.1$ & $  1471.8$ & $   1.31 \pm   0.54$ & $-0.40 \pm  0.07$ & $-0.29 \pm  0.08$ \\   
e012 & E,PN &  5 36  9.4 &  10  1 24.8 &      0.4 & $     6.7$ & $  2028.5$ & $   1.61 \pm   0.61$ & $-0.26 \pm  0.22$ & $-0.22 \pm  0.23$ \\   
e013 & E,PN &  5 36 23.1 &   9 45 14.9 &      0.4 & $    10.0$ & $  1040.7$ & $   0.74 \pm   0.38$ & $-0.66 \pm  0.06$ & $-0.66 \pm  0.08$ \\   
e014 & E,PN &  5 35 51.3 &   9 55 10.6 &      0.4 & $     7.0$ & $   907.2$ & $   0.63 \pm   0.32$ & $-0.40 \pm  0.08$ & $-0.34 \pm  0.10$ \\   
e015 & E,PN &  5 36 10.1 &  10  1 57.6 &      0.4 & $     7.2$ & $   809.9$ & $   0.74 \pm   0.36$ & $ 0.24 \pm  0.09$ & $ 0.46 \pm  0.08$ \\   
e016 & E,PN &  5 36 16.7 &   9 50 47.8 &      0.4 & $     4.5$ & $   783.1$ & $   0.44 \pm   0.23$ & $-0.35 \pm  0.09$ & $-0.24 \pm  0.11$ \\   
e017 & E,PN &  5 36 18.9 &   9 51 35.5 &      0.4 & $     3.6$ & $   905.2$ & $   0.50 \pm   0.25$ & $-0.46 \pm  0.07$ & $-0.41 \pm  0.09$ \\   
e018 & E,PN &  5 36 47.0 &  10  5 52.3 &      0.5 & $    12.6$ & $   854.9$ & $   0.77 \pm   0.41$ & $-0.25 \pm  0.10$ & $-0.21 \pm  0.11$ \\   
e019 & E,PN &  5 36 57.6 &   9 53 28.2 &      0.4 & $     9.4$ & $   803.0$ & $   0.55 \pm   0.29$ & $-0.47 \pm  0.09$ & $-0.43 \pm  0.11$ \\   
e020 & E,PN &  5 36 16.3 &   9 59 24.1 &      0.5 & $     4.3$ & $   665.0$ & $   0.36 \pm   0.21$ & $-0.29 \pm  0.10$ & $-0.23 \pm  0.12$ \\   
e021 & E,PN &  5 36 40.0 &  10  4 33.5 &      0.5 & $    10.6$ & $   608.9$ & $   0.65 \pm   0.37$ & $ 0.51 \pm  0.11$ & $ 0.69 \pm  0.08$ \\   
e022 & E,PN &  5 36  9.4 &   9 47  2.0 &      0.5 & $     8.6$ & $   518.8$ & $   0.46 \pm   0.30$ & $-0.33 \pm  0.11$ & $-0.28 \pm  0.13$ \\   
e023 & E,PN &  5 35 47.5 &   9 45 50.0 &      0.6 & $    12.3$ & $   416.0$ & $   0.62 \pm   0.48$ & $-0.52 \pm  0.12$ & $-0.47 \pm  0.14$ \\   
e024 & E,PN &  5 35 43.3 &   9 59 55.9 &      0.6 & $    10.2$ & $   364.0$ & $   0.61 \pm   0.41$ & $ 0.23 \pm  0.14$ & $ 0.51 \pm  0.11$ \\   
e025 & E,PN &  5 36 26.3 &   9 51 14.1 &      0.5 & $     4.3$ & $   390.7$ & $   0.28 \pm   0.20$ & $-0.28 \pm  0.11$ & $-0.24 \pm  0.14$ \\   
e026 & E,PN &  5 36 33.4 &   9 58  2.9 &      0.5 & $     4.4$ & $   416.1$ & $   0.37 \pm   0.23$ & $ 0.51 \pm  0.13$ & $ 0.72 \pm  0.08$ \\   
e027 & E,PN &  5 36 28.9 &   9 54 28.1 &      0.6 & $     2.3$ & $   273.1$ & $   0.27 \pm   0.23$ & $-0.03 \pm  0.23$ & $ 0.16 \pm  0.23$ \\   
e028 & E,PN &  5 35 57.7 &   9 47 33.3 &      0.8 & $     9.4$ & $   276.3$ & $   0.32 \pm   0.28$ & $-0.46 \pm  0.14$ & $-0.44 \pm  0.18$ \\   
e029 & E,PN &  5 35 34.8 &  10  0 34.9 &      0.9 & $    12.4$ & $   336.6$ & $   0.42 \pm   0.34$ & $-0.53 \pm  0.12$ & $-0.46 \pm  0.16$ \\   
e030 & E,PN &  5 36  2.8 &   9 42  8.2 &      0.7 & $    13.7$ & $   391.3$ & $   0.61 \pm   0.47$ & $-0.39 \pm  0.12$ & $-0.28 \pm  0.16$ \\   
e031 & E,PN &  5 36 52.7 &   9 52 57.3 &      0.6 & $     8.4$ & $   277.2$ & $   0.25 \pm   0.21$ & $-0.26 \pm  0.14$ & $-0.26 \pm  0.17$ \\   
e032 & E,PN &  5 35 37.9 &   9 44 10.1 &      0.7 & $    15.1$ & $   303.8$ & $   1.09 \pm   0.83$ & $ 0.32 \pm  0.11$ & $ 0.52 \pm  0.09$ \\   
e033 & E,PN &  5 35 59.6 &   9 50 17.3 &      0.7 & $     7.0$ & $   257.3$ & $   0.50 \pm   0.40$ & $ 0.48 \pm  0.13$ & $ 0.72 \pm  0.08$ \\   
e034 & E,PN &  5 35 22.3 &   9 52 26.2 &      0.9 & $    14.5$ & $   259.0$ & $   0.71 \pm   0.66$ & $-0.36 \pm  0.11$ & $-0.31 \pm  0.14$ \\   
e035 & E,PN &  5 35 39.4 &   9 50 33.1 &      0.8 & $    11.0$ & $   198.7$ & $   0.27 \pm   0.27$ & $-0.51 \pm  0.16$ & $-0.51 \pm  0.25$ \\   
e036 & E,PN &  5 35 30.4 &   9 50 35.5 &      0.9 & $    13.0$ & $   156.0$ & $   0.37 \pm   0.38$ & $-0.23 \pm  0.18$ & $-0.13 \pm  0.22$ \\   
e037 & E,PN &  5 35 52.0 &   9 50 28.2 &      0.8 & $     8.3$ & $   171.2$ & $   0.21 \pm   0.22$ & $-0.34 \pm  0.16$ & $-0.21 \pm  0.19$ \\   
e038 & E,PN &  5 35 27.1 &   9 53 11.8 &      0.9 & $    13.2$ & $   157.1$ & $   0.49 \pm   0.46$ & $ 0.42 \pm  0.19$ & $ 0.60 \pm  0.14$ \\   
e039 & E,PN &  5 36 56.4 &   9 53 38.1 &      0.8 & $     9.1$ & $   155.5$ & $   0.25 \pm   0.22$ & $ 0.54 \pm  0.20$ & $ 0.71 \pm  0.14$ \\   
e040 & E,PN &  5 36 50.2 &   9 54 23.1 &      0.8 & $     7.5$ & $   155.9$ & $   0.16 \pm   0.17$ & $-0.06 \pm  0.17$ & $ 0.07 \pm  0.19$ \\   
e041 & E,M1 &  5 37 10.1 &   9 55 26.1 &      1.1 & $    12.3$ & $   143.2$ & $   0.36 \pm   0.44$ & $-0.33 \pm  0.24$ & $-0.25 \pm  0.29$ \\   
e042 & E,PN &  5 36 46.1 &   9 53 14.3 &      1.1 & $     6.7$ & $    98.3$ & $   0.09 \pm   0.12$ & $-0.45 \pm  0.22$ & $-0.43 \pm  0.27$ \\   
e043 & E,PN &  5 35 50.0 &   9 57 50.6 &      1.0 & $     7.8$ & $    90.6$ & $   0.18 \pm   0.21$ & $ 0.17 \pm  0.23$ & $ 0.44 \pm  0.18$ \\   
e044 & E,PN &  5 36 26.8 &   9 49 19.0 &      0.9 & $     6.1$ & $   100.5$ & $   0.17 \pm   0.18$ & $ 0.67 \pm  0.17$ & $ 0.81 \pm  0.11$ \\   
e045 & E,PN &  5 35 30.6 &   9 54 29.2 &      0.9 & $    12.2$ & $   118.7$ & $   0.37 \pm   0.37$ & $ 0.09 \pm  0.23$ & $ 0.43 \pm  0.17$ \\   
e046 & E,PN &  5 35 33.4 &   9 51 47.1 &      1.0 & $    12.0$ & $   105.4$ & $   0.32 \pm   0.36$ & $-0.12 \pm  0.85$ & $ 0.87 \pm  0.10$ \\   
e047 & E,PN &  5 36 54.5 &   9 53 26.7 &      1.0 & $     8.7$ & $    91.6$ & $   0.14 \pm   0.17$ & $-0.35 \pm  0.24$ & $-0.20 \pm  0.29$ \\   
e048 & E,M2 &  5 36 50.7 &   9 54  5.3 &      1.1 & $     7.7$ & $    68.2$ & $   0.12 \pm   0.15$ & $ 0.79 \pm  0.26$ & $ 0.88 \pm  0.14$ \\   
e049 & E,PN &  5 35 48.8 &   9 49 24.1 &      0.9 & $     9.6$ & $    83.5$ & $   0.22 \pm   0.27$ & $ 0.78 \pm  0.18$ & $ 0.87 \pm  0.12$ \\   
e050 & E,PN &  5 36 17.4 &  10  1 38.2 &      0.9 & $     6.5$ & $    96.5$ & $   0.14 \pm   0.16$ & $ 0.32 \pm  0.18$ & $ 0.50 \pm  0.15$ \\   
e051 & E,PN &  5 35 57.3 &   9 58 25.3 &      0.9 & $     6.4$ & $    99.5$ & $   0.17 \pm   0.19$ & $-0.16 \pm  0.23$ & $ 0.03 \pm  0.25$ \\   
e052 & E,PN &  5 35 21.0 &   9 47 28.1 &      1.3 & $    16.4$ & $   127.3$ & $   0.47 \pm   0.68$ & $-0.05 \pm  0.18$ & $ 0.02 \pm  0.22$ \\   
e053 & E,PN &  5 36 14.0 &   9 53  4.9 &      1.0 & $     2.6$ & $    71.4$ & $   0.11 \pm   0.14$ & $ 0.22 \pm  0.22$ & $ 0.44 \pm  0.19$ \\   
e054 & E,PN &  5 36 12.2 &  10  0 56.9 &      1.2 & $     6.1$ & $    61.7$ & $   0.07 \pm   0.12$ & $-0.49 \pm  0.23$ & $-0.37 \pm  0.30$ \\   
e055 & E,PN &  5 35 52.6 &   9 48 34.2 &      1.2 & $     9.4$ & $   101.5$ & $   0.20 \pm   0.24$ & $-0.16 \pm  0.22$ & $-0.11 \pm  0.26$ \\   
e056 & E,PN &  5 36 44.1 &   9 52  5.4 &      0.9 & $     6.7$ & $    73.9$ & $   0.15 \pm   0.19$ & $-0.06 \pm  0.25$ & $ 0.29 \pm  0.21$ \\   
e057 & E,PN &  5 36 49.2 &   9 58 20.5 &      1.1 & $     7.9$ & $    68.5$ & $   0.08 \pm   0.13$ & $-0.66 \pm  0.19$ & $-0.63 \pm  0.25$ \\   
e058 & E,PN &  5 36 54.8 &   9 50 22.9 &      1.0 & $     9.8$ & $    71.0$ & $   0.13 \pm   0.20$ & $-0.17 \pm  0.27$ & $-0.14 \pm  0.32$ \\   
e059 & E,PN &  5 36 14.9 &   9 52  5.6 &      0.9 & $     3.3$ & $    60.5$ & $   0.12 \pm   0.16$ & $ 0.65 \pm  0.29$ & $ 0.89 \pm  0.09$ \\   
e060 & E,PN &  5 36  9.1 &   9 56 19.9 &      1.0 & $     2.9$ & $    77.0$ & $   0.12 \pm   0.14$ & $ 0.55 \pm  0.25$ & $ 0.74 \pm  0.15$ \\   
\hline \end{tabular}
\\
Flag indicating the instrument used for the spectral analysis;   see text in Sect.2.1.\\
 \end{center} \end{table*}

\addtocounter{table}{-1}

\begin{table*} \begin{center}
\caption{X-ray sources detected with $ML > $15.0 in the EPIC observation of Collinder 69, East field (C69E).}
\label{tab:x-sources} \newcolumntype{d}[1]{D{.}{.}{#1}}
\begin{tabular}{lcrrrrrrrr} \\ \hline
Designation  &
Instr.$^1$   &
$\alpha_{\rm x,2000}$ & 
$\delta_{\rm x,2000}$ &
Poser               & 
Offax               & 
$ML$                & 
\multicolumn{1}{c}{Rate} & 
\multicolumn{1}{c}{${\rm HR_1}$} &  
\multicolumn{1}{c}{${\rm HR_2}$}  \\  
C69-X-      &         &                      &                       & [$^{\prime\prime}$] & [$^{\prime}$] &      & \multicolumn{1}{c}{[$\cdot 10^{-3}$\,cts/s]} &    &  \\ 
\hline
e061 & E,PN &  5 35 56.6 &  10  1 51.6 &      1.1 & $     8.8$ & $    68.9$ & $   0.15 \pm   0.19$ & $ 0.26 \pm  0.22$ & $ 0.45 \pm  0.19$ \\ 
e062 & E,PN &  5 36  2.6 &  10  5 37.8 &      1.3 & $    11.3$ & $    67.2$ & $   0.14 \pm   0.22$ & $-0.96 \pm  0.13$ & $-0.76 \pm  0.27$ \\ 
e063 & E,PN &  5 35 54.6 &  10  1 51.9 &      1.1 & $     9.1$ & $    58.7$ & $   0.17 \pm   0.23$ & $ 0.27 \pm  0.34$ & $ 0.62 \pm  0.20$ \\ 
e064 & E,PN &  5 35 22.2 &   9 53 58.0 &      1.5 & $    14.3$ & $    57.8$ & $   0.26 \pm   0.40$ & $-0.31 \pm  0.21$ & $-0.31 \pm  0.25$ \\ 
e065 & E,M1 &  5 35 38.0 &   9 53 16.5 &      1.3 & $    10.5$ & $    47.7$ & $   0.20 \pm   0.31$ & $ 0.48 \pm  0.31$ & $ 0.62 \pm  0.24$ \\ 
e066 & E,PN &  5 35 38.1 &  10  0 59.7 &      1.1 & $    11.8$ & $    48.1$ & $   0.20 \pm   0.28$ & $ 0.21 \pm  0.28$ & $ 0.45 \pm  0.23$ \\ 
e067 & E,PN &  5 36  8.6 &   9 53 22.0 &      1.4 & $     3.3$ & $    25.0$ & $   0.07 \pm   0.13$ & $ 0.67 \pm  0.23$ & $ 0.78 \pm  0.16$ \\ 
e068 & E,PN &  5 36 20.3 &   9 56  4.3 &      1.2 & $     0.9$ & $    34.8$ & $   0.06 \pm   0.10$ & $-0.70 \pm  0.34$ & $-0.01 \pm  0.34$ \\ 
e069 & E,M1 &  5 35 51.8 &   9 53 34.4 &      1.6 & $     7.1$ & $    41.7$ & $   0.11 \pm   0.21$ & $ 0.02 \pm  0.36$ & $ 0.02 \pm  0.41$ \\ 
e070 & E,PN &  5 36 13.5 &  10  1 16.3 &      1.5 & $     6.3$ & $    31.3$ & $   0.09 \pm   0.15$ & $ 0.65 \pm  0.32$ & $ 0.83 \pm  0.15$ \\ 
e071 & E,PN &  5 35 29.3 &   9 46 36.4 &      1.4 & $    15.1$ & $    52.1$ & $   0.37 \pm   0.57$ & $ 0.35 \pm  0.21$ & $ 0.49 \pm  0.18$ \\ 
e072 & E,PN &  5 36 27.5 &   9 45 26.1 &      1.7 & $     9.9$ & $    25.0$ & $   0.07 \pm   0.16$ & $-0.66 \pm  0.42$ & $-0.66 \pm  0.56$ \\ 
e073 & E,M2 &  5 36 11.3 &   9 59 39.7 &      1.4 & $     5.0$ & $    28.2$ & $   0.08 \pm   0.14$ & $ 0.78 \pm  0.30$ & $ 0.90 \pm  0.14$ \\ 
e074 & E,PN &  5 36 35.6 &   9 54 23.9 &      1.5 & $     3.9$ & $    22.0$ & $   0.05 \pm   0.10$ & $-0.26 \pm  0.68$ & $ 0.66 \pm  0.22$ \\ 
e075 & E,PN &  5 35 32.5 &   9 57 56.5 &      1.3 & $    12.0$ & $    45.2$ & $   0.20 \pm   0.29$ & $ 0.25 \pm  0.28$ & $ 0.43 \pm  0.24$ \\ 
e076 & E,PN &  5 36 46.3 &   9 54 46.8 &      1.3 & $     6.5$ & $    41.5$ & $   0.09 \pm   0.14$ & $ 0.13 \pm  0.35$ & $ 0.46 \pm  0.25$ \\ 
e077 & E,PN &  5 36  1.2 &   9 58 44.0 &      1.3 & $     5.8$ & $    28.8$ & $   0.09 \pm   0.16$ & $ 0.19 \pm  0.43$ & $ 0.49 \pm  0.31$ \\ 
e078 & E,PN &  5 36 27.1 &   9 51 36.8 &      1.4 & $     4.0$ & $    33.0$ & $   0.07 \pm   0.13$ & $-0.71 \pm  0.29$ & $-0.33 \pm  0.39$ \\ 
e079 & E,PN &  5 35 21.4 &   9 49 56.1 &      1.5 & $    15.3$ & $    52.4$ & $   0.23 \pm   0.45$ & $-0.46 \pm  0.23$ & $-0.30 \pm  0.30$ \\ 
e080 & E,PN &  5 36 57.6 &  10  3 16.0 &      1.3 & $    12.3$ & $    43.7$ & $   0.14 \pm   0.22$ & $-0.34 \pm  0.41$ & $ 0.22 \pm  0.32$ \\ 
e081 & E,M1 &  5 35 55.6 &   9 50 54.1 &      1.5 & $     7.4$ & $    41.3$ & $   0.11 \pm   0.20$ & $-0.69 \pm  0.44$ & $-0.64 \pm  0.55$ \\ 
e082 & E,M1 &  5 35 56.9 &   9 46 51.2 &      2.5 & $    10.1$ & $    29.6$ & $   0.15 \pm   0.32$ & $-1.00 \pm  0.22$ & $-0.40 \pm  0.49$ \\ 
e083 & E,PN &  5 36 46.9 &   9 56 22.0 &      1.5 & $     6.7$ & $    29.1$ & $   0.08 \pm   0.14$ & $ 0.40 \pm  0.25$ & $ 0.53 \pm  0.22$ \\ 
e084 & E,PN &  5 35 35.8 &   9 44 34.4 &      1.8 & $    15.2$ & $    15.3$ & $   0.13 \pm   0.37$ & $-0.25 \pm  0.34$ & $-0.25 \pm  0.40$ \\ 
e085 & E,PN &  5 36 13.9 &   9 52 31.0 &      1.9 & $     3.1$ & $    17.9$ & $   0.06 \pm   0.13$ & $ 0.44 \pm  0.31$ & $ 0.72 \pm  0.17$ \\ 
e086 & E,PN &  5 36 53.2 &   9 50 18.4 &      1.5 & $     9.5$ & $    25.8$ & $   0.07 \pm   0.16$ & $-0.02 \pm  0.38$ & $ 0.35 \pm  0.29$ \\ 
e087 & E,M1 &  5 35 49.5 &  10  4 35.1 &      1.5 & $    12.0$ & $    31.9$ & $   0.20 \pm   0.36$ & $ 0.82 \pm  0.18$ & $ 0.88 \pm  0.11$ \\ 
e088 & E,PN &  5 36 30.3 &   9 43 34.1 &      2.1 & $    11.9$ & $    16.2$ & $   0.11 \pm   0.25$ & $ 0.64 \pm  0.31$ & $ 0.70 \pm  0.26$ \\ 
e089 &      &  5 36 56.9 &   9 52 53.8 &      1.6 & $     9.4$ & $    17.9$ & $   0.09 \pm   0.17$ & $ 0.00 \pm  0.00$ & $ 0.00 \pm  0.00$ \\ 
e090 & E,PN &  5 35 43.8 &   9 56  0.8 &      1.5 & $     9.0$ & $    37.8$ & $   0.13 \pm   0.19$ & $ 0.57 \pm  0.32$ & $ 0.79 \pm  0.16$ \\ 
e091 & E,PN &  5 36 31.4 &   9 45  2.2 &      2.0 & $    10.5$ & $    21.7$ & $   0.09 \pm   0.20$ & $-0.18 \pm  0.39$ & $ 0.07 \pm  0.39$ \\ 
e092 & E,M2 &  5 37 10.3 &   9 51 27.1 &      2.2 & $    13.0$ & $    21.7$ & $   0.14 \pm   0.37$ & $-0.60 \pm  0.43$ & $-0.44 \pm  0.55$ \\ 
e093 & E,PN &  5 35 25.5 &   9 47 39.0 &      1.7 & $    15.4$ & $    19.6$ & $   0.23 \pm   0.49$ & $ 0.74 \pm  0.28$ & $ 0.86 \pm  0.15$ \\ 
e094 & E,M2 &  5 35 51.5 &  10  6 34.7 &      1.7 & $    13.4$ & $    45.0$ & $   0.42 \pm   0.67$ & $ 0.53 \pm  0.37$ & $ 0.74 \pm  0.21$ \\ 
e095 & E,PN &  5 36 56.0 &   9 52  4.9 &      1.6 & $     9.4$ & $    21.8$ & $   0.09 \pm   0.17$ & $-0.14 \pm  0.60$ & $ 0.53 \pm  0.29$ \\ 
e096 & E,M1 &  5 36 10.0 &  10  0 18.2 &      2.0 & $     5.7$ & $    15.6$ & $   0.06 \pm   0.16$ & $-1.00 \pm  0.46$ & $-0.28 \pm  0.49$ \\ 
e097 & E,M2 &  5 37 15.5 &   9 57 49.6 &      2.3 & $    13.9$ & $    15.3$ & $   0.14 \pm   0.36$ & $-1.00 \pm  2.48$ & $ 0.80 \pm  0.24$ \\ 
e098 & E,PN &  5 36 25.2 &   9 41 29.8 &      2.3 & $    13.7$ & $    21.8$ & $   0.17 \pm   0.34$ & $-0.22 \pm  0.31$ & $ 0.06 \pm  0.31$ \\ 
e099 &      &  5 36 14.4 &   9 47 47.3 &      1.9 & $     7.5$ & $    18.3$ & $   0.09 \pm   0.22$ & $ 0.00 \pm  0.00$ & $ 0.00 \pm  0.00$ \\ 
e100 &      &  5 36 34.9 &   9 58 40.3 &      4.3 & $     5.1$ & $    19.0$ & $   0.08 \pm   0.16$ & $ 0.00 \pm  0.00$ & $ 0.00 \pm  0.00$ \\ 
e101 & E,PN &  5 36 40.1 &   9 48 36.2 &      1.8 & $     8.2$ & $    17.2$ & $   0.06 \pm   0.14$ & $ 0.82 \pm  0.26$ & $ 0.88 \pm  0.17$ \\ 
e102 & E,PN &  5 35 30.6 &   9 56 15.0 &      2.1 & $    12.2$ & $    15.9$ & $   0.11 \pm   0.24$ & $ 0.25 \pm  0.51$ & $ 0.62 \pm  0.27$ \\ 
e103 &      &  5 35 32.6 &   9 56 10.4 &      2.4 & $    11.7$ & $    17.3$ & $   0.11 \pm   0.23$ & $ 0.00 \pm  0.00$ & $ 0.00 \pm  0.00$ \\ 
e104 & E,PN &  5 35 55.7 &   9 52 11.0 &      1.7 & $     6.7$ & $    18.1$ & $   0.06 \pm   0.14$ & $-0.52 \pm  0.35$ & $-0.20 \pm  0.39$ \\ 
e105 &      &  5 36 55.6 &  10  1  8.0 &      2.3 & $    10.6$ & $    18.0$ & $   0.06 \pm   0.16$ & $ 0.00 \pm  0.00$ & $ 0.00 \pm  0.00$ \\ 
e106 & E,PN &  5 36  0.4 &   9 41 12.5 &      1.5 & $    14.8$ & $    58.1$ & $   0.47 \pm   0.89$ & $ 0.38 \pm  0.26$ & $ 0.54 \pm  0.22$ \\ 
e107 &      &  5 35 59.4 &   9 53 32.3 &      2.2 & $     5.3$ & $    18.2$ & $   0.08 \pm   0.18$ & $ 0.00 \pm  0.00$ & $ 0.00 \pm  0.00$ \\ 
e108 &      &  5 36 33.9 &  10  1  8.9 &      2.1 & $     6.9$ & $    18.3$ & $   0.11 \pm   0.24$ & $ 0.00 \pm  0.00$ & $ 0.00 \pm  0.00$ \\ 
e109 &      &  5 36 10.9 &  10  0 51.8 &      1.7 & $     6.1$ & $    18.6$ & $   0.25 \pm   0.51$ & $-0.12 \pm  0.36$ & $ 0.23 \pm  0.31$ \\ 
e110 &      &  5 36 44.5 &   9 59 26.1 &      2.6 & $     6.4$ & $    47.5$ & $   0.18 \pm   0.43$ & $ 0.95 \pm  0.17$ & $ 0.98 \pm  0.08$ \\ 
e111 &      &  5 36 21.7 &   9 53 49.5 &      2.6 & $     1.9$ & $    25.0$ & $   0.11 \pm   0.31$ & $-0.03 \pm  0.71$ & $ 0.56 \pm  0.32$ \\ 
e112 &      &  5 37  5.5 &   9 56  9.9 &      3.2 & $    10.3$ & $    17.0$ & $   0.22 \pm   0.61$ & $ 0.47 \pm  0.47$ & $ 0.75 \pm  0.22$ \\ 
\hline \end{tabular}
\\
Flag indicating the instrument used for the spectral analysis;   see text in Sect.2.1.\\
 \end{center} \end{table*}

\clearpage

%
\begin{table*} \begin{center}
\caption{X-ray sources detected with $ML > $15.0 in the EPIC observation of Collinder 69, West field (C69W).}
\label{tab:XsourcesW} \newcolumntype{d}[1]{D{.}{.}{#1}}
\begin{tabular}{lcrrrrrrrr} \\ \hline
Designation  &
Instr.$^1$   &
$\alpha_{\rm x,2000}$ & 
$\delta_{\rm x,2000}$ &
Poser               & 
Offax               & 
$ML$                & 
\multicolumn{1}{c}{Rate} & 
\multicolumn{1}{c}{${\rm HR_1}$} &  
\multicolumn{1}{c}{${\rm HR_2}$}  \\  
C69-X-      &         &                      &                       & [$^{\prime\prime}$] & [$^{\prime}$] &      & \multicolumn{1}{c}{[$\cdot 10^{-3}$\,cts/s]} &    &  \\ 
\hline
w001 & E,PN &  5 34  6.9 &  10  1  0.7 &      0.2 & $     6.1$ & $  6535.1$ & $   6.15 \pm   1.55$ & $-0.14 \pm  0.07$ & $-0.05 \pm  0.08$  \\  
w002 & E,PN &  5 34 36.2 &   9 53 45.8 &      0.3 & $    11.6$ & $  3393.6$ & $   4.61 \pm   1.44$ & $-0.08 \pm  0.07$ & $ 0.05 \pm  0.07$  \\  
w003 & E,PN &  5 33 56.4 &   9 53 57.4 &      0.3 & $     3.0$ & $  1766.4$ & $   1.28 \pm   0.53$ & $-0.34 \pm  0.07$ & $-0.26 \pm  0.09$  \\  
w004 & E,PN &  5 33 47.3 &   9 55 37.2 &      0.4 & $     1.2$ & $  1705.1$ & $   1.83 \pm   0.78$ & $-0.47 \pm  0.16$ & $-0.43 \pm  0.20$  \\  
w005 & E,PN &  5 34 32.8 &   9 59 30.6 &      0.4 & $    10.9$ & $  1010.2$ & $   1.38 \pm   0.74$ & $-0.29 \pm  0.10$ & $-0.24 \pm  0.12$  \\  
w006 & E,M1 &  5 34 39.2 &   9 52 56.0 &      0.5 & $    12.6$ & $   672.4$ & $   2.84 \pm   1.76$ & $ 0.11 \pm  0.14$ & $ 0.29 \pm  0.13$  \\  
w007 & E,PN &  5 34 28.0 &   9 48 48.2 &      0.4 & $    12.1$ & $   811.9$ & $   1.56 \pm   0.88$ & $-0.14 \pm  0.11$ & $-0.03 \pm  0.12$  \\  
w008 & E,PN &  5 33 50.1 &   9 50 35.8 &      0.6 & $     5.9$ & $   626.7$ & $   0.73 \pm   0.47$ & $-0.19 \pm  0.11$ & $-0.12 \pm  0.13$  \\  
w009 & E,PN &  5 34  4.2 &  10  0 47.4 &      0.7 & $     5.5$ & $   253.4$ & $   0.39 \pm   0.34$ & $-0.35 \pm  0.15$ & $-0.34 \pm  0.18$  \\  
w010 & E,PN &  5 34  2.0 &   9 58 17.1 &      0.6 & $     3.4$ & $   429.5$ & $   0.78 \pm   0.55$ & $ 0.66 \pm  0.16$ & $ 0.83 \pm  0.08$  \\  
w011 & E,PN &  5 34  5.5 &   9 42 46.7 &      0.8 & $    14.2$ & $   387.2$ & $   1.29 \pm   1.08$ & $-0.42 \pm  0.11$ & $-0.32 \pm  0.13$  \\  
w012 & E,PN &  5 34  5.0 &   9 57  4.2 &      0.7 & $     3.7$ & $   312.2$ & $   0.35 \pm   0.30$ & $-0.57 \pm  0.14$ & $-0.41 \pm  0.19$  \\  
w013 & E,PN &  5 34 35.5 &   9 59 44.5 &      0.8 & $    11.6$ & $   268.0$ & $   0.60 \pm   0.52$ & $-0.21 \pm  0.15$ & $-0.17 \pm  0.18$  \\  
w014 & E,PN &  5 33 50.4 &  10  4 19.8 &      0.8 & $     7.8$ & $   279.6$ & $   0.35 \pm   0.35$ & $-0.70 \pm  0.10$ & $-0.69 \pm  0.13$  \\  
w015 & E,PN &  5 34  8.5 &   9 51 25.3 &      0.8 & $     6.8$ & $   180.6$ & $   0.29 \pm   0.30$ & $-0.44 \pm  0.17$ & $-0.37 \pm  0.22$  \\  
w016 & E,PN &  5 33 13.0 &   9 49 38.9 &      1.0 & $    11.5$ & $   123.7$ & $   0.73 \pm   0.82$ & $ 0.25 \pm  0.34$ & $ 0.58 \pm  0.21$  \\  
w017 & E,M1 &  5 33 37.0 &   9 56 35.7 &      1.0 & $     3.3$ & $    92.2$ & $   0.27 \pm   0.34$ & $ 0.48 \pm  0.30$ & $ 0.62 \pm  0.23$  \\  
w018 & E,PN &  5 34 18.4 &   9 52 38.4 &      1.1 & $     7.9$ & $   139.4$ & $   0.41 \pm   0.49$ & $-0.06 \pm  0.27$ & $-0.01 \pm  0.29$  \\  
w019 & E,M2 &  5 33 42.7 &  10  3 48.5 &      1.1 & $     7.5$ & $   109.2$ & $   0.34 \pm   0.44$ & $ 1.00 \pm  0.20$ & $ 1.00 \pm  0.02$  \\  
w020 & E,PN &  5 33 51.3 &   9 46 41.4 &      1.5 & $     9.8$ & $   121.8$ & $   0.37 \pm   0.49$ & $-0.50 \pm  0.28$ & $-0.50 \pm  0.35$  \\  
w021 & E,PN &  5 33 40.6 &  10  2 39.1 &      1.2 & $     6.6$ & $   117.1$ & $   0.25 \pm   0.28$ & $ 0.43 \pm  0.21$ & $ 0.67 \pm  0.14$  \\  
w022 & E,M2 &  5 33 43.1 &  10 10 12.4 &      2.2 & $    13.8$ & $    78.9$ & $   0.51 \pm   0.82$ & $ 0.22 \pm  0.21$ & $ 0.29 \pm  0.22$  \\  
w023 & E,M1 &  5 34 39.3 &  10  1 29.5 &      1.4 & $    13.0$ & $    71.8$ & $   0.48 \pm   0.77$ & $-0.32 \pm  0.36$ & $-0.27 \pm  0.42$  \\  
w024 & E,PN &  5 34 11.8 &   9 57  3.8 &      1.0 & $     5.3$ & $    87.2$ & $   0.17 \pm   0.24$ & $-0.24 \pm  0.26$ & $-0.14 \pm  0.32$  \\  
w025 & E,PN &  5 34  2.5 &  10  7  7.0 &      1.4 & $    11.0$ & $    60.7$ & $   0.31 \pm   0.47$ & $ 0.30 \pm  0.36$ & $ 0.49 \pm  0.29$  \\  
w026 & E,PN &  5 33  3.9 &  10  2  6.4 &      1.0 & $    12.7$ & $    62.0$ & $   0.37 \pm   0.53$ & $ 0.15 \pm  0.37$ & $ 0.50 \pm  0.25$  \\  
w027 & E,M1 &  5 33 24.1 &   9 57  5.4 &      1.5 & $     6.5$ & $    44.8$ & $   0.18 \pm   0.33$ & $ 0.69 \pm  0.23$ & $ 0.73 \pm  0.21$  \\  
w028 & E,PN &  5 33 35.3 &  10  8 25.7 &      1.3 & $    12.5$ & $    78.9$ & $   0.34 \pm   0.45$ & $ 0.20 \pm  0.29$ & $ 0.54 \pm  0.19$  \\  
w029 & E,PN &  5 33 15.2 &   9 50 29.2 &      1.3 & $    10.5$ & $    59.5$ & $   0.23 \pm   0.38$ & $-0.23 \pm  0.26$ & $-0.23 \pm  0.31$  \\  
w030 & E,PN &  5 34  1.5 &   9 56 31.8 &      1.5 & $     2.8$ & $    22.7$ & $   0.11 \pm   0.23$ & $-0.02 \pm  0.44$ & $ 0.41 \pm  0.30$  \\  
w031 & E,PN &  5 33 27.0 &  10  1 38.9 &      1.6 & $     7.7$ & $    43.9$ & $   0.16 \pm   0.26$ & $ 0.72 \pm  0.48$ & $ 0.97 \pm  0.06$  \\  
w032 & E,PN &  5 33 48.3 &  10  1 43.9 &      1.6 & $     5.2$ & $    35.3$ & $   0.08 \pm   0.16$ & $-0.55 \pm  0.34$ & $-0.55 \pm  0.39$  \\  
w033 & E,M1 &  5 33 42.7 &   9 58  2.2 &      1.6 & $     2.4$ & $    37.1$ & $   0.16 \pm   0.28$ & $ 0.92 \pm  0.28$ & $ 0.98 \pm  0.06$  \\  
w034 & E,PN &  5 33 27.8 &  10  4 20.3 &      1.6 & $     9.6$ & $    46.0$ & $   0.14 \pm   0.26$ & $-0.40 \pm  0.31$ & $-0.40 \pm  0.39$  \\  
w035 & E,M2 &  5 33 14.1 &   9 55 45.3 &      1.5 & $     8.9$ & $    51.8$ & $   0.29 \pm   0.47$ & $ 0.82 \pm  0.18$ & $ 0.87 \pm  0.13$  \\  
w036 & E,M2 &  5 33 40.6 &   9 51 60.0 &      1.9 & $     5.1$ & $    24.2$ & $   0.09 \pm   0.22$ & $-0.79 \pm  0.50$ & $-0.29 \pm  0.60$  \\  
w037 & E,PN &  5 33 33.0 &   9 57 35.8 &      1.6 & $     4.4$ & $    29.7$ & $   0.06 \pm   0.15$ & $-0.48 \pm  0.42$ & $-0.48 \pm  0.60$  \\  
w038 & E,PN &  5 33 44.7 &  10  1 45.8 &      3.8 & $     5.4$ & $    24.8$ & $   0.10 \pm   0.20$ & $ 0.87 \pm  0.23$ & $ 0.90 \pm  0.16$  \\  
w039 & E,M1 &  5 33 47.6 &   9 52 57.2 &      2.2 & $     3.6$ & $    16.3$ & $   0.05 \pm   0.15$ & $-0.56 \pm  0.53$ & $-0.56 \pm  0.62$  \\  
w040 & E,PN &  5 33 39.2 &  10  0 13.0 &      3.3 & $     4.6$ & $    31.8$ & $   0.08 \pm   0.17$ & $-0.56 \pm  0.33$ & $-0.56 \pm  0.42$  \\  
w041 & E,PN &  5 33 43.1 &   9 43 42.2 &      2.5 & $    12.9$ & $    24.7$ & $   0.17 \pm   0.40$ & $-0.84 \pm  0.36$ & $-0.84 \pm  0.54$  \\  
w042 &      &  5 33 51.3 &   9 48 14.5 &      2.1 & $     8.3$ & $    17.9$ & $   0.13 \pm   0.31$ & $ 0.00 \pm  0.00$ & $ 0.00 \pm  0.00$  \\  
w043 & E,PN &  5 33 52.2 &   9 53  4.0 &      2.4 & $     3.5$ & $    17.1$ & $   0.11 \pm   0.26$ & $ 1.00 \pm  0.10$ & $ 1.00 \pm  0.04$  \\  
w044 & E,PN &  5 33 34.0 &   9 50 23.2 &      1.9 & $     7.3$ & $    29.8$ & $   0.13 \pm   0.24$ & $-0.16 \pm  0.48$ & $ 0.38 \pm  0.32$  \\  
w045 &      &  5 33 12.9 &   9 57 58.0 &      2.0 & $     9.3$ & $    15.4$ & $   0.13 \pm   0.30$ & $ 0.00 \pm  0.00$ & $ 0.00 \pm  0.00$  \\  
w046 & E,PN &  5 33  3.8 &   9 58 41.5 &      2.3 & $    11.7$ & $    27.5$ & $   0.12 \pm   0.29$ & $-0.12 \pm  0.43$ & $ 0.09 \pm  0.42$  \\  
w047 & E,M2 &  5 33 11.6 &   9 59  1.3 &      2.1 & $     9.8$ & $    22.3$ & $   0.10 \pm   0.26$ & $-0.12 \pm  0.54$ & $-0.01 \pm  0.56$  \\  
w048 & E,PN &  5 34 24.6 &   9 56 58.1 &      1.9 & $     8.5$ & $    18.1$ & $   0.12 \pm   0.26$ & $ 1.00 \pm  0.44$ & $ 1.00 \pm  0.15$  \\  
w049 &      &  5 34 34.3 &  10  3 23.0 &      1.8 & $    12.8$ & $    16.7$ & $   0.17 \pm   0.40$ & $ 0.13 \pm  0.49$ & $ 0.35 \pm  0.41$  \\  
w050 &      &  5 33 28.9 &   9 51  8.6 &      3.8 & $     7.5$ & $    28.6$ & $   0.11 \pm   0.38$ & $ 0.00 \pm  0.00$ & $ 0.00 \pm  0.00$  \\  
w051 &      &  5 33 39.6 &   9 51  3.7 &      2.4 & $     6.1$ & $    53.0$ & $   0.21 \pm   0.55$ & $ 0.00 \pm  0.00$ & $ 0.00 \pm  0.00$  \\  
w052 &      &  5 33 23.1 &   9 56 49.3 &      2.5 & $     7.7$ & $    18.4$ & $   0.29 \pm   0.84$ & $ 1.00 \pm  0.22$ & $ 1.00 \pm  0.07$  \\  
\hline \end{tabular}
\\
Flag indicating the instrument used for the spectral analysis;   see text in Sect.2.1.\\
 \end{center} \end{table*}

\clearpage

\clearpage

\setcounter{table}{4}
%
%
\begin{landscape}
\begin{table}
\tiny
\caption{Optical and infrared photometry for all the 205  counterparts for 164 X-ray sources of Collinder 69 (eastern {\rm XMM-Newton} pointing or Col 69\,E).} 
\label{tab:photometryE} 
\begin{tabular}{l l               l               l              l              l               l               l               l               l               l               l               l               l                         l           l} \\ 
Name    &   $B$    &   $V$       & $Rc$        &    $Ic$     &   $J$         &   $H$         &   $K$         &   [$3.6$]   &   [$4.5$]   &   [$5.8$]   &   [$8.0$]   &  [$24$]    &  Notes                   & R.A.      & DEC        \\   %
C69-X-\#   &   [mag]  &   [mag]     & [mag]       &    [mag]    &   [mag]       &   [mag]       &   [mag]       &   [mag]     &   [mag]     &   [mag]     &   [mag]     &  [mag]     &                          & (2000)    & (2000)     \\   
\hline      
\hline      
e001   & 15.73 0.01  & 14.43 0.00  & 13.50 0.00  & 12.57 0.00  & 11.416 0.023  & 10.725 0.022  & 10.524 0.023  & 10.26 0.00  & 10.32 0.00  & 10.24 0.01  & 10.17 0.01  &    --   -- & 2 7 7 7 6 6 6 6 6 6 6 - & 83.980968 & +09.9420736 \\  
e002   &    --   --  &    --   --  & 20.04 0.05  & 19.44 0.05  &     --    --  &     --    --  &     --    --  & 16.47 0.07  & 15.59 0.05  & 14.45 0.13  & 13.19 0.06  &    --   -- & - - 5 5 - - - 3 3 3 3 - & 84.133460 & +09.7389598 \\  
e003   &    --   --  & 12.41 0.00  & 11.87 0.00  & 11.36 0.00  & 10.585 0.024  & 10.152 0.025  &  9.990 0.023  &  9.93 0.00  &  9.92 0.00  &  9.91 0.00  &  9.85 0.00  &    --   -- & - 8 8 8 1 1 1 3 3 3 3 - & 84.253517 & +09.8179200 \\  
e004   & 16.14 0.01  & 14.85 0.01  & 13.91 0.01  & 12.94 0.01  & 11.548 0.029  & 10.859 0.023  & 10.651 0.024  & 10.50 0.00  & 10.49 0.00  & 10.44 0.01  & 10.26 0.01  &    --   -- & 2 8 8 8 6 6 6 6 6 6 6 - & 83.991539 & +09.9091240 \\  
e005c  &    --   --  &    --   --  & 21.56 0.02  & 20.69 0.02  & 19.311 0.056  & 18.085 0.046  & 17.472 0.079  & 15.89 0.04  & 15.61 0.05  & 14.68 0.12  & 14.01 0.13  &    --   -- & - - 5 5 4 4 4 3 3 3 3 - & 84.116250 & +09.9242930 \\  
e005w  &    --   --  &    --   --  & 22.39 0.05  & 21.40 0.04  & 20.258 0.144  & 19.194 0.121  & 19.055 0.345  & 16.63 0.09  & 16.82 0.15  &    --   --  &    --   --  &    --   -- & - - 5 5 4 4 4 3 3 - - - & 84.115390 & +09.9241960 \\  
e006   & 16.47 0.01  & 15.04 0.00  & 14.16 0.00  & 13.33 0.00  & 12.188 0.024  & 11.482 0.023  & 11.323 0.021  & 11.16 0.00  & 11.21 0.01  & 11.17 0.02  & 11.07 0.02  &    --   -- & 2 8 8 8 6 6 6 6 6 6 6 - & 84.085380 & +09.8720186 \\  
e007c  &    --   --  & 14.87 0.00  & 13.98 0.00  & 13.10 0.00  & 11.941 0.024  & 11.278 0.027  & 11.092 0.023  & 10.90 0.00  & 10.90 0.00  & 10.84 0.01  & 10.80 0.01  &    --   -- & - 8 8 8 6 6 6 6 6 6 6 - & 84.079263 & +10.0641340 \\  
e008   &    --   --  & 13.26 0.00  & 12.68 0.00  & 12.13 0.00  & 11.340 0.023  & 10.796 0.022  & 10.676 0.019  & 10.54 0.00  & 10.57 0.00  & 10.52 0.01  & 10.48 0.01  &    --   -- & - 8 8 8 1 1 1 3 3 3 3 - & 84.077312 & +09.7526149 \\  
e009   &    --   --  &    --   --  &    --   --  &    --   --  &  9.887 0.027  &  9.378 0.026  &  9.141 0.024  &  8.89 0.00  &  8.73 0.00  &  8.63 0.00  &  8.35 0.00  &  5.87 0.01 & - - - - 1 1 1 3 3 3 3 3 & 83.829475 & +09.9151335 \\  
e010c  &    --   --  & 12.89 0.00  & 12.28 0.00  & 11.69 0.00  & 10.936 0.023  & 10.378 0.022  & 10.235 0.021  & 10.14 0.00  & 10.13 0.00  & 10.10 0.01  & 10.05 0.01  &    --   -- & - 7 7 7 1 1 1 3 3 3 3 - & 83.976307 & +10.0731510 \\  
e011c  &    --   --  & 15.19 0.00  & 14.20 0.00  & 13.12 0.00  & 12.163 0.044  & 11.409 0.051  & 11.090 0.033  & 10.67 0.00  & 10.67 0.00  & 10.61 0.01  & 10.58 0.01  &    --   -- & - 8 8 8 6 6 6 6 6 6 6 - & 84.084070 & +09.7339210 \\  
e011e  &    --   --  &    --   --  & 15.31 0.00  & 14.18 0.00  &     --    --  &     --    --  &     --    --  &    --   --  &    --   --  &    --   --  &    --   --  &    --   -- & - - 5 5 - - - - - - - - & 84.084620 & +09.7340960 \\  
e012   & 14.62 0.01  & 13.02 0.01  &    --   --  &    --   --  &  9.110 0.021  &  8.826 0.046  &  8.690 0.021  &  8.71 0.00  &  8.55 0.00  &  8.53 0.00  &  8.33 0.00  &  6.70 0.02 & 2 2 - - 1 1 1 3 3 3 3 3 & 84.039236 & +10.0239120 \\  
e013   &    --   --  &    --   --  &    --   --  &    --   --  &  6.900 0.024  &  6.468 0.047  &  6.345 0.021  &    --   --  &    --   --  &    --   --  &    --   --  &  6.31 0.01 & - - - - 1 1 1 - - - - 3 & 84.096227 & +09.7543767 \\  
e014   & 17.50 0.01  & 15.69 0.00  & 14.64 0.00  & 13.45 0.00  & 12.102 0.023  & 11.411 0.022  & 11.156 0.019  & 11.01 0.00  & 10.98 0.00  & 10.90 0.01  & 10.68 0.01  &    --   -- & 2 7 7 7 6 6 6 6 6 6 6 - & 83.963918 & +09.9196904 \\  
e015   &    --   --  &    --   --  & 21.34 0.01  & 20.61 0.01  & 19.223 0.053  & 18.590 0.072  & 17.896 0.115  & 15.44 0.03  & 15.39 0.05  & 13.13 0.09  & 13.27 0.11  &    --   -- & - - 5 5 4 4 4 3 3 3 3 - & 84.042061 & +10.0328960 \\  
e016   & 16.81 0.01  & 15.31 0.00  & 14.37 0.00  & 13.36 0.00  & 11.991 0.024  & 11.284 0.022  & 11.090 0.023  & 10.80 0.00  & 10.80 0.00  & 10.72 0.01  & 10.64 0.01  &    --   -- & 2 8 8 8 6 6 6 6 6 6 6 - & 84.069113 & +09.8468478 \\  
e017   & 17.80 0.01  & 16.61 0.00  & 15.45 0.00  & 14.03 0.00  & 12.488 0.024  & 11.872 0.022  & 11.687 0.021  & 11.44 0.00  & 11.42 0.01  & 11.35 0.02  & 11.30 0.02  &    --   -- & 2 8 8 8 6 6 6 6 6 6 6 - & 84.078553 & +09.8598079 \\  
e018   &    --   --  & 15.74 0.00  & 14.76 0.00  & 13.75 0.00  & 12.454 0.026  & 11.744 0.022  & 11.546 0.026  & 11.36 0.00  & 11.36 0.00  & 11.33 0.01  & 11.28 0.02  &    --   -- & - 7 7 7 1 1 1 3 3 3 3 - & 84.196284 & +10.0977120 \\  
e019   &    --   --  & 16.05 0.00  & 15.02 0.00  & 13.91 0.00  & 12.695 0.028  & 11.958 0.023  & 11.767 0.021  & 11.57 0.00  & 11.58 0.00  & 11.55 0.01  & 11.48 0.02  &    --   -- & - 7 7 7 1 1 1 3 3 3 3 - & 84.239585 & +09.8905510 \\  
e020   & 14.13 0.01  & 11.78 0.00  & 11.17 0.00  & 10.57 0.00  &  9.729 0.024  &  9.303 0.023  &  9.147 0.019  &  9.15 0.00  &  9.04 0.00  &  9.01 0.00  &  8.99 0.00  &    --   -- & 2 8 8 8 1 1 1 3 3 3 3 - & 84.068451 & +09.9903775 \\  
e021c  &    --   --  &    --   --  &    --   --  &    --   --  & 20.181 0.111  & 19.036 0.107  & 18.478 0.212  & 16.20 0.05  & 15.73 0.06  & 14.53 0.10  &    --   --  &    --   -- & - - - - ? ? ? 3 3 3 - - & 84.166840 & +10.0761670 \\  
e021w  &    --   --  &    --   --  &    --   --  &    --   --  & 19.944 0.090  & 19.313 0.141  & 19.705 0.620  & 17.07 0.09  & 16.60 0.11  &    --   --  &    --   --  &    --   -- & - - - - 4 4 4 3 3 - - - & 84.165860 & +10.0760410 \\  
e022   & 18.22 0.01  & 16.46 0.00  & 15.33 0.00  & 13.94 0.00  & 12.410 0.029  & 11.714 0.023  & 11.493 0.021  & 11.25 0.00  & 11.22 0.01  & 11.18 0.02  & 11.08 0.02  &    --   -- & 2 7 7 7 6 6 6 6 6 6 6 - & 84.038801 & +09.7843548 \\  
e023   &    --   --  &    --   --  &    --   --  &    --   --  & 11.359 0.022  & 10.780 0.023  & 10.548 0.021  & 10.29 0.00  & 10.25 0.00  & 10.19 0.01  & 10.13 0.01  &    --   -- & - - - - 6 6 6 6 6 6 6 - & 83.948123 & +09.7640605 \\  
e024   &    --   --  &    --   --  & 20.73 0.01  & 19.90 0.01  & 18.920 0.020  & 18.048 0.025  & 17.320 0.016  & 16.02 0.07  & 14.82 0.03  &    --   --  & 13.01 0.06  &    --   -- & - - 5 5 9 9 9 3 3 - 3 - & 83.930262 & +09.9989864 \\  
e025   & 18.04 0.01  & 16.31 0.00  & 15.29 0.00  & 14.19 0.00  & 12.866 0.026  & 12.153 0.022  & 11.931 0.027  & 11.82 0.00  & 11.81 0.00  & 11.74 0.01  & 11.79 0.02  &    --   -- & 2 8 8 8 1 1 1 3 3 3 3 - & 84.109597 & +09.8539695 \\  
e026   &    --   --  &    --   --  &    --   --  &    --   --  & 19.568 0.069  & 18.219 0.054  & 17.390 0.073  & 15.66 0.04  & 15.27 0.04  & 14.36 0.12  &    --   --  &    --   -- & - - - - 4 4 4 3 3 3 - - & 84.139100 & +09.9674730 \\  
e027c  & 18.22 0.01  & 16.57 0.00  & 15.46 0.00  & 14.20 0.00  & 12.893 0.029  & 12.108 0.023  & 11.945 0.024  & 11.77 0.00  & 11.76 0.00  & 11.67 0.02  & 11.70 0.02  &    --   -- & 2 8 8 8 1 1 1 3 3 3 3 - & 84.120530 & +09.9075260 \\  
e027n  &    --   --  &    --   --  &    --   --  & 20.65 0.18  & 19.121 0.050  & 18.540 0.071  & 18.141 0.153  &    --   --  &    --   --  &    --   --  &    --   --  &    --   -- & - - - 5 4 4 4 - - - - - & 84.120420 & +09.9089960 \\  
e028c  & 17.13 0.01  & 15.68 0.00  & 14.65 0.00  & 13.55 0.00  & 12.221 0.027  & 11.471 0.022  & 11.290 0.024  & 11.09 0.00  & 11.11 0.00  & 11.07 0.01  & 10.93 0.02  &    --   -- & 2 7 7 7 6 6 6 6 6 6 6 - & 83.990196 & +09.7929641 \\  
e029c  &    --   --  & 15.87 0.00  & 14.80 0.00  & 13.61 0.00  & 12.455 0.033  & 11.800 0.042  & 11.502 0.027  & 11.15 0.00  & 11.15 0.00  & 11.06 0.01  & 11.02 0.02  &    --   -- & - 8 8 8 6 6 6 6 6 6 6 - & 83.895130 & +10.0097960 \\  
e030   &    --   --  & 15.22 0.00  & 14.30 0.00  & 13.39 0.00  & 12.151 0.023  & 11.455 0.022  & 11.250 0.023  & 11.03 0.00  & 11.01 0.00  & 10.92 0.01  & 10.75 0.01  &  8.64 0.08 & - 7 7 7 1 1 1 3 3 3 3 3 & 84.012193 & +09.7017196 \\  
e031   &    --   --  & 16.69 0.00  & 15.50 0.00  & 14.06 0.00  & 12.512 0.023  & 11.841 0.022  & 11.665 0.024  & 11.41 0.00  & 11.38 0.00  & 11.37 0.01  & 11.38 0.02  &    --   -- & - 7 7 7 1 1 1 3 3 3 3 - & 84.219484 & +09.8823130 \\  
e032   &    --   --  &    --   --  &    --   --  &    --   --  & 18.058 0.026  & 17.211 0.022  & 16.320 0.025  & 15.04 0.04  & 14.28 0.03  & 13.77 0.11  & 12.61 0.05  &  8.58 0.08 & - - - - 4 4 4 3 3 3 3 3 & 83.907920 & +09.7361380 \\  
e033   &    --   --  &    --   --  & 23.11 0.07  & 22.16 0.06  & 19.881 0.037  & 19.564 0.081  & 18.767 0.054  & 17.44 0.12  & 16.42 0.10  &    --   --  &    --   --  &    --   -- & - - 5 5 9 9 9 3 3 - - - & 83.997733 & +09.8385000 \\  
e034c  &    --   --  & 15.43 0.00  & 14.48 0.00  & 13.61 0.00  & 12.309 0.023  & 11.629 0.026  & 11.459 0.024  & 11.31 0.00  & 11.37 0.00  & 11.31 0.01  & 11.24 0.02  &    --   -- & - 7 7 7 1 1 1 3 3 3 3 - & 83.842550 & +09.8743420 \\  
e034w  &    --   --  &    --   --  &    --   --  &    --   --  & 18.780 0.017  & 18.268 0.028  & 18.219 0.042  &    --   --  &    --   --  &    --   --  &    --   --  &    --   -- & - - - - 9 9 9 - - - - - & 83.841760 & +09.8738260 \\  
e035c  &    --   --  & 16.76 0.00  & 15.54 0.00  & 14.05 0.00  & 12.553 0.024  & 11.877 0.022  & 11.594 0.024  & 11.36 0.00  & 11.32 0.01  & 11.23 0.02  & 11.22 0.03  &    --   -- & - 7 7 7 6 6 6 6 6 6 6 - & 83.914530 & +09.8425200 \\  
e036   &    --   --  & 17.22 0.00  & 15.87 0.00  & 14.23 0.00  & 12.500 0.024  & 11.856 0.023  & 11.587 0.027  & 11.26 0.00  & 11.19 0.01  & 11.13 0.01  & 11.12 0.02  &    --   -- & - 7 7 7 6 6 6 6 6 6 6 - & 83.876938 & +09.8429168 \\  
e037c  & 20.12 0.01  & 18.40 0.01  & 16.78 0.01  & 15.34 0.01  & 13.782 0.026  & 13.098 0.025  & 12.846 0.029  & 12.49 0.01  & 12.49 0.01  & 12.38 0.03  & 12.24 0.05  &    --   -- & 2 2 6 6 6 6 6 6 6 6 6 - & 83.966620 & +09.8416020 \\  
e038c  &    --   --  &    --   --  & 20.52 0.09  & 19.74 0.06  & 18.333 0.014  & 17.425 0.014  & 16.621 0.010  & 15.25 0.03  & 14.90 0.04  & 14.35 0.10  & 13.80 0.11  &    --   -- & - - 5 5 9 9 9 3 3 3 3 - & 83.863320 & +09.8863180 \\  
e038w  &    --   --  &    --   --  &    --   --  &    --   --  & 19.986 0.054  & 19.474 0.084  & 18.714 0.062  &    --   --  &    --   --  &    --   --  &    --   --  &    --   -- & - - - - 9 9 9 - - - - - & 83.862750 & +09.8862350 \\  
e039   &    --   --  &    --   --  &    --   --  &    --   --  &     --    --  &     --    --  &     --    --  & 16.28 0.06  & 15.90 0.07  &    --   --  &    --   --  &    --   -- & - - - - - - - 3 3 - - - & 84.235030 & +09.8939090 \\  
e040   & 19.04 0.09  & 17.43 0.00  & 16.21 0.00  & 14.76 0.00  & 13.237 0.024  & 12.534 0.022  & 12.304 0.024  & 12.10 0.00  & 12.06 0.00  & 12.01 0.02  & 12.09 0.03  &    --   -- & 2 8 8 8 1 1 1 3 3 3 3 - & 84.209405 & +09.9066000 \\  
\hline 
\end{tabular} 
$\,$\\
Catalog IDs:  1 -- 2MASS, 2 -- Optical Monitor (XMM), 3 -- Spitzer, 4 -- O2000/2005, 5 -- CFHT1999, 6 -- Barrado 2004$+$2007, 7 -- Dolan\&Mathieu 1999+2001, 8 -- Dolan\&Mathieu 2002, 9 -- O2000/2007
\end{table}
\end{landscape}


\newpage
\addtocounter{table}{-1}
\begin{landscape}
\begin{table}
\tiny
\caption{Optical and infrared photometry for all the 205  counterparts for 164 X-ray sources of Collinder 69 (eastern {\rm XMM-Newton} pointing or Col 69\,E).}
\label{tab:photometry} 
\begin{tabular}{l l               l               l              l              l               l               l               l               l               l               l               l               l                         l           l} \\ 
Name    &   $B$    &   $V$       & $Rc$        &    $Ic$     &   $J$         &   $H$         &   $K$         &   [$3.6$]   &   [$4.5$]   &   [$5.8$]   &   [$8.0$]   &  [$24$]    &  Notes                   & R.A.      & DEC        \\   %
C69-X-\#   &   [mag]  &   [mag]     & [mag]       &    [mag]    &   [mag]       &   [mag]       &   [mag]       &   [mag]     &   [mag]     &   [mag]     &   [mag]     &  [mag]     &                          & (2000)    & (2000)     \\   
\hline      
\hline
 e041  &    --   -- & 16.72 0.00 & 15.64 0.00 & 14.30 0.00 & 12.857 0.024 & 12.081 0.022 & 11.768 0.024 &    --   -- &    --   -- &     --   -- &    --   -- &  6.99 0.05 &  - 7 7 7 1 1 1 - - - - 3 & 84.292398 & +09.9241550 \\  
 e042  &    --   -- &    --   -- &    --   -- &    --   -- &  7.358 0.019 &  7.130 0.033 &  7.050 0.026 &    --   -- &    --   -- &   6.99 0.00 &  7.00 0.00 &  7.08 0.02 &  - - - - 1 1 1 - - 3 3 3 & 84.192595 & +09.8873230 \\  
 e043  &    --   -- &    --   -- & 22.31 0.03 & 21.61 0.04 & 20.341 0.052 & 19.307 0.069 & 18.702 0.053 & 16.78 0.11 & 16.27 0.11 &     --   -- &    --   -- &    --   -- &  - - 5 5 9 9 9 3 3 - - - & 83.958290 & +09.9645592 \\  
 e044  &    --   -- &    --   -- &    --   -- &    --   -- &     --    -- &     --    -- &     --    -- & 16.75 0.07 & 15.87 0.07 &     --   -- &    --   -- &    --   -- &  - - - - - - - 3 3 - - - & 84.111610 & +09.8219420 \\  
 e045  &    --   -- &    --   -- & 22.75 0.00 & 22.75 0.00 & 21.855 0.216 & 20.511 0.207 & 19.410 0.105 &    --   -- &    --   -- &     --   -- &    --   -- &    --   -- &  - - 5 5 9 9 9 - - - - - & 83.877360 & +09.9081050 \\  
 e046  &    --   -- &    --   -- & 22.75 0.00 & 22.18 0.06 & 20.503 0.062 & 19.273 0.061 & 18.213 0.035 & 16.48 0.06 & 16.05 0.10 &     --   -- &    --   -- &    --   -- &  - - 5 5 9 9 9 3 3 - - - & 83.889187 & +09.8632789 \\  
 e047  &    --   -- &    --   -- &    --   -- &    --   -- & 14.524 0.032 & 13.928 0.045 & 13.668 0.043 & 13.27 0.01 & 13.18 0.01 &  13.22 0.05 & 13.07 0.06 &    --   -- &  - - - - 1 1 1 3 3 3 3 - & 84.227258 & +09.8907420 \\  
 e048  &    --   -- &    --   -- &    --   -- &    --   -- &    $>2$1.000 &    $>$19.750 &    $>$18.750 &    --   -- &    --   -- &     --   -- &    --   -- &    --   -- &  - - - - 4 4 4 - - - - - & 84.211400 & +09.9014610 \\  
 e049  &    --   -- &    --   -- & 21.80 0.02 & 20.86 0.02 & 20.069 0.045 & 19.398 0.068 & 18.788 0.059 & 17.03 0.10 & 16.59 0.14 &     --   -- &    --   -- &    --   -- &  - - 5 5 9 9 9 3 3 - - - & 83.953577 & +09.8233012 \\  
 e050  &    --   -- &    --   -- & 21.64 0.02 & 20.71 0.01 &    $>$21.000 &    $>$19.750 &    $>$18.750 &    --   -- &    --   -- &     --   -- &    --   -- &    --   -- &  - - 5 5 4 4 4 - - - - - & 84.072762 & +10.0276070 \\  
 e051  &    --   -- &    --   -- & 20.69 0.01 & 19.96 0.01 & 19.669 0.029 & 18.867 0.041 & 18.139 0.030 & 15.71 0.04 & 15.56 0.05 &  14.19 0.13 & 13.24 0.09 &    --   -- &  - - 5 5 9 9 9 3 3 3 3 - & 83.988965 & +09.9741924 \\  
 e052  &    --   -- & 17.90 0.00 & 17.24 0.00 & 16.51 0.00 & 15.105 0.003 & 14.393 0.002 & 14.184 0.004 & 14.14 0.01 & 14.10 0.02 &  14.02 0.07 &    --   -- &    --   -- &  - 8 8 8 4 4 4 3 3 3 - - & 83.836844 & +09.7908774 \\  
 e053  &    --   -- &    --   -- & 23.63 0.11 & 22.75 0.10 & 21.359 0.128 &    $>$20.750 & 19.524 0.103 & 17.62 0.13 & 16.79 0.13 &     --   -- &    --   -- &    --   -- &  - - 5 5 9 9 9 3 3 - - - & 84.058310 & +09.8848590 \\  
 e054  & 19.05 0.01 & 17.45 0.01 & 15.91 0.01 & 14.60 0.01 & 13.266 0.024 & 12.559 0.022 & 12.285 0.021 & 12.02 0.01 & 11.99 0.01 &  11.97 0.02 & 11.92 0.04 &    --   -- &  2 2 6 6 6 6 6 6 6 6 6 - & 84.050619 & +10.0159390 \\  
 e055  & 19.35 0.01 & 17.71 0.01 & 16.19 0.01 & 14.73 0.01 & 13.189 0.024 & 12.509 0.022 & 12.271 0.027 & 11.97 0.01 & 11.95 0.01 &  11.81 0.03 & 11.86 0.04 &    --   -- &  2 2 6 6 6 6 6 6 6 6 6 - & 83.968848 & +09.8087980 \\  
 e056  &    --   -- &    --   -- &    --   -- &    --   -- & 21.011 0.315 & 19.800 0.214 & 18.230 0.151 &    --   -- & 16.00 0.14 &     --   -- &    --   -- &    --   -- &  - - - - 4 4 4 - 3 - - - & 84.183930 & +09.8681680 \\  
 e057  & 13.30 0.01 & 12.03 0.01 & 11.30 0.00 & 10.71 0.00 &  9.897 0.026 &  9.294 0.022 &  9.191 0.024 &  9.17 0.00 &  9.15 0.00 &   9.10 0.00 &  9.09 0.00 &    --   -- &  2 8 8 8 1 1 1 3 3 3 3 - & 84.205351 & +09.9722170 \\  
 e058  &    --   -- & 15.54 0.00 & 14.57 0.00 & 13.63 0.00 & 12.418 0.024 & 11.717 0.023 & 11.522 0.021 & 11.37 0.00 & 11.42 0.00 &  11.31 0.01 & 11.26 0.02 &    --   -- &  - 7 7 7 1 1 1 3 3 3 3 - & 84.228583 & +09.8402840 \\  
 e059  &    --   -- &    --   -- & 22.75 0.00 & 22.75 0.00 &    $>$21.250 &    $>$20.750 &    $>$19.750 &    --   -- &    --   -- &     --   -- &    --   -- &    --   -- &  - - 5 5 9 9 9 - - - - - & 84.061930 & +09.8682130 \\  
 e060c &    --   -- &    --   -- &    --   -- &    --   -- & 12.206 0.026 & 11.572 0.022 & 11.439 0.021 & 11.38 0.00 & 11.39 0.00 &  11.36 0.01 & 11.30 0.02 &    --   -- &  - - - - 1 1 1 3 3 3 3 - & 84.036570 & +09.9392720 \\  
 e060n &    --   -- &    --   -- & 20.50 0.02 & 19.68 0.01 & 20.293 0.058 & 20.148 0.152 & 19.048 0.088 &    --   -- &    --   -- &     --   -- &    --   -- &    --   -- &  - - 5 5 9 9 9 - - - - - & 84.038300 & +09.9396290 \\  
 e061  &    --   -- &    --   -- & 21.16 0.01 & 20.42 0.01 & 19.995 0.093 & 18.893 0.089 & 18.448 0.194 & 16.11 0.05 & 15.39 0.05 &     --   -- &    --   -- &    --   -- &  - - 5 5 4 4 4 3 3 - - - & 83.986235 & +10.0312760 \\  
 e062  &    --   -- &    --   -- &    --   -- &    --   -- & 11.386 0.024 & 11.018 0.026 & 10.923 0.021 & 10.86 0.00 & 10.89 0.00 &  10.84 0.01 & 10.86 0.01 &    --   -- &  - - - - 1 1 1 3 3 3 3 - & 84.011607 & +10.0942480 \\  
 e063  &    --   -- &    --   -- & 23.01 0.06 & 21.99 0.05 & 20.671 0.176 & 19.917 0.218 & 18.508 0.194 & 16.77 0.09 & 16.06 0.10 &     --   -- &    --   -- &    --   -- &  - - 5 5 4 4 4 3 3 - - - & 83.977594 & +10.0311300 \\  
 e064  &    --   -- &    --   -- & 15.61 0.00 & 14.43 0.00 & 13.083 0.027 & 12.380 0.027 & 12.137 0.024 & 12.02 0.00 & 11.89 0.01 &  11.86 0.02 & 11.85 0.03 &    --   -- &  - - 5 5 1 1 1 3 3 3 3 - & 83.842427 & +09.8995644 \\  
 e065  &    --   -- &    --   -- &    --   -- &    --   -- & 13.201 0.026 & 12.625 0.027 & 12.499 0.026 & 12.25 0.00 & 12.45 0.01 &  13.34 0.04 & 12.41 0.05 &    --   -- &  - - - - 1 1 1 3 3 3 3 - & 83.908829 & +09.8878650 \\  
 e066  &    --   -- &    --   -- & 20.55 0.10 & 19.87 0.07 &     --    -- &     --    -- &     --    -- & 15.71 0.04 & 14.85 0.03 &  13.83 0.07 & 12.72 0.06 &  8.50 0.08 &  - - 5 5 - - - 3 3 3 3 3 & 83.909390 & +10.0174980 \\  
 e067  & 13.75 0.01 & 12.94 0.01 &    --   -- &    --   -- & 11.184 0.026 & 10.943 0.022 & 10.847 0.023 & 10.77 0.00 & 10.77 0.00 &  10.77 0.01 & 10.76 0.01 &    --   -- &  2 2 - - 1 1 1 3 3 3 3 - & 84.035970 & +09.8889444 \\  
 e068  & 14.76 0.01 & 13.81 0.01 & 13.36 0.00 & 12.91 0.00 & 12.281 0.027 & 12.013 0.023 & 11.910 0.023 & 11.85 0.00 & 11.88 0.00 &  11.52 0.01 & 11.81 0.03 &    --   -- &  2 8 8 8 1 1 1 3 3 3 3 - & 84.084210 & +09.9341020 \\  
 e069c & 16.63 0.01 & 15.50 0.01 & 14.73 0.00 & 14.11 0.00 & 13.268 0.023 & 12.761 0.025 & 12.601 0.029 & 12.57 0.00 & 12.59 0.01 &  12.53 0.02 & 12.54 0.04 &    --   -- &  2 2 5 5 1 1 1 3 3 3 3 - & 83.965470 & +09.8927750 \\  
 e070  &    --   -- &    --   -- &    --   -- &    --   -- &    $>$21.000 &    $>$19.750 &    $>$18.750 &    --   -- &    --   -- &     --   -- &    --   -- &    --   -- &  - - - - 4 4 4 - - - - - & 84.056270 & +10.0212060 \\  
 e071c &    --   -- &    --   -- & 13.95 0.00 & 13.28 0.00 & 12.313 0.026 & 11.691 0.025 & 11.556 0.026 & 11.45 0.00 & 11.48 0.00 &  11.44 0.01 & 11.35 0.02 &  7.56 0.04 &  - - 5 5 1 1 1 3 3 3 3 3 & 83.871690 & +09.7756380 \\  
 e072  &    --   -- & 17.60 0.00 & 16.44 0.00 & 14.92 0.00 & 13.436 0.032 & 12.763 0.023 & 12.537 0.027 & 12.28 0.00 & 12.27 0.01 &  12.17 0.02 & 12.23 0.03 &    --   -- &  - 8 8 8 1 1 1 3 3 3 3 - & 84.114436 & +09.7571574 \\  
 e073  & 17.11 0.01 & 16.24 0.01 & 15.70 0.00 & 15.14 0.00 & 14.696 0.002 & 14.322 0.002 & 14.217 0.004 & 13.97 0.01 & 13.93 0.02 &  13.80 0.06 & 13.60 0.10 &    --   -- &  2 2 5 5 4 4 4 3 3 3 3 - & 84.048085 & +09.9945899 \\  
 e074  &    --   -- &    --   -- &    --   -- &    --   -- & 18.371 0.028 & 17.257 0.022 & 16.554 0.034 & 15.54 0.04 & 15.58 0.08 &     --   -- & 13.75 0.14 &    --   -- &  - - - - 4 4 4 3 3 - 3 - & 84.148170 & +09.9066340 \\  
 e075  &    --   -- &    --   -- & 22.17 0.03 & 21.42 0.03 & 20.197 0.047 & 19.689 0.091 & 19.118 0.077 & 17.16 0.09 & 16.54 0.11 &     --   -- &    --   -- &    --   -- &  - - 5 5 9 9 9 3 3 - - - & 83.885503 & +09.9652674 \\  
 e076  &    --   -- &    --   -- &    --   -- &    --   -- & 20.068 0.127 & 18.940 0.093 & 18.651 0.239 &    --   -- &    --   -- &     --   -- &    --   -- &    --   -- &  - - - - 4 4 4 - - - - - & 84.192920 & +09.9130050 \\  
 e077  & 19.57 0.01 & 17.66 0.01 & 16.67 0.00 & 15.74 0.00 & 14.806 0.073 & 13.871 0.096 & 13.490 0.074 & 12.68 0.01 & 12.70 0.01 &  12.59 0.03 &    --   -- &    --   -- &  2 8 8 8 1 1 1 3 3 3 - - & 84.004794 & +09.9793416 \\  
 e078  &    --   -- &    --   -- & 17.77 0.01 & 16.17 0.01 & 14.515 0.041 & 13.881 0.023 & 13.651 0.051 & 13.23 0.01 & 13.12 0.01 &  12.93 0.05 & 13.13 0.10 &    --   -- &  - - 6 6 6 6 6 6 6 6 6 - & 84.112749 & +09.8597427 \\  
 e079  &    --   -- &    --   -- & 16.12 0.01 & 14.76 0.01 & 13.184 0.026 & 12.477 0.026 & 12.253 0.026 & 12.04 0.01 & 12.04 0.01 &  12.02 0.03 & 11.90 0.04 &    --   -- &  - - 6 6 6 6 6 6 6 6 6 - & 83.839291 & +09.8324832 \\  
 e080  &    --   -- &    --   -- &    --   -- &    --   -- &     --    -- &     --    -- &     --    -- & 15.37 0.03 & 14.69 0.03 &  14.34 0.13 & 12.99 0.07 &  8.94 0.10 &  - - - - - - - 3 3 3 3 3 & 84.240200 & +10.0544470 \\  
\hline 
\end{tabular} 
$\,$\\
Catalog IDs:  1 -- 2MASS, 2 -- Optical Monitor (XMM), 3 -- Spitzer, 4 -- O2000/2005, 5 -- CFHT1999, 6 -- Barrado 2004$+$2007, 7 -- Dolan\&Mathieu 1999+2001, 8 -- Dolan\&Mathieu 2002, 9 -- O2000/2007
\end{table}
\end{landscape}


\addtocounter{table}{-1}
\begin{landscape}
\begin{table}
\tiny
\caption{Optical and infrared photometry for all the 205  counterparts for 164 X-ray sources of Collinder 69 (eastern {\rm XMM-Newton} pointing or Col 69\,E). } 
\label{tab:photometry} 
\begin{tabular}{l l               l               l              l              l               l               l               l               l               l               l               l               l                         l           l} \\ 
Name    &   $B$    &   $V$       & $Rc$        &    $Ic$     &   $J$         &   $H$         &   $K$         &   [$3.6$]   &   [$4.5$]   &   [$5.8$]   &   [$8.0$]   &  [$24$]    &  Notes                   & R.A.      & DEC        \\   %
C69-X-\#   &   [mag]  &   [mag]     & [mag]       &    [mag]    &   [mag]       &   [mag]       &   [mag]       &   [mag]     &   [mag]     &   [mag]     &   [mag]     &  [mag]     &                          & (2000)    & (2000)     \\   
\hline      
\hline      
 e081c & 19.07 0.01 & 17.41 0.00 & 16.08 0.00 & 14.41 0.00 & 12.732 0.026 & 12.097 0.031 & 11.827 0.026 & 11.47 0.00 & 11.40 0.01 &  11.30 0.02 & 11.34 0.03 &    --   -- &  2 7 7 7 6 6 6 6 6 6 6 - & 83.981820 & +09.8482590 \\  
 e081e &    --   -- &    --   -- & 19.46 0.00 & 18.31 0.00 &     --    -- &     --    -- &     --    -- &    --   -- &    --   -- &     --   -- &    --   -- &    --   -- &  - - 5 5 - - - - - - - - & 83.982440 & +09.8475150 \\  
 e082  & 18.32 0.01 & 16.79 0.00 & 15.63 0.00 & 14.18 0.00 & 12.755 0.030 & 12.004 0.023 & 11.775 0.023 & 11.52 0.00 & 11.53 0.01 &  11.43 0.02 & 11.37 0.03 &    --   -- &  2 7 7 7 6 6 6 6 6 6 6 - & 83.987755 & +09.7813924 \\  
 e083c &    --   -- &    --   -- &    --   -- &    --   -- & 19.663 0.075 & 18.949 0.092 & 18.475 0.181 & 17.25 0.09 & 16.53 0.10 &  15.30 0.28 & 13.71 0.10 &    --   -- &  - - - - 4 4 4 3 3 3 3 - & 84.195440 & +09.9395310 \\  
 e083n &    --   -- &    --   -- &    --   -- &    --   -- & 19.204 0.050 & 18.541 0.063 & 18.845 0.253 & 17.51 0.13 &    --   -- &     --   -- &    --   -- &    --   -- &  - - - - 4 4 4 3 - - - - & 84.195150 & +09.9403400 \\  
 e084  &    --   -- & 17.23 0.00 & 16.01 0.00 & 14.66 0.00 & 13.230 0.026 & 12.518 0.025 & 12.270 0.026 & 12.08 0.00 & 12.08 0.01 &  12.02 0.02 & 12.12 0.04 &    --   -- &  - 7 7 7 1 1 1 3 3 3 3 - & 83.899019 & +09.7431743 \\  
 e085  &    --   -- &    --   -- &    --   -- &    --   -- & 22.373 0.293 & 21.927 0.694 & 21.067 0.409 &    --   -- &    --   -- &     --   -- &    --   -- &    --   -- &  - - - - 9 9 9 - - - - - & 84.057760 & +09.8752750 \\  
 e086  & 20.11 0.22 & 18.50 0.01 &    --   -- &    --   -- & 13.760 0.024 & 13.099 0.027 & 12.841 0.029 & 12.60 0.00 & 12.54 0.01 &  12.39 0.03 & 12.41 0.04 &    --   -- &  2 2 - - 1 1 1 3 3 3 3 - & 84.221250 & +09.8387320 \\  
 e087c &    --   -- &    --   -- & 23.03 0.06 & 22.10 0.06 &     --    -- &     --    -- &     --    -- & 16.71 0.05 & 15.75 0.04 &  15.04 0.17 & 13.88 0.11 &    --   -- &  - - 5 5 - - - 3 3 3 3 - & 83.956080 & +10.0762610 \\  
 e087s &    --   -- &    --   -- & 23.80 0.14 & 22.58 0.09 &     --    -- &     --    -- &     --    -- &    --   -- &    --   -- &     --   -- &    --   -- &    --   -- &  - - 5 5 - - - - - - - - & 83.955540 & +10.0755990 \\  
 e088n &    --   -- &    --   -- & 22.07 0.03 & 21.28 0.03 &     --    -- &     --    -- &     --    -- & 17.48 0.10 & 17.47 0.20 &     --   -- &    --   -- &    --   -- &  - - 5 5 - - - 3 3 - - - & 84.125470 & +09.7276300 \\  
 e088s &    --   -- &    --   -- & 18.72 0.00 & 17.97 0.00 &     --    -- &     --    -- &     --    -- & 16.17 0.04 & 16.12 0.08 &     --   -- &    --   -- &    --   -- &  - - 5 5 - - - 3 3 - - - & 84.127460 & +09.7248470 \\  
 e089  &    --   -- &    --   -- &    --   -- &    --   -- &     --    -- &     --    -- &     --    -- & 16.99 0.11 & 16.45 0.13 &     --   -- &    --   -- &    --   -- &  - - - - - - - 3 3 - - - & 84.237180 & +09.8816230 \\  
 e090  &    --   -- &    --   -- & 22.06 0.03 & 21.45 0.04 & 20.083 0.043 & 19.129 0.057 & 18.406 0.039 & 16.65 0.06 & 15.92 0.06 &     --   -- &    --   -- &    --   -- &  - - 5 5 9 9 9 3 3 - - - & 83.932880 & +09.9341180 \\  
 e091  &    --   -- &    --   -- & 18.12 0.01 & 16.40 0.01 & 14.647 0.037 & 13.985 0.045 & 13.682 0.039 & 13.39 0.01 & 13.30 0.02 &  13.28 0.08 & 13.18 0.12 &    --   -- &  - - 6 6 6 6 6 6 6 6 6 - & 84.131387 & +09.7503815 \\  
 e092  &    --   -- &    --   -- &    --   -- &    --   -- & 13.053 0.024 & 12.482 0.023 & 12.161 0.021 &    --   -- &    --   -- &     --   -- &    --   -- &    --   -- &  - - - - 1 1 1 - - - - - & 84.293353 & +09.8575880 \\  
 e093  &    --   -- &    --   -- & 20.72 0.01 & 20.17 0.01 & 19.240 0.023 & 18.403 0.028 & 18.036 0.030 & 16.34 0.04 & 15.97 0.05 &  14.66 0.09 & 13.73 0.07 &    --   -- &  - - 5 5 9 9 9 3 3 3 3 - & 83.856160 & +09.7949190 \\  
 e094  &    --   -- &    --   -- & 21.11 0.01 & 20.07 0.01 &     --    -- &     --    -- &     --    -- & 15.97 0.04 & 15.65 0.07 &     --   -- &    --   -- &    --   -- &  - - 5 5 - - - 3 3 - - - & 83.965231 & +10.1099520 \\  
 e095  &    --   -- &    --   -- &    --   -- &    --   -- &     --    -- &     --    -- &     --    -- & 16.33 0.05 & 15.43 0.05 &     --   -- & 13.72 0.14 &    --   -- &  - - - - - - - 3 3 - 3 - & 84.233290 & +09.8680370 \\  
 e096c & 19.91 0.01 & 18.23 0.01 & 16.57 0.01 & 15.06 0.01 & 13.521 0.024 & 12.935 0.022 & 12.643 0.027 & 12.33 0.01 & 12.27 0.01 &  12.17 0.03 & 12.64 0.07 &    --   -- &  2 2 6 6 6 6 6 6 6 6 6 - & 84.043210 & +10.0053220 \\  
 e097  &    --   -- &    --   -- &    --   -- &    --   -- & 16.374 0.130 & 15.676 0.168 & 14.744 0.123 &    --   -- &    --   -- &     --   -- &    --   -- &    --   -- &  - - - - 1 1 1 - - - - - & 84.315251 & +09.9638890 \\  
 e098  &    --   -- &    --   -- &    --   -- &    --   -- & 15.168 0.054 & 14.439 0.049 & 14.183 0.065 & 13.94 0.01 & 13.86 0.01 &  13.76 0.07 & 13.78 0.13 &    --   -- &  - - - - 1 1 1 3 3 3 3 - & 84.104194 & +09.6920190 \\  
 e099  &    --   -- &    --   -- & 21.94 0.03 & 20.85 0.02 & 19.216 0.022 & 18.164 0.022 & 17.320 0.016 & 15.90 0.04 & 15.40 0.05 &     --   -- &    --   -- &    --   -- &  - - 5 5 9 9 9 3 3 - - - & 84.059388 & +09.7965685 \\  
 e100  &    --   -- &    --   -- &    --   -- &    --   -- &    $>$21.000 &    $>$19.750 &    $>$18.750 &    --   -- &    --   -- &     --   -- &    --   -- &    --   -- &  - - - - 4 4 4 - - - - - & 84.145230 & +09.9778570 \\  
 e101  &    --   -- &    --   -- &    --   -- &    --   -- &     --    -- &     --    -- &     --    -- &    --   -- &    --   -- &     --   -- &    --   -- &    --   -- &  - - - - - - - - - - - - & 84.167280 & +09.8100580 \\  
 e102c &    --   -- &    --   -- & 18.82 0.00 & 18.03 0.00 & 17.069 0.005 & 16.338 0.005 & 16.240 0.007 & 15.53 0.04 & 15.48 0.06 &  15.04 0.22 & 15.10 0.48 &    --   -- &  - - 5 5 9 9 9 3 3 3 3 - & 83.877360 & +09.9374070 \\  
 e102e &    --   -- &    --   -- & 21.06 0.02 & 19.82 0.01 & 18.385 0.012 & 17.670 0.015 & 17.411 0.018 &    --   -- &    --   -- &     --   -- &    --   -- &    --   -- &  - - 5 5 9 9 9 - - - - - & 83.877930 & +09.9372070 \\  
 e102s &    --   -- &    --   -- & 20.93 0.01 & 20.28 0.01 & 19.339 0.027 & 18.598 0.035 & 18.057 0.032 & 15.36 0.03 & 15.03 0.04 &  14.33 0.12 & 13.56 0.10 &    --   -- &  - - 5 5 9 9 9 3 3 3 3 - & 83.877830 & +09.9367550 \\  
 e103c &    --   -- &    --   -- & 22.57 0.07 & 21.41 0.03 & 20.170 0.052 & 19.023 0.060 & 18.285 0.046 & 17.44 0.30 & 17.26 0.35 &     --   -- &    --   -- &    --   -- &  - - 5 5 9 9 9 3 3 - - - & 83.885830 & +09.9363730 \\  
 e103s &    --   -- &    --   -- &    --   -- & 22.46 0.09 & 20.737 0.079 & 19.816 0.097 & 18.789 0.061 & 16.99 0.12 & 17.05 0.25 &     --   -- &    --   -- &    --   -- &  - - - 5 9 9 9 3 3 - - - & 83.885870 & +09.9351430 \\  
 e103w &    --   -- &    --   -- & 22.25 0.05 & 20.78 0.02 & 19.180 0.021 & 18.156 0.024 & 17.266 0.015 & 15.61 0.04 & 15.70 0.08 &     --   -- &    --   -- &    --   -- &  - - 5 5 9 9 9 3 3 - - - & 83.884460 & +09.9354380 \\  
 e104c & 19.19 0.01 & 17.56 0.01 & 16.18 0.00 & 14.86 0.00 & 13.431 0.001 & 12.732 0.001 & 12.436 0.001 & 12.29 0.00 & 12.26 0.01 &  12.19 0.02 & 12.11 0.03 &    --   -- &  2 2 5 5 4 4 4 3 3 3 3 - & 83.981540 & +09.8694630 \\  
 e104e &    --   -- &    --   -- &    --   -- &    --   -- & 19.331 0.032 & 18.951 0.058 & 18.637 0.057 &    --   -- &    --   -- &     --   -- &    --   -- &    --   -- &  - - - - 9 9 9 - - - - - & 83.982382 & +09.8691280 \\  
 e105  &    --   -- &    --   -- &    --   -- &    --   -- & 14.058 0.028 & 13.439 0.036 & 13.099 0.031 & 12.81 0.01 & 12.79 0.01 &  12.81 0.03 & 12.66 0.05 &    --   -- &  - - - - 1 1 1 3 3 3 3 - & 84.230945 & +10.0196270 \\  
 e106  &    --   -- &    --   -- &    --   -- &    --   -- & 19.958 0.116 & 18.538 0.083 & 17.601 0.086 & 16.35 0.04 & 15.02 0.03 &  13.94 0.06 & 12.98 0.06 &    --   -- &  - - - - 4 4 4 3 3 3 3 - & 84.002820 & +09.6865570 \\  
 e107  &    --   -- &    --   -- & 19.69 0.00 & 18.72 0.00 & 17.646 0.012 & 16.816 0.013 & 16.554 0.035 & 16.25 0.05 & 16.09 0.08 &     --   -- &    --   -- &    --   -- &  - - 5 5 4 4 4 3 3 - - - & 83.997378 & +09.8923268 \\  
 e108  &    --   -- &    --   -- &    --   -- &    --   -- &    $>$21.000 &    $>$19.750 & 18.805 0.270 & 17.08 0.14 & 16.13 0.08 &     --   -- &    --   -- &    --   -- &  - - - - 4 4 4 3 3 - - - & 84.141150 & +10.0191500 \\  
 e109c &    --   -- &    --   -- & 22.22 0.03 & 21.60 0.03 &    $>$21.000 &    $>$19.750 &    $>$18.750 & 17.08 0.08 & 16.57 0.11 &  15.93 0.29 &    --   -- &    --   -- &  - - 5 5 4 4 4 3 3 3 - - & 84.045290 & +10.0147090 \\  
 e109e &    --   -- &    --   -- & 22.95 0.09 & 21.35 0.03 & 20.146 0.114 & 19.332 0.141 & 18.551 0.204 & 17.73 0.12 &    --   -- &     --   -- &    --   -- &    --   -- &  - - 5 5 4 4 4 3 - - - - & 84.045740 & +10.0147620 \\  
 e109w &    --   -- &    --   -- & 20.58 0.01 & 19.62 0.01 & 19.018 0.043 & 18.112 0.048 &    $>$18.750 & 17.19 0.09 &    --   -- &     --   -- &    --   -- &    --   -- &  - - 5 5 4 4 4 3 - - - - & 84.044360 & +10.0148120 \\  
 e110  &    --   -- &    --   -- &    --   -- &    --   -- &    $>$21.000 &    $>$19.750 &    $>$18.750 &    --   -- &    --   -- &     --   -- &    --   -- &    --   -- &  - - - - 4 4 4 - - - - - & 84.185590 & +09.9905940 \\  
 e111  &    --   -- & 14.40 0.00 & 14.07 0.00 & 13.72 0.00 & 13.231 0.001 & 13.010 0.001 & 12.932 0.002 & 12.82 0.01 & 12.81 0.01 &  13.00 0.04 & 13.21 0.06 &    --   -- &  - 8 8 8 4 4 4 3 3 3 3 - & 84.088850 & +09.8971860 \\  
 e112  &    --   -- &    --   -- &    --   -- &    --   -- &     --    -- &     --    -- &     --    -- &    --   -- &    --   -- &     --   -- &    --   -- &    --   -- &  - - - - - - - - - - - - & 84.272730 & +09.9360910 \\  
\hline 
\end{tabular} 
$\,$\\
Catalog IDs:  1 -- 2MASS, 2 -- Optical Monitor (XMM), 3 -- Spitzer, 4 -- O2000/2005, 5 -- CFHT1999, 6 -- Barrado 2004$+$2007, 7 -- Dolan\&Mathieu 1999+2001, 8 -- Dolan\&Mathieu 2002, 9 -- O2000/2007
\end{table}
\end{landscape}


\clearpage

\setcounter{table}{5}
%
%
\begin{landscape}
\begin{table}
\tiny
\caption{Optical and infrared photometry for all the 205 counterparts for 164 X-ray sources of Collinder 69  (western {\rm XMM-Newton} pointing or Col 69\,W). } 
\label{tab:photometryW} 
\begin{tabular}{l l               l               l              l              l               l               l               l               l               l               l               l               l                         l           l} \\ 
Name   &   $B$   &   $V$      & $Rc$       &    $Ic$    &   $J$        &   $H$        &   $K$        &   [$3.6$]  &   [$4.5$]  &   [$5.8$]  &   [$8.0$]  &  [$24$]   &  Notes                  & R.A.     & DEC        \\   %
C69-X-\#  &   [mag] &   [mag]    & [mag]      &    [mag]   &   [mag]      &   [mag]      &   [mag]      &   [mag]    &   [mag]    &   [mag]    &   [mag]    &  [mag]    &                         & (2000)   & (2000)     \\   
\hline      
\hline
 w001c &    --   -- &    --   -- &    --   -- &    --   -- & 10.194 0.024 &  9.736 0.027 &  9.589 0.021 &  9.53 0.00 &  9.55 0.00 &  9.51 0.00 &  9.45 0.00 &    --   -- &  - - - - 1 1 1 3 3 3 3 - & 83.528630 & +10.0171150 \\  
 w001n &    --   -- &    --   -- &    --   -- &    --   -- &     --    -- &     --    -- &     --    -- &    --   -- &    --   -- &    --   -- &    --   -- &    --   -- &  - - - - - - - - - - - - & 83.527540 & +10.0178790 \\  
 w002c &    --   -- & 15.91 0.00 & 14.77 0.00 & 13.53 0.00 & 12.046 0.028 & 11.324 0.024 & 11.092 0.025 & 10.88 0.00 & 10.83 0.00 & 10.81 0.01 & 10.74 0.01 &    --   -- &  - 7 7 7 6 6 6 6 6 6 6 - & 83.650830 & +09.8955050 \\  
 w002n &    --   -- &    --   -- & 18.61 0.02 & 17.94 0.01 & 17.008 0.008 & 16.562 0.015 & 16.526 0.029 &    --   -- & 70.00 0.00 &    --   -- &    --   -- &    --   -- &  - - 5 5 4 4 4 - - - - - & 83.652020 & +09.8968490 \\  
 w003  &    --   -- & 15.69 0.01 & 14.49 0.00 & 13.13 0.00 & 11.656 0.022 & 10.918 0.022 & 10.719 0.023 & 10.51 0.00 & 10.48 0.00 & 10.47 0.01 & 10.34 0.01 &    --   -- &  - 7 7 7 6 6 6 6 6 6 6 - & 83.484741 & +09.8990646 \\  
 w004  &    --   -- & 14.16 0.01 & 13.34 0.00 & 12.51 0.01 & 11.297 0.022 & 10.595 0.022 & 10.426 0.021 & 10.23 0.00 & 10.26 0.00 & 10.21 0.01 & 10.21 0.01 &    --   -- &  - 8 8 8 6 6 6 6 6 6 6 - & 83.446590 & +09.9273299 \\  
 w005c &    --   -- & 14.21 0.00 & 13.46 0.00 & 12.77 0.00 & 11.839 0.027 & 11.227 0.026 & 11.059 0.021 & 10.94 0.00 & 10.96 0.00 & 10.92 0.01 & 10.92 0.01 &    --   -- &  - 7 7 7 1 1 1 3 3 3 3 - & 83.636690 & +09.9921350 \\  
 w005s &    --   -- &    --   -- & 16.27 0.00 & 15.72 0.00 & 15.163 0.002 & 14.874 0.004 & 14.749 0.007 &    --   -- &    --   -- &    --   -- &    --   -- &    --   -- &  - - 5 5 4 4 4 - - - - - & 83.635910 & +09.9908070 \\  
 w005w &    --   -- &    --   -- & 18.06 0.01 & 17.46 0.01 & 16.722 0.006 & 16.370 0.014 & 16.236 0.025 &    --   -- &    --   -- &    --   -- &    --   -- &    --   -- &  - - 5 5 4 4 4 - - - - - & 83.634850 & +09.9919500 \\  
 w006  &    --   -- & 15.09 0.00 & 14.17 0.00 & 13.31 0.00 & 12.072 0.027 & 11.360 0.023 & 11.186 0.022 & 11.01 0.00 & 11.04 0.00 & 10.96 0.01 & 10.91 0.01 &    --   -- &  - 7 7 7 1 1 1 3 3 3 3 - & 83.663238 & +09.8819439 \\  
 w007c &    --   -- & 17.42 0.00 & 16.16 0.00 & 14.80 0.00 & 13.319 0.022 & 12.588 0.023 & 12.397 0.027 & 12.22 0.00 & 12.23 0.01 & 12.16 0.02 & 12.28 0.03 &    --   -- &  - 8 8 8 1 1 1 3 3 3 3 - & 83.617170 & +09.8133610 \\  
 w007e &    --   -- &    --   -- &    --   -- &    --   -- & 19.300 0.048 & 18.510 0.073 & 17.606 0.070 & 15.81 0.05 & 15.97 0.09 &    --   -- &    --   -- &    --   -- &  - - - - 4 4 4 3 3 - - - & 83.618610 & +09.8133080 \\  
 w008c &    --   -- & 16.03 0.00 & 15.05 0.00 & 13.94 0.00 & 12.684 0.030 & 11.954 0.029 &     --    -- & 11.45 0.00 & 11.32 0.01 & 10.97 0.01 &  9.86 0.01 &  6.21 0.01 &  - 7 7 7 6 6 - 6 6 6 6 6 & 83.458150 & +09.8436640 \\  
 w008n &    --   -- &    --   -- & 16.52 0.00 & 15.87 0.00 & 14.852 0.002 & 14.319 0.003 & 14.139 0.005 & 13.56 0.01 & 13.58 0.01 & 13.07 0.04 & 11.83 0.02 &    --   -- &  - - 5 5 4 4 4 3 3 3 3 - & 83.457530 & +09.8445950 \\  
 w009c &    --   -- & 13.06 0.01 &    --   -- &    --   -- & 11.496 0.024 & 11.166 0.027 & 11.120 0.027 & 11.08 0.00 & 11.08 0.00 & 11.06 0.01 & 11.05 0.01 &    --   -- &  - 2 - - 1 1 1 3 3 3 3 - & 83.517360 & +10.0134200 \\  
 w009s &    --   -- &    --   -- & 15.08 0.00 & 14.39 0.00 & 13.364 0.001 & 12.707 0.001 & 12.490 0.001 & 12.72 0.01 & 12.72 0.01 & 12.79 0.04 & 12.68 0.04 &    --   -- &  - - 5 5 4 4 4 3 3 3 3 - & 83.518440 & +10.0116000 \\  
 w010c &    --   -- &    --   -- & 19.50 0.00 & 18.68 0.00 & 17.602 0.015 &     --    -- &     --    -- & 14.75 0.02 & 14.17 0.02 & 13.55 0.04 & 12.45 0.03 &  8.62 0.07 &  - - 5 5 4 - - 3 3 3 3 3 & 83.508380 & +09.9717320 \\  
 w010e &    --   -- &    --   -- & 20.59 0.01 & 19.80 0.01 & 18.672 0.038 &     --    -- &     --    -- &    --   -- &    --   -- &    --   -- &    --   -- &    --   -- &  - - 5 5 4 - - - - - - - & 83.508870 & +09.9711810 \\  
 w011c &    --   -- & 12.75 0.00 & 12.16 0.00 & 11.61 0.00 & 10.801 0.023 & 10.262 0.024 & 10.125 0.021 & 10.08 0.00 & 10.11 0.00 & 10.01 0.01 & 10.01 0.01 &    --   -- &  - 7 7 7 1 1 1 3 3 3 3 - & 83.523070 & +09.7129320 \\  
 w011s &    --   -- &    --   -- &    --   -- &    --   -- & 17.709 0.020 & 17.021 0.026 & 16.687 0.048 &    --   -- &    --   -- &    --   -- &    --   -- &    --   -- &  - - - - 4 4 4 - - - - - & 83.524280 & +09.7113960 \\  
 w012  &    --   -- & 16.06 0.00 & 15.05 0.00 & 13.95 0.00 & 12.500 0.022 & 11.727 0.024 & 11.336 0.021 & 10.88 0.00 & 10.60 0.00 & 10.23 0.01 &  9.54 0.00 &  6.93 0.02 &  - 7 7 7 1 1 1 3 3 3 3 3 & 83.520295 & +09.9517170 \\  
 w013c &    --   -- & 16.42 0.00 & 15.25 0.00 & 13.93 0.00 & 12.459 0.024 & 11.727 0.026 & 11.492 0.021 & 11.30 0.00 & 11.31 0.01 & 11.20 0.02 & 11.18 0.02 &    --   -- &  - 7 7 7 1 1 1 6 6 6 6 - & 83.648260 & +09.9956380 \\  
 w013n &    --   -- &    --   -- & 21.53 0.02 & 20.79 0.02 & 19.498 0.059 & 18.650 0.097 & 18.759 0.225 &    --   -- &    --   -- &    --   -- &    --   -- &    --   -- &  - - 5 5 4 4 4 - - - - - & 83.648550 & +09.9977390 \\  
 w014c &    --   -- &    --   -- &    --   -- &    --   -- &  9.232 0.030 &  9.021 0.030 &  8.894 0.024 &  8.95 0.00 &  8.88 0.00 &  8.85 0.00 &  8.84 0.00 &  7.96 0.05 &  - 2 - - 1 1 1 3 3 3 3 3 & 83.460180 & +10.0724590 \\  
 w014w &    --   -- &    --   -- &    --   -- &    --   -- & 14.884 0.308 & 14.401 0.308 & 14.166 0.236 & 13.82 0.06 & 13.84 0.07 & 13.57 0.07 & 13.74 0.17 &    --   -- &  - - - - 1 1 1 3 3 3 3 - & 83.457800 & +10.0724880 \\  
 w015  &    --   -- & 16.83 0.02 & 15.39 0.01 & 14.17 0.01 & 12.924 0.024 & 12.318 0.024 & 12.065 0.023 & 11.84 0.01 & 11.80 0.01 & 11.75 0.02 & 11.67 0.02 &    --   -- &  - 2 6 6 6 6 6 6 6 6 6 - & 83.534911 & +09.8569582 \\  
 w016  &    --   -- &    --   -- &    --   -- &    --   -- &     --    -- &     --    -- &     --    -- & 16.82 0.09 & 16.35 0.12 &    --   -- &    --   -- &    --   -- &  - - - - - - - 3 3 - - - & 83.303970 & +09.8274840 \\  
 w017  &    --   -- &    --   -- &    --   -- &    --   -- &     --    -- &     --    -- &     --    -- & 17.26 0.13 & 16.60 0.11 &    --   -- &    --   -- &    --   -- &  - - - - - - - 3 3 - - - & 83.404040 & +09.9432430 \\  
 w018  &    --   -- & 15.14 0.01 & 14.31 0.00 & 13.61 0.00 & 12.723 0.022 & 12.207 0.023 & 12.022 0.024 & 11.94 0.00 & 11.93 0.01 & 11.87 0.02 & 11.86 0.03 &    --   -- &  - 2 5 5 1 1 1 3 3 3 3 - & 83.576309 & +09.8772189 \\  
 w019  &    --   -- & 19.40 0.17 &    --   -- &    --   -- & 16.667 0.155 & 15.675 0.149 & 14.770 0.112 & 13.81 0.01 & 13.17 0.01 & 12.57 0.03 & 11.11 0.02 &  8.06 0.05 &  - 2 - - 1 1 1 3 3 3 3 3 & 83.427366 & +10.0632810 \\  
 w020  &    --   -- &    --   -- &    --   -- &    --   -- &  9.479 0.023 &  9.224 0.022 &  9.153 0.023 &  9.24 0.00 &  9.11 0.00 &  9.10 0.00 &  9.08 0.00 &    --   -- &  - - - - 1 1 1 3 3 3 3 - & 83.463238 & +09.7784576 \\  
 w021  &    --   -- &    --   -- &    --   -- &    --   -- &     --    -- &     --    -- &     --    -- & 15.98 0.04 & 15.50 0.05 &    --   -- & 13.85 0.14 &    --   -- &  - - - - - - - 3 3 - 3 - & 83.419080 & +10.0442010 \\  
 w022  &    --   -- &    --   -- &    --   -- &    --   -- &     --    -- &     --    -- &     --    -- &    --   -- &    --   -- &    --   -- &    --   -- &    --   -- &  - - - - - - - - - - - - & 83.429710 & +10.1701040 \\  
 w023  &    --   -- &    --   -- & 15.47 0.01 & 13.98 0.01 & 12.576 0.024 & 11.936 0.023 & 11.706 0.021 & 11.40 0.00 & 11.38 0.01 & 11.29 0.02 & 11.26 0.02 &    --   -- &  - - 6 6 6 6 6 6 6 6 6 - & 83.663531 & +10.0248160 \\  
 w024c &    --   -- & 17.41 0.03 & 15.93 0.01 & 14.63 0.01 & 13.117 0.023 & 12.454 0.024 & 12.192 0.019 & 11.92 0.01 & 11.86 0.01 & 11.76 0.02 & 11.79 0.03 &    --   -- &  - 2 6 6 6 6 6 6 6 6 6 - & 83.549040 & +09.9510840 \\  
 w024s &    --   -- &    --   -- & 21.05 0.16 & 20.03 0.09 & 18.543 0.027 & 17.855 0.045 & 17.879 0.105 & 15.87 0.05 &    --   -- &    --   -- &    --   -- &    --   -- &  - - 5 5 4 4 4 3 - - - - & 83.549380 & +09.9496580 \\  
 w025c &    --   -- &    --   -- & 21.63 0.02 & 20.52 0.01 & 19.291 0.058 & 18.562 0.085 & 18.621 0.205 & 16.90 0.09 &    --   -- &    --   -- &    --   -- &    --   -- &  - - 5 5 4 4 4 3 - - - - & 83.510050 & +10.1182540 \\  
 w025e &    --   -- &    --   -- & 21.77 0.02 & 21.15 0.03 & 19.684 0.082 & 18.729 0.097 & 17.743 0.093 & 16.41 0.05 & 15.64 0.05 & 15.56 0.23 &    --   -- &    --   -- &  - - 5 5 4 4 4 3 3 3 - - & 83.511410 & +10.1185150 \\  
 w025w &    --   -- &    --   -- & 20.75 0.01 & 19.43 0.01 & 18.087 0.021 & 17.348 0.031 & 16.831 0.040 & 16.60 0.06 & 16.78 0.16 &    --   -- &    --   -- &    --   -- &  - - 5 5 4 4 4 3 3 - - - & 83.509380 & +10.1181110 \\  
 w026  &    --   -- &    --   -- &    --   -- &    --   -- &     --    -- &     --    -- &     --    -- &    --   -- &    --   -- &    --   -- &    --   -- &    --   -- &  - - - - - - - - - - - - & 83.266370 & +10.0351130 \\  
 w027  &    --   -- & 16.58 0.01 & 15.97 0.00 & 15.43 0.00 & 14.582 0.035 & 14.134 0.047 & 13.996 0.046 & 13.94 0.01 &    --   -- & 13.86 0.06 & 13.72 0.13 &    --   -- &  - 8 8 8 1 1 1 3 - 3 3 - & 83.349933 & +09.9521250 \\  
 w028  &    --   -- &    --   -- &    --   -- &    --   -- & 16.138 0.101 & 15.389 0.120 & 14.876 0.098 & 14.40 0.01 & 14.22 0.02 & 13.88 0.07 & 13.36 0.08 &    --   -- &  - - - - 1 1 1 3 3 3 3 - & 83.397147 & +10.1405910 \\  
 w029  &    --   -- & 17.50 0.03 & 16.29 0.00 & 14.71 0.00 & 12.949 0.028 & 12.287 0.029 & 11.876 0.028 & 11.45 0.00 & 11.24 0.00 & 11.07 0.01 & 10.61 0.01 &  8.11 0.05 &  - 8 8 8 1 1 1 3 3 3 3 3 & 83.313132 & +09.8417150 \\  
 w030c &    --   -- &    --   -- & 19.80 0.00 & 19.12 0.00 & 18.890 0.050 &     --    -- &     --    -- & 15.74 0.03 & 14.60 0.02 & 13.54 0.05 & 12.60 0.05 &    --   -- &  - - 5 5 4 - - 3 3 3 3 - & 83.505450 & +09.9431290 \\  
 w030e &    --   -- &    --   -- & 20.83 0.01 & 19.73 0.01 & 18.589 0.036 &     --    -- &     --    -- & 17.14 0.10 &    --   -- &    --   -- &    --   -- &    --   -- &  - - 5 5 4 - - 3 - - - - & 83.507810 & +09.9425150 \\  
\hline 
\end{tabular} 
$\,$\\
Catalog IDs:  1 -- 2MASS, 2 -- Optical Monitor (XMM), 3 -- Spitzer, 4 -- O2000/2005, 5 -- CFHT1999, 6 -- Barrado 2004$+$2007, 7 -- Dolan\&Mathieu 1999+2001, 8 -- Dolan\&Mathieu 2002, 9 -- O2000/2007
\end{table}
\end{landscape}


\addtocounter{table}{-1}
\begin{landscape}
\begin{table}
\tiny 
\caption{Optical and infrared photometry for all the 205 counterparts for 164 X-ray sources of Collinder 69  (western {\rm XMM-Newton} pointing or Col 69\,W). } 
\label{tab:photometry} 
\begin{tabular}{l l               l               l              l              l               l               l               l               l               l               l               l               l                         l           l} \\ 
Name   &   $B$   &   $V$      & $Rc$       &    $Ic$    &   $J$        &   $H$        &   $K$        &   [$3.6$]  &   [$4.5$]  &   [$5.8$]  &   [$8.0$]  &  [$24$]   &  Notes                  & R.A.     & DEC        \\   %
C69-X-\#  &   [mag] &   [mag]    & [mag]      &    [mag]   &   [mag]      &   [mag]      &   [mag]      &   [mag]    &   [mag]    &   [mag]    &   [mag]    &  [mag]    &                         & (2000)   & (2000)     \\   
\hline      
\hline
 w031   &    --   --  &    --   --  &    --   --  &    --   --  & 16.826 0.179  & 16.006 0.202  & 15.238 0.135  & 14.38 0.02  & 13.84 0.02  & 13.31 0.05  & 12.31 0.04  &    --   -- &  - - - - 1 1 1 3 3 3 3 - & 83.362344 & +10.0276360 \\  
 w032c  &    --   --  & 18.41 0.03  & 17.23 0.00  & 16.06 0.00  & 14.471 0.054  & 13.827 0.060  & 13.465 0.063  & 13.08 0.01  & 13.05 0.01  & 13.07 0.04  & 13.35 0.09  &    --   -- &  - 2 5 5 1 1 1 3 3 3 3 - & 83.451270 & +10.0287590 \\  
 w032e  &    --   --  &    --   --  & 16.80 0.00  & 15.93 0.00  &     --    --  &     --    --  &     --    --  & 13.27 0.01  & 13.18 0.01  & 13.20 0.05  & 13.31 0.09  &    --   -- &  - - 5 5 - - - 3 3 3 3 - & 83.451590 & +10.0283530 \\  
 w032s  &    --   --  &    --   --  & 23.16 0.10  & 21.99 0.07  &     --    --  &     --    --  &     --    --  &    --   --  &    --   --  &    --   --  &    --   --  &    --   -- &  - - 5 5 - - - - - - - - & 83.451570 & +10.0267440 \\  
 w032w  &    --   --  &    --   --  & 18.71 0.01  & 16.99 0.01  & 15.162 0.048  & 14.576 0.060  & 14.268 0.082  & 13.72 0.02  & 13.58 0.02  & 13.26 0.07  & 12.45 0.05  &    --   -- &  - - 6 6 6 6 6 6 6 6 6 - & 83.449640 & +10.0276670 \\  
 w032z  &    --   --  &    --   --  &    --   --  & 22.53 0.12  &     --    --  &     --    --  &     --    --  &    --   --  &    --   --  &    --   --  &    --   --  &    --   -- &  - - - 5 - - - - - - - - & 83.450360 & +10.0267300 \\  
 w033   &    --   --  &    --   --  &    --   --  &    --   --  &     --    --  &     --    --  &     --    --  &    --   --  &    --   --  &    --   --  &    --   --  &    --   -- &  - - - - - - - - - - - - & 83.427960 & +09.9672690 \\  
 w034   &    --   --  & 16.41 0.01  & 15.52 0.00  & 14.72 0.00  & 13.647 0.028  & 12.948 0.032  & 12.752 0.030  & 12.60 0.00  & 12.64 0.01  & 12.65 0.03  & 12.62 0.05  &    --   -- &  - 8 8 8 1 1 1 3 3 3 3 - & 83.365972 & +10.0722730 \\  
 w035   &    --   --  &    --   --  &    --   --  &    --   --  &     --    --  &     --    --  &     --    --  & 17.13 0.13  &    --   --  &    --   --  &    --   --  &    --   -- &  - - - - - - - 3 - - - - & 83.308680 & +09.9292630 \\  
 w036   &    --   --  & 14.60 0.01  & 14.09 0.00  & 13.58 0.03  & 12.799 0.024  & 12.435 0.022  & 12.392 0.027  & 12.20 0.00  & 12.22 0.01  & 12.23 0.02  & 12.32 0.04  &    --   -- &  - 8 8 8 1 1 1 3 3 3 3 - & 83.418051 & +09.8664630 \\  
 w037   &    --   --  & 17.35 0.03  & 16.39 0.00  & 15.23 0.01  & 14.046 0.029  & 13.293 0.036  & 13.070 0.032  & 12.90 0.01  & 12.94 0.01  & 12.88 0.03  &    --   --  &    --   -- &  - 8 8 8 1 1 1 3 3 3 - - & 83.387396 & +09.9606980 \\  
 w038c  &    --   --  &    --   --  &    --   --  &    --   --  &     --    --  &     --    --  &     --    --  &    --   --  &    --   --  &    --   --  &    --   --  &    --   -- &  - - - - - - - - - - - - & 83.436040 & +10.0293920 \\  
 w038s  &    --   --  &    --   --  & 18.42 0.01  & 17.86 0.01  &     --    --  &     --    --  &     --    --  & 15.94 0.03  & 15.68 0.05  &    --   --  &    --   --  &    --   -- &  - - 5 5 - - - 3 3 - - - & 83.435420 & +10.0275230 \\  
 w039c  &    --   --  &    --   --  & 20.45 0.08  & 19.70 0.01  &     --    --  &     --    --  &     --    --  & 17.57 0.16  &    --   --  &    --   --  &    --   --  &    --   -- &  - - 5 5 - - - 3 - - - - & 83.449370 & +09.8822020 \\  
 w039w  &    --   --  & 14.62 0.01  & 14.04 0.00  & 13.53 0.00  & 12.972 0.024  & 12.661 0.023  & 12.611 0.032  & 13.00 0.01  & 13.05 0.01  & 13.24 0.04  & 13.07 0.07  &    --   -- &  - 2 5 5 1 1 1 3 3 3 3 - & 83.447470 & +09.8832160 \\  
 w040   &    --   --  & 16.21 0.01  & 15.26 0.00  & 14.34 0.02  & 13.234 0.025  & 12.531 0.031  & 12.352 0.026  & 12.22 0.00  & 12.26 0.01  & 12.15 0.02  & 12.17 0.03  &    --   -- &  - 8 8 8 1 1 1 3 3 3 3 - & 83.413316 & +10.0036900 \\  
 w041   &    --   --  & 17.35 0.10  & 16.34 0.00  & 14.99 0.05  & 13.752 0.029  & 13.129 0.029  & 12.896 0.030  & 12.61 0.01  & 12.60 0.01  & 12.48 0.03  & 12.55 0.05  &    --   -- &  - 8 8 8 1 1 1 3 3 3 3 - & 83.429965 & +09.7279361 \\  
 w042c  &    --   --  &    --   --  & 23.16 0.08  & 22.10 0.06  &    $>$20.750  &    $>$19.750  &    $>$19.000  & 17.18 0.11  & 16.94 0.14  &    --   --  &    --   --  &    --   -- &  - - 5 5 4 4 4 3 3 - - - & 83.463710 & +09.8034280 \\  
 w042e  &    --   --  &    --   --  & 21.02 0.01  & 19.78 0.01  & 18.409 0.028  & 17.659 0.038  & 17.477 0.080  & 16.90 0.10  &    --   --  &    --   --  &    --   --  &    --   -- &  - - 5 5 4 4 4 3 - - - - & 83.464820 & +09.8034270 \\  
 w042w  &    --   --  &    --   --  & 23.08 0.07  & 22.24 0.07  & 20.323 0.162  & 20.617 0.625  & 18.833 0.289  & 17.56 0.12  & 16.76 0.12  &    --   --  &    --   --  &    --   -- &  - - 5 5 4 4 4 3 3 - - - & 83.462520 & +09.8044530 \\  
 w043   &    --   --  & 12.35 0.01  &    --   --  &    --   --  & 11.150 0.023  & 10.860 0.023  & 10.798 0.023  & 10.73 0.00  & 10.75 0.00  & 10.72 0.01  & 10.74 0.01  &    --   -- &  - 2 - - 1 1 1 3 3 3 3 - & 83.466746 & +09.8840859 \\  
 w044   &    --   --  &    --   --  &    --   --  &    --   --  &     --    --  &     --    --  &     --    --  & 16.34 0.07  & 15.78 0.06  & 14.47 0.14  & 13.72 0.14  &    --   -- &  - - - - - - - 3 3 3 3 - & 83.391590 & +09.8397680 \\  
 w045   &    --   --  &    --   --  &    --   --  &    --   --  &     --    --  &     --    --  &     --    --  & 15.66 0.06  &    --   --  &    --   --  &    --   --  &    --   -- &  - - - - - - - 3 - - - - & 83.303580 & +09.9661210 \\  
 w046   &    --   --  &    --   --  &    --   --  &    --   --  & 14.015 0.035  & 13.359 0.033  & 13.053 0.039  &    --   --  &    --   --  &    --   --  &    --   --  &    --   -- &  - - - - 1 1 1 - - - - - & 83.265781 & +09.9784980 \\  
 w047   &    --   --  &    --   --  &    --   --  &    --   --  & 14.716 0.042  & 14.086 0.049  & 13.815 0.051  &    --   --  &    --   --  &    --   --  &    --   --  &    --   -- &  - - - - 1 1 1 - - - - - & 83.298360 & +09.9841260 \\  
 w048   &    --   --  &    --   --  & 22.05 0.05  & 20.31 0.01  & 18.595 0.024  & 17.937 0.039  & 17.782 0.083  & 16.77 0.07  & 16.74 0.13  &    --   --  &    --   --  &    --   -- &  - - 5 5 4 4 4 3 3 - - - & 83.602812 & +09.9491913 \\  
 w049   &    --   --  &    --   --  & 22.90 0.06  & 21.83 0.05  &    $>$20.000  &    $>$19.000  &    $>$18.250  & 16.44 0.05  & 15.94 0.06  &    --   --  &    --   --  &    --   -- &  - - 5 5 4 4 4 3 3 - - - & 83.643163 & +10.0553090 \\  
 w050   &    --   --  &    --   --  &    --   --  &    --   --  &     --    --  &     --    --  &     --    --  &    --   --  &    --   --  &    --   --  &    --   --  &    --   -- &  - - - - - - - - - - - - & 83.370230 & +09.8523960 \\  
 w051   &    --   --  &    --   --  &    --   --  &    --   --  &     --    --  &     --    --  &     --    --  &    --   --  &    --   --  &    --   --  &    --   --  &    --   -- &  - - - - - - - - - - - - & 83.415110 & +09.8510330 \\  
 w052   &    --   --  &    --   --  &    --   --  &    --   --  &     --    --  &     --    --  &     --    --  &    --   --  &    --   --  &    --   --  &    --   --  &    --   -- &  - - - - - - - - - - - - & 83.346180 & +09.9470260 \\  
\hline 
\end{tabular} 
$\,$\\
Catalog IDs:  1 -- 2MASS, 2 -- Optical Monitor (XMM), 3 -- Spitzer, 4 -- O2000/2005, 5 -- CFHT1999, 6 -- Barrado 2004$+$2007, 7 -- Dolan\&Mathieu 1999+2001, 8 -- Dolan\&Mathieu 2002, 9 -- O2000/2007
\end{table}
\end{landscape}


\clearpage

\clearpage

\setcounter{table}{6}
%
%
\begin{table}
\begin{center}
\caption{
Summary with the source of the optical and infrared photometry or the 205 counterparts of the 164 XMM-{\it Newton} sources, 
and some properties of the possible and probable members.
}
\label{tab:summaryTAB}
\begin{tabular}{l r r }\hline
Survey                  &All counterparts& Probable/Possible members \\
                        & Number         & Number                    \\
                        & (205 total)    & (69 total)$^1$            \\

 \hline
$B$ or $V$ OM           & 37$^2$    & 26$^4$ \\
$RI$ Dolan and Mathieu  & 53        & 42     \\
$RI$ CFHT or ByN        & 84        & 13     \\
$JHK$ 2MASS             & 63        & 35     \\
$JHK$ Omega2000         & 76        &  2     \\
$JHK$ ByN2007           & 31        & 33     \\
IRAC                    & 163       & 64     \\
MIPS                    &16 (2)$^3$ & 10 (2)$^3$\\
\hline
Probable members        & --        & 61      \\
Possible members        & --        &  5      \\
Possible companions     & --        &  3      \\
Listed in DM1999        & --        & 35$^5$  \\ 
Listed in ByN2004       & --        & 31$^6$  \\ 
Class II$^9$            & --        &  4$^7$  \\
Class III$^9$           & --        & 62$^8$  \\
\hline
\end{tabular}
\end{center}
$^1$ 61 probable members, 5 possible members, and 3 possible companions which might be also members.\\
$^2$ 27 in $B$, 20 in $V$.\\
$^3$ Two objects with MIPS data but no IRAC photometry.\\
$^4$ 22 in $B$, 11 in $V$.\\
$^5$ 16 not in ByN2004.\\
$^6$ 19 in common with DM99.\\
$^7$ One among these four has been classified as Class I/II based on IRAC CCD.\\
$^8$ 55 Class III and no disk. Other two with transition disks (24 micron excess),
 other two with thin disks (based on IRAC slope),
and another two with possible Class III based on SED (incomplete IRAC data). 
Finally, one has no IRAC data, but its 24 micron flux indicates it is a Class III with a transition disk.\\
$^9$  Including probable and possible members. 
\end{table}

\setcounter{table}{7}
%
%
\begin{table*}
\tiny 
\begin{center}
\caption{Membership criteria and other information for X-ray sources of Collinder 69  (eastern {\rm XMM-Newton} pointing or Col 69\,E).}
\label{tab:membershipcriteriaE} 
\newcolumntype{d}[1]{D{.}{.}{#1}}
\begin{tabular}{l  l   l             l      l      l       l     l        l      l       l      l               l      l}     \\  
\hline                                                                                                                                               
 XMMID & Alpha     & Delta      &QSO$^1$&HRD$^2$&CMD$^3$&CCD$^3$&Final$^4$&IRAC$^5$& ByN   & DM   & DM             &Sacco$^6$ & Sp.Type$^6$      \\
 C69-X-& (2000)    & (2000)     &       &     &       &     &        &      & 2004  & 1999 & 2002           &     &        \\        
\hline                                                                                                                                               
e001  & 83.980968 & +09.9420736 &  No   & Y   & YY-YY & Y   & Y      & III  & 003   & 046  & --             & Y?  & K8.5   \\ 
e002  & 84.133460 & +09.7389598 &  ?    & N   & NN-N- & N   & NM     & I    & --    & --   & --             & --  & --     \\ 
e003  & 84.253517 & +09.8179200 &  No   & Y   & YY-YY & Y   & Y      & III  & --    & --   & 0975-01780292  & --  & --     \\ 
e004  & 83.991539 & +09.9091240 &  No   & Y?  & YY-YY & Y   & Y      & III  & 008   & 051  & 0975-01770231  & --  & --     \\ 
e005c & 84.116250 & +09.9242930 &  Yes  & N   & NN-N- & ?   & NM     & II   & --    & --   & --             & --  & --     \\ 
e005w & 84.115390 & +09.9241960 &  Yes  & N   & N--N- & --  & NM     & --   & --    & --   & --             & --  & --     \\ 
e006  & 84.085380 & +09.8720186 &  No   & Y   & YY-YY & Y   & Y      & III  & 017   & 060  & 0975-01766593  & --  & --     \\ 
e007c & 84.079263 & +10.0641340 &  No   & Y   & YY-YY & Y   & Y      & III  & 014   & 058  & 0975-01766070  & --  & --     \\ 
e008  & 84.077312 & +09.7526149 &  No   & Y   & YY-YY & Y   & Y      & III  & --    & 057  & 0975-01765910  & --  & --     \\ 
e009  & 83.829475 & +09.9151335 &  ?    & Y   & YYY-Y & Y   & Y      & III  & --    & --   & --             & --  & --     \\ 
e010c & 83.976307 & +10.0731510 &  No   & Y   & YY-YY & Y   & Y      & III  & --    & 045  & --             & --  & --     \\ 
e011c & 84.084070 & +09.7339210 &  pos  & Y   & YY-YY & Y   & Y      & III  & 025   & 059  & 0975-01766491  & --  & --     \\ 
e011e & 84.084620 & +09.7340960 &  ?    & --  & ---Y- & --  & Y?     & --   & --    & --   & --             & --  & --     \\ 
e012  & 84.039236 & +10.0239120 &  ?    & Y?  & YYY-Y & Y   & Y      & III  & --    & --   & --             & --  & --     \\ 
e013  & 84.096227 & +09.7543767 &  ?    & Y?  & ---   & --  & Y?     & --   & --    & --   & --             & --  & --     \\ 
e014  & 83.963918 & +09.9196904 &  No   & Y   & YY-YY & Y   & Y      & III  & 022   & 044  & --             & Y   & M1.5   \\ 
e015  & 84.042061 & +10.0328960 &  Yes  & N   & NN-N- & N   & NM     & III  & --    & --   & --             & --  & --     \\ 
e016  & 84.069113 & +09.8468478 &  No   & Y   & YY-YY & Y   & Y      & III  & 018   & 056  & 0975-01765210  & --  & --     \\ 
e017  & 84.078553 & +09.8598079 &  No   & Y   & YY-YY & Y   & Y      & III  & 028   & --   & 0975-01766058  & --  & --     \\ 
e018  & 84.196284 & +10.0977120 &  No   & Y   & YY-YY & Y   & Y      & III  & --    & 065  & 0975-01775590  & --  & --     \\ 
e019  & 84.239585 & +09.8905510 &  No   & Y   & YY-YY & Y   & Y      & III  & --    & 070  & --             & --  & --     \\ 
e020  & 84.068451 & +09.9903775 &  No   & Y   & YY-YY & Y   & Y      & III  & --    & --   & 0975-01765144  & --  & --     \\ 
e021c & 84.166840 & +10.0761670 &  ?    & Y   & N---Y & --  & NM?    & --   & --    & --   & --             & --  & --     \\ 
e021w & 84.165860 & +10.0760410 &  ?    & N   & N---Y & --  & NM     & --   & --    & --   & --             & --  & --     \\ 
e022  & 84.038801 & +09.7843548 &  No   & Y   & YY-YY & Y   & Y      & III  & 032   & 055  & --             & --  & --     \\ 
e023  & 83.948123 & +09.7640605 &  ?    & Y   & YY-YY & Y   & Y      & III  & 004   & --   & --             & --  & --     \\ 
e024  & 83.930262 & +09.9989864 &  Yes  & N   & NN-N- & --  & NM     & --   & --    & --   & --             & --  & --     \\ 
e025  & 84.109597 & +09.8539695 &  No   & Y   & YY-YY & Y   & Y      & III  & --    & 061  & 0975-01768603  & --  & --     \\ 
e026  & 84.139100 & +09.9674730 &  ?    & Y   & N---Y & --  & NM?    & --   & --    & --   & --             & --  & --     \\ 
e027c & 84.120530 & +09.9075260 &  No   & Y   & YY-YY & Y   & Y      & III  & --    & 062  & 0975-01769502  & --  & --     \\ 
e027n & 84.120420 & +09.9089960 &  ?    & N   & ---N- & --  & NM     & --   & --    & --   & --             & --  & --     \\ 
e028c & 83.990196 & +09.7929641 &  No   & Y   & YY-YY & Y   & Y      & III  & 023   & 050  & --             & --  & --     \\ 
e029c & 83.895130 & +10.0097960 &  No   & Y   & YY-YY & Y   & Y      & III  & 033   & 039  & 0975-01751340  & --  & --     \\ 
e030  & 84.012193 & +09.7017196 &  No   & Y   & YYYYY & Y   & Y      & III  & --    & 053  & --             & --  & --     \\ 
e031  & 84.219484 & +09.8823130 &  No   & Y   & YY-YY & Y   & Y      & III  & --    & 067  & --             & --  & --     \\ 
e032  & 83.907920 & +09.7361380 &  ?    & Y?  & NN?-? & N   & NM     & I    & --    & --   & --             & --  & --     \\ 
e033  & 83.997733 & +09.8385000 &  No   & N   & N--NY & --  & NM     & --   & --    & --   & --             & --  & --     \\ 
e034c & 83.842550 & +09.8743420 &  No   & Y   & YY-YY & Y   & Y      & III  & --    & 035  & --             & --  & --     \\ 
e034w & 83.841760 & +09.8738260 &  ?    & N   & ----  & --  & NM?    & --   & --    & --   & --             & --  & --     \\ 
e035c & 83.914530 & +09.8425200 &  No   & Y   & YY-YY & Y   & Y      & III  & 040   & 041  & --             & Y   & M3.0   \\ 
e036  & 83.876938 & +09.8429168 &  No   & Y?  & YY-YY & Y   & Y      & III  & 041   & 038  & --             & Y   & M3.5   \\ 
e037c & 83.966620 & +09.8416020 &  No   & Y   & ?Y-YY & Y   & Y      & III  & 064   & --   & --             & --  & --     \\ 
e038c & 83.863320 & +09.8863180 &  Yes  & N   & NN-N- & ?   & NM     & II   & --    & --   & --             & --  & --     \\ 
e038w & 83.862750 & +09.8862350 &  ?    & Y?  & ---   & --  & NM?    & --   & --    & --   & --             & --  & --     \\ 
e039  & 84.235030 & +09.8939090 &  ?    & --  & N---  & --  & NM?    & --   & --    & --   & --             & --  & --     \\ 
e040  & 84.209405 & +09.9066000 &  No   & Y   & YY-YY & Y   & Y      & III  & --    & --   & 0975-01776670  & --  & --     \\ 
\hline 
\end{tabular} 
\end{center} 
$\,$ \\ %
$^1$ ``Yes'' stands for objects located in the area of the CCD ($I-J$ versus $J-[3.6]$) where quasars are located, after \cite{Bouy09.1}.  See Fig. 5.\\
$^2$ Membership based on the HR diagram, see Fig. 6.\\
$^3$ Membership based on several color-magnitude and color-color diagrams, see Fig. 7.\\
$^4$ Final membership assigment: Probable member = Y, Possible member = Y?, Possible non-member = NM?, Probable non-member=NM,  noData=--\\ %
$^5$ Disk Class based on the Spitzer/IRAC color-color diagram, after \cite{Lada87.1} and \cite{Allen04}.\\
$^6$ Membership and spectra types after \cite{Sacco08.1}.\\
\end{table*}



\addtocounter{table}{-1}
\begin{table*}
\tiny 
\begin{center}
\caption{Membership criteria and other information for X-ray sources of Collinder 69  (eastern {\rm XMM-Newton} pointing or Col 69\,E).}
\label{tab:membershipcriteria} 
\newcolumntype{d}[1]{D{.}{.}{#1}}
\begin{tabular}{l  l   l             l      l      l       l     l        l      l       l      l               l      l}   
\hline                                                                                                                                               
 XMMID & Alpha     & Delta      &QSO$^1$&HRD$^2$&CMD$^3$&CCD$^3$&Final$^4$&IRAC$^5$& ByN   & DM   & DM             &Sacco$^6$ & Sp.Type$^6$      \\
 C69-X-& (2000)    & (2000)     &       &     &       &     &        &      & 2004  & 1999 & 2002           &     &        \\         
\hline                                                                                                                                               
e041  & 84.292398 & +09.9241550 &  ?    & Y   & ---Y- & --  & Y?     & --   & --    & 071  & --             & --  & --     \\ 
e042  & 84.192595 & +09.8873230 &  ?    & Y?  & --Y-  & --  & Y?     & --   & --    & --   & --             & --  & --     \\ 
e043  & 83.958290 & +09.9645592 &  Yes  & N   & N--N- & --  & NM     & --   & --    & --   & --             & --  & --     \\ 
e044  & 84.111610 & +09.8219420 &  ?    & --  & N---  & --  & NM?    & --   & --    & --   & --             & --  & --     \\ 
e045  & 83.877360 & +09.9081050 &  ?    & N   & ----  & --  & NM?    & --   & --    & --   & --             & --  & --     \\ 
e046  & 83.889187 & +09.8632789 &  Yes  & N   & N---  & --  & NM     & --   & --    & --   & --             & --  & --     \\ 
e047  & 84.227258 & +09.8907420 &  ?    & N   & YY--Y & Y   & NM?    & III  & --    & --   & --             & --  & --     \\ 
e048  & 84.211400 & +09.9014610 &  ?    & --  & ----  & --  & NM?    & --   & --    & --   & --             & --  & --     \\ 
e049  & 83.953577 & +09.8233012 &  Yes  & N   & N--N- & --  & NM     & --   & --    & --   & --             & --  & --     \\ 
e050  & 84.072762 & +10.0276070 &  ?    & --  & ---N- & --  & NM?    & --   & --    & --   & --             & --  & --     \\ 
e051  & 83.988965 & +09.9741924 &  Yes  & N   & NN-N- & N   & NM     & II   & --    & --   & --             & --  & --     \\ 
e052  & 83.836844 & +09.7908774 &  No   & N   & ?--NN & --  & NM     & --   & --    & --   & 0975-01746751  & --  & --     \\ 
e053  & 84.058310 & +09.8848590 &  Yes  & N   & N--N- & --  & NM     & --   & --    & --   & --             & --  & --     \\ 
e054  & 84.050619 & +10.0159390 &  No   & Y   & YY-YY & Y   & Y      & III  & 051   & --   & --             & --  & --     \\ 
e055  & 83.968848 & +09.8087980 &  No   & Y   & YY-YY & Y   & Y      & III  & 054   & --   & --             & --  & --     \\ 
e056  & 84.183930 & +09.8681680 &  ?    & Y?  & ---   & --  & NM?    & --   & --    & --   & --             & --  & --     \\ 
e057  & 84.205351 & +09.9722170 &  No   & Y   & YY-YY & Y   & Y      & III  & --    & --   & 0975-01776371  & --  & --     \\ 
e058  & 84.228583 & +09.8402840 &  No   & Y   & YY-YY & Y   & Y      & III  & --    & 069  & --             & --  & --     \\ 
e059  & 84.061930 & +09.8682130 &  ?    & --  & ----  & --  & noD    & --   & --    & --   & --             & --  & --     \\ 
e060c & 84.036570 & +09.9392720 &  ?    & N?  & YY--Y & Y   & NM?    & III  & --    & --   & --             & --  & --     \\ 
e060n & 84.038300 & +09.9396290 &  ?    & N   & ---N- & --  & NM     & --   & --    & --   & --             & --  & --     \\ 
e061  & 83.986235 & +10.0312760 &  Yes  & N   & N--N- & --  & NM     & --   & --    & --   & --             & --  & --     \\ 
e062  & 84.011607 & +10.0942480 &  ?    & N?  & YY--Y & Y   & NM?    & III  & --    & --   & --             & --  & --     \\ 
e063  & 83.977594 & +10.0311300 &  Yes  & N   & N--N- & --  & NM     & --   & --    & --   & --             & --  & --     \\ 
e064  & 83.842427 & +09.8995644 &  No   & Y   & YY-YY & Y   & Y      & III  & --    & --   & --             & --  & --     \\ 
e065  & 83.908829 & +09.8878650 &  ?    & N   & ?Y--Y & Y   & NM?    & I/II & --    & --   & --             & --  & --     \\ 
e066  & 83.909390 & +10.0174980 &  ?    & Y?  & NN?N- & N   & NM     & I    & --    & --   & --             & --  & --     \\ 
e067  & 84.035970 & +09.8889444 &  ?    & N?  & YY--Y & Y   & NM?    & III  & --    & --   & --             & --  & --     \\ 
e068  & 84.084210 & +09.9341020 &  No   & N   & YY-?N & Y   & NM     & III  & --    & --   & 0975-01766490  & --  & --     \\ 
e069c & 83.965470 & +09.8927750 &  No   & N   & YY-?N & Y   & NM     & III  & --    & --   & --             & --  & --     \\ 
e070  & 84.056270 & +10.0212060 &  ?    & --  & ----  & --  & NM?    & --   & --    & --   & --             & --  & --     \\ 
e071c & 83.871690 & +09.7756380 &  No   & N?  & YY??Y & Y   & NM?    & III  & --    & --   & --             & --  & --     \\ 
e072  & 84.114436 & +09.7571574 &  No   & Y   & YY-YY & Y   & Y      & III  & --    & --   & 0975-01768998  & --  & --     \\ 
e073  & 84.048085 & +09.9945899 &  No   & N   & ?Y-NN & --  & NM     & III  & --    & --   & --             & --  & --     \\ 
e074  & 84.148170 & +09.9066340 &  ?    & Y?  & NN--Y & --  & NM     & --   & --    & --   & --             & --  & --     \\ 
e075  & 83.885503 & +09.9652674 &  Yes  & N   & N--N- & --  & NM     & --   & --    & --   & --             & --  & --     \\ 
e076  & 84.192920 & +09.9130050 &  ?    & N   & ----  & --  & NM?    & --   & --    & --   & --             & --  & --     \\ 
e077  & 84.004794 & +09.9793416 &  Yes  & N   & Y--N- & --  & NM     & --   & --    & --   & 0975-01759942  & --  & --     \\ 
e078  & 84.112749 & +09.8597427 &  No   & Y?  & ?Y-YY & Y   & Y      & III  & 090   & --   & --             & --  & --     \\ 
e079  & 83.839291 & +09.8324832 &  No   & Y   & YY-YY & Y   & Y      & III  & 055   & --   & --             & Y   & M2.5   \\ 
e080  & 84.240200 & +10.0544470 &  ?    & --  & NNN   & ?   & NM     & I    & --    & --   & --             & --  & --     \\ 
\hline 
\end{tabular} 
\end{center} 
$\,$ \\ %
$\,$ \\ %
$^1$ ``Yes'' stands for objects located in the area of the CCD ($I-J$ versus $J-[3.6]$) where quasars are located, after \cite{Bouy09.1}.  See Fig. 5.\\
$^2$ Membership based on the HR diagram, see Fig. 6.\\
$^3$ Membership based on several color-magnitude and color-color diagrams, see Fig. 7.\\
$^4$ Final membership assigment: Probable member = Y, Possible member = Y?, Possible non-member = NM?, Probable non-member=NM,  noData=--\\ %
$^5$ Disk Class based on the Spitzer/IRAC color-color diagram, after \cite{Lada87.1} and \cite{Allen04}.\\
$^6$ Membership and spectra types after \cite{Sacco08.1}.\\
\end{table*}



\addtocounter{table}{-1}
\begin{table*}
\tiny 
\begin{center}
\caption{Membership criteria and other information for X-ray sources of Collinder 69  (eastern {\rm XMM-Newton} pointing or Col 69\,E).}
\label{tab:membershipcriteria} 
\newcolumntype{d}[1]{D{.}{.}{#1}}
\begin{tabular}{l  l   l             l      l      l       l     l        l      l       l      l               l      l}     \\    
\hline                                                                                                                                               
 XMMID & Alpha     & Delta      &QSO$^1$&HRD$^2$&CMD$^3$&CCD$^3$&Final$^4$&IRAC$^5$& ByN   & DM   & DM             &Sacco$^6$ & Sp.Type$^6$      \\
 C69-X-& (2000)    & (2000)     &       &     &       &     &        &      & 2004  & 1999 & 2002           &     &        \\           
\hline                                                                                                                                               
e081c & 83.981820 & +09.8482590 &  No   & Y   & YY-YY & Y   & Y      & III  & 047   & 047  & --             & --  & --     \\ 
e081e & 83.982440 & +09.8475150 &  ?    & --  & ---N- & --  & NM?    & --   & --    & --   & --             & --  & --     \\ 
e082  & 83.987755 & +09.7813924 &  No   & Y   & YY-YY & Y   & Y      & III  & 039   & 049  & --             & --  & --     \\ 
e083c & 84.195440 & +09.9395310 &  ?    & N   & NN--Y & N   & NM     & I    & --    & --   & --             & --  & --     \\ 
e083n & 84.195150 & +09.9403400 &  ?    & N   & ----Y & --  & NM?    & --   & --    & --   & --             & --  & --     \\ 
e084  & 83.899019 & +09.7431743 &  No   & Y   & YY-YY & Y   & Y      & III  & --    & 040  & 0975-01751635  & --  & --     \\ 
e085  & 84.057760 & +09.8752750 &  ?    & N?  & ----  & --  & NM?    & --   & --    & --   & --             & --  & --     \\ 
e086  & 84.221250 & +09.8387320 &  ?    & Y   & YY--Y & Y   & Y      & III  & --    & --   & --             & --  & --     \\ 
e087c & 83.956080 & +10.0762610 &  ?    & Y?  & NN-N- & N   & NM     & I    & --    & --   & --             & --  & --     \\ 
e087s & 83.955540 & +10.0755990 &  ?    & --  & ---N- & --  & NM?    & --   & --    & --   & --             & --  & --     \\ 
e088n & 84.125470 & +09.7276300 &  ?    & N   & N--N- & --  & NM     & --   & --    & --   & --             & --  & --     \\ 
e088s & 84.127460 & +09.7248470 &  ?    & N   & N--N- & --  & NM     & --   & --    & --   & --             & --  & --     \\ 
e089  & 84.237180 & +09.8816230 &  ?    & --  & N---  & --  & NM?    & --   & --    & --   & --             & --  & --     \\ 
e090  & 83.932880 & +09.9341180 &  Yes  & N   & N--N- & --  & NM     & --   & --    & --   & --             & --  & --     \\ 
e091  & 84.131387 & +09.7503815 &  No   & Y   & ?Y-YY & Y   & Y      & III  & 098   & --   & --             & --  & --     \\ 
e092  & 84.293353 & +09.8575880 &  ?    & N?  & ----  & --  & NM?    & --   & --    & --   & --             & --  & --     \\ 
e093  & 83.856160 & +09.7949190 &  Yes  & N   & NN-N- & N   & NM     & II   & --    & --   & --             & --  & --     \\ 
e094  & 83.965231 & +10.1099520 &  ?    & N   & N--N- & --  & NM     & --   & --    & --   & --             & --  & --     \\ 
e095  & 84.233290 & +09.8680370 &  ?    & --  & NN    & --  & NM     & --   & --    & --   & --             & --  & --     \\ 
e096c & 84.043210 & +10.0053220 &  No   & Y   & YY-YY & Y   & Y?     & III  & 058   & --   & --             & --  & --     \\ 
e097  & 84.315251 & +09.9638890 &  ?    & Y?  & ---   & --  & NM?    & --   & --    & --   & --             & --  & --     \\ 
e098  & 84.104194 & +09.6920190 &  ?    & N   & ?Y--Y & --  & NM?    & III  & --    & --   & --             & --  & --     \\ 
e099  & 84.059388 & +09.7965685 &  Yes  & N   & N--N- & --  & NM     & --   & --    & --   & --             & --  & --     \\ 
e100  & 84.145230 & +09.9778570 &  ?    & --  & ----  & --  & NM?    & --   & --    & --   & --             & --  & --     \\ 
e101  & 84.167280 & +09.8100580 &  ?    & --  & ----  & --  & noD    & --   & --    & --   & --             & --  & --     \\ 
e102c & 83.877360 & +09.9374070 &  pos  & N   & NY-NY & --  & NM     & III  & --    & --   & --             & --  & --     \\ 
e102e & 83.877930 & +09.9372070 &  ?    & N   & ---N- & --  & NM     & --   & --    & --   & --             & --  & --     \\ 
e102s & 83.877830 & +09.9367550 &  Yes  & N   & NN-N- & ?   & NM     & II   & --    & --   & --             & --  & --     \\ 
e103c & 83.885830 & +09.9363730 &  Yes  & N   & N--N- & --  & NM     & --   & --    & --   & --             & --  & --     \\ 
e103s & 83.885870 & +09.9351430 &  Yes  & N   & N---  & --  & NM     & --   & --    & --   & --             & --  & --     \\ 
e103w & 83.884460 & +09.9354380 &  Yes  & N   & N--N- & --  & NM     & --   & --    & --   & --             & --  & --     \\ 
e104c & 83.981540 & +09.8694630 &  No   & Y?  & YY-YY & Y   & Y      & III  & --    & --   & --             & --  & --     \\ 
e104e & 83.982382 & +09.8691280 &  ?    & Y?  & ---   & --  & Yco    & --   & --    & --   & --             & --  & --     \\ 
e105  & 84.230945 & +10.0196270 &  ?    & Y   & ?Y--Y & Y   & Y      & III  & --    & --   & --             & --  & --     \\ 
e106  & 84.002820 & +09.6865570 &  ?    & Y   & NN--Y & N   & NM?    & I    & --    & --   & --             & --  & --     \\ 
e107  & 83.997378 & +09.8923268 &  No   & N   & N--NY & --  & NM     & --   & --    & --   & --             & --  & --     \\ 
e108  & 84.141150 & +10.0191500 &  ?    & --  & N---  & --  & NM?    & --   & --    & --   & --             & --  & --     \\ 
e109c & 84.045290 & +10.0147090 &  Yes  & N   & N--N- & --  & NM     & --   & --    & --   & --             & --  & --     \\ 
e109e & 84.045740 & +10.0147620 &  Yes  & N   & ---N- & --  & NM     & --   & --    & --   & --             & --  & --     \\ 
e109w & 84.044360 & +10.0148120 &  Yes  & N   & ---N- & --  & NM     & --   & --    & --   & --             & --  & --     \\ 
e110  & 84.185590 & +09.9905940 &  ?    & --  & ----  & --  & NM?    & --   & --    & --   & --             & --  & --     \\ 
e111  & 84.088850 & +09.8971860 &  No   & N   & YY-NN & Y   & NM     & III  & --    & --   & 0975-01766890  & --  & --     \\ 
e112  & 84.272730 & +09.9360910 &  ?    & --  & ----  & --  & noD    & --   & --    & --   & --             & --  & --     \\ 
\hline 
\end{tabular} 
\end{center} 
$\,$ \\ %
$^1$ ``Yes'' stands for objects located in the area of the CCD ($I-J$ versus $J-[3.6]$) where quasars are located, after \cite{Bouy09.1}.  See Fig. 5.\\
$^2$ Membership based on the HR diagram, see Fig. 6.\\
$^3$ Membership based on several color-magnitude and color-color diagrams, see Fig. 7.\\
$^4$ Final membership assigment: Probable member = Y, Possible member = Y?, Possible non-member = NM?, Probable non-member=NM,  noData=--\\ %
$^5$ Disk Class based on the Spitzer/IRAC color-color diagram, after \cite{Lada87.1} and \cite{Allen04}.\\
$^6$ Membership and spectra types after \cite{Sacco08.1}.\\
\end{table*}



\setcounter{table}{8}
%
%
\begin{table*}
\tiny 
\begin{center}
\caption{Membership criteria and other information for X-ray sources of Collinder 69  (western {\rm XMM-Newton} pointing or Col 69\,W).}
\label{tab:membershipcriteriaW} 
\newcolumntype{d}[1]{D{.}{.}{#1}}
\begin{tabular}{l  l   l             l      l      l       l     l        l      l       l      l               l      l}     \\   
\hline                                                                                                                                               
 XMMID & Alpha     & Delta      &QSO$^1$&HRD$^2$&CMD$^3$&CCD$^3$&Final$^4$&IRAC$^5$& ByN   & DM   & DM             &Sacco$^6$ & Sp.Type$^6$      \\
 C69-X-& (2000)    & (2000)     &       &     &       &     &        &      & 2004  & 1999 & 2002           &     &        \\          
\hline                                                                                                                                               
w001c & 83.528630 & +10.0171150 &  ?    & Y   & YY--Y & Y   & Y      & III  & --    & --   & --             & --  & --     \\ 
w001n & 83.527540 & +10.0178790 &  ?    & --  & ----  & --  & ?co    & --   & --    & --   & --             & --  & --     \\ 
w002c & 83.650830 & +09.8955050 &  No   & Y   & YY-YY & Y   & Y      & III  & 026   & 012  & --             & Y   & K9.5   \\ 
w002n & 83.652020 & +09.8968490 &  ?    & N   & ---N- & --  & NM     & --   & --    & --   & --             & --  & --     \\ 
w003  & 83.484741 & +09.8990646 &  No   & Y?  & YY-YY & Y   & Y      & III  & 013   & 004  & --             & --  & --     \\ 
w004  & 83.446590 & +09.9273299 &  No   & Y?  & YY-YY & Y   & Y      & III  & 001   & 001  & 0975-01716988  & --  & --     \\ 
w005c & 83.636690 & +09.9921350 &  No   & Y?  & YY-YY & Y   & Y      & III  & --    & 009  & --             & Y   & K4.5   \\ 
w005s & 83.635910 & +09.9908070 &  ?    & N   & ---N- & --  & NM     & --   & --    & --   & --             & --  & --     \\ 
w005w & 83.634850 & +09.9919500 &  ?    & N   & ---N- & --  & NM     & --   & --    & --   & --             & --  & --     \\ 
w006  & 83.663238 & +09.8819439 &  No   & Y   & YY-YY & Y   & Y      & III  & --    & 014  & --             & Y   & K7.0   \\ 
w007c & 83.617170 & +09.8133610 &  No   & Y   & YY-YY & Y   & Y      & III  & --    & --   & 0975-01729702  & --  & --     \\ 
w007e & 83.618610 & +09.8133080 &  ?    & Y   & N---Y & --  & NM?    & --   & --    & --   & --             & --  & --     \\ 
w008c & 83.458150 & +09.8436640 &  No   & Y   & Y?YYY & N   & Y      & I/II & 038   & 002  & --             & --  & --     \\ 
w008n & 83.457530 & +09.8445950 &  No   & N   & ?N-NY & N   & NM     & I/II & --    & --   & --             & --  & --     \\ 
w009c & 83.517360 & +10.0134200 &  ?    & N   & YY--Y & Y   & NM?    & III  & --    & --   & --             & --  & --     \\ 
w009s & 83.518440 & +10.0116000 &  No   & N   & YY-?N & Y   & NM     & III  & --    & --   & --             & --  & --     \\ 
w010c & 83.508380 & +09.9717320 &  Yes  & N   & NNNN- & ?   & NM     & II   & --    & --   & --             & --  & --     \\ 
w010e & 83.508870 & +09.9711810 &  ?    & N   & ---N- & --  & NM     & --   & --    & --   & --             & --  & --     \\ 
w011c & 83.523070 & +09.7129320 &  No   & Y   & YY-YY & Y   & Y      & III  & --    & 007  & --             & --  & --     \\ 
w011s & 83.524280 & +09.7113960 &  ?    & N   & ----  & --  & NM?    & --   & --    & --   & --             & --  & --     \\ 
w012  & 83.520295 & +09.9517170 &  No   & Y   & YYYYY & ?   & Y      & II   & --    & 006  & --             & --  & --     \\ 
w013c & 83.648260 & +09.9956380 &  No   & Y   & YY-YY & Y   & Y      & III  & 037   & 011  & --             & Y   & M2.0   \\ 
w013n & 83.648550 & +09.9977390 &  ?    & N   & ---N- & --  & NM     & --   & --    & --   & --             & --  & --     \\ 
w014c & 83.460180 & +10.0724590 &  ?    & Y?  & YYY-Y & Y   & Y      & III  & --    & --   & --             & --  & --     \\ 
w014w & 83.457800 & +10.0724880 &  ?    & N   & ?Y--? & --  & NM?    & III  & --    & --   & --             & --  & --     \\ 
w015  & 83.534911 & +09.8569582 &  No   & Y   & YY-YY & Y   & Y      & III  & 044   & --   & --             & --  & --     \\ 
w016  & 83.303970 & +09.8274840 &  ?    & --  & N---  & --  & NM?    & --   & --    & --   & --             & --  & --     \\ 
w017  & 83.404040 & +09.9432430 &  ?    & --  & N---  & --  & NM?    & --   & --    & --   & --             & --  & --     \\ 
w018  & 83.576309 & +09.8772189 &  No   & N?  & YY-?Y & Y   & NM?    & III  & --    & --   & --             & --  & --     \\ 
w019  & 83.427366 & +10.0632810 &  ?    & N   & NN?-? & N   & NM     & I    & --    & --   & --             & --  & --     \\ 
w020  & 83.463238 & +09.7784576 &  ?    & Y?  & YY--Y & Y   & Y      & III  & --    & --   & --             & --  & --     \\ 
w021  & 83.419080 & +10.0442010 &  ?    & --  & NN    & --  & NM     & --   & --    & --   & --             & --  & --     \\ 
w022  & 83.429710 & +10.1701040 &  ?    & --  & ----  & --  & noD    & --   & --    & --   & --             & --  & --     \\ 
w023  & 83.663531 & +10.0248160 &  No   & Y   & YY-YY & Y   & Y      & III  & 036   & --   & --             & N   & M3.0   \\ 
w024c & 83.549040 & +09.9510840 &  No   & Y   & YY-YY & Y   & Y      & III  & 052   & --   & --             & --  & --     \\ 
w024s & 83.549380 & +09.9496580 &  Yes  & N   & ---N- & --  & NM     & --   & --    & --   & --             & --  & --     \\ 
w025c & 83.510050 & +10.1182540 &  Yes  & N   & ---N- & --  & NM     & --   & --    & --   & --             & --  & --     \\ 
w025e & 83.511410 & +10.1185150 &  Yes  & N   & N--N- & --  & NM     & --   & --    & --   & --             & --  & --     \\ 
w025w & 83.509380 & +10.1181110 &  No   & N   & N--NY & --  & NM     & --   & --    & --   & --             & --  & --     \\ 
w026  & 83.266370 & +10.0351130 &  ?    & --  & ----  & --  & noD    & --   & --    & --   & --             & --  & --     \\ 
w027  & 83.349933 & +09.9521250 &  No   & N   & ---NN & --  & NM     & --   & --    & --   & 0975-01709853  & --  & --     \\ 
w028  & 83.397147 & +10.1405910 &  ?    & Y   & ??--Y & ?   & Y?     & II   & --    & --   & --             & --  & --     \\ 
w029  & 83.313132 & +09.8417150 &  No   & Y?  & YY?YY & ?   & Y      & II   & --    & --   & 0975-01707181  & --  & --     \\ 
w030c & 83.505450 & +09.9431290 &  Yes  & N   & NN-N- & N   & NM     & I    & --    & --   & --             & --  & --     \\ 
w030e & 83.507810 & +09.9425150 &  No   & N   & ---NY & --  & NM     & --   & --    & --   & --             & --  & --     \\ 
\hline 
\end{tabular} 
\end{center} 
$\,$ \\ %
$^1$ ``Yes'' stands for objects located in the area of the CCD ($I-J$ versus $J-[3.6]$) where quasars are located, after \cite{Bouy09.1}.  See Fig. 5.\\
$^2$ Membership based on the HR diagram, see Fig. 6.\\
$^3$ Membership based on several color-magnitude and color-color diagrams, see Fig. 7.\\
$^4$ Final membership assigment: Probable member = Y, Possible member = Y?, Possible non-member = NM?, Probable non-member=NM,  noData=--\\ %
$^5$ Disk Class based on the Spitzer/IRAC color-color diagram, after \cite{Lada87.1} and \cite{Allen04}.\\
$^6$ Membership and spectra types after \cite{Sacco08.1}.\\
\end{table*}



\addtocounter{table}{-1}
\begin{table*}
\tiny 
\begin{center}
\caption{Membership criteria and other information for X-ray sources of Collinder 69  (western {\rm XMM-Newton} pointing or Col 69\,W).}
\label{tab:membershipcriteria} 
\newcolumntype{d}[1]{D{.}{.}{#1}}
\begin{tabular}{l  l   l             l      l      l       l     l        l      l       l      l               l      l}     \\    
\hline                                                                                                                                               
 XMMID & Alpha     & Delta      &QSO$^1$&HRD$^2$&CMD$^3$&CCD$^3$&Final$^4$&IRAC$^5$& ByN   & DM   & DM             &Sacco$^6$ & Sp.Type$^6$      \\
 C69-X-& (2000)    & (2000)     &       &     &       &     &        &      & 2004  & 1999 & 2002           &     &        \\         
\hline                                                                                                                                               
w031  & 83.362344 & +10.0276360 &  ?    & Y?  & NN--? & ?   & NM     & II   & --    & --   & --             & --  & --     \\ 
w032c & 83.451270 & +10.0287590 &  No   & N?  & YY-NY & Y   & NM     & III  & --    & --   & --             & --  & --     \\ 
w032e & 83.451590 & +10.0283530 &  ?    & N?  & YY-N- & Y   & NM     & III  & --    & --   & --             & --  & --     \\ 
w032s & 83.451570 & +10.0267440 &  ?    & --  & ---N- & --  & NM?    & --   & --    & --   & --             & --  & --     \\ 
w032w & 83.449640 & +10.0276670 &  No   & Y   & ??-YY & ?   & Y      & II   & 113   & --   & --             & --  & --     \\ 
w032z & 83.450360 & +10.0267300 &  ?    & --  & ---N- & --  & NM?    & --   & --    & --   & --             & --  & --     \\ 
w033  & 83.427960 & +09.9672690 &  ?    & --  & ----  & --  & noD    & --   & --    & --   & --             & --  & --     \\ 
w034  & 83.365972 & +10.0722730 &  No   & N?  & ?Y-?Y & Y   & NM?    & III  & --    & --   & 0975-01711038  & --  & --     \\ 
w035  & 83.308680 & +09.9292630 &  ?    & --  & ----  & --  & noD    & --   & --    & --   & --             & --  & --     \\ 
w036  & 83.418051 & +09.8664630 &  No   & N   & YY-?N & Y   & NM     & III  & --    & --   & 0975-01714944  & --  & --     \\ 
w037  & 83.387396 & +09.9606980 &  No   & N?  & ?--?Y & --  & NM?    & --   & --    & --   & 0975-01712636  & --  & --     \\ 
w038c & 83.436040 & +10.0293920 &  ?    & --  & ----  & --  & noD    & --   & --    & --   & --             & --  & --     \\ 
w038s & 83.435420 & +10.0275230 &  ?    & N   & N--N- & --  & NM     & --   & --    & --   & --             & --  & --     \\ 
w039c & 83.449370 & +09.8822020 &  ?    & N   & ---N- & --  & NM     & --   & --    & --   & --             & --  & --     \\ 
w039w & 83.447470 & +09.8832160 &  No   & N   & ?Y-?N & Y   & NM     & III  & --    & --   & --             & --  & --     \\ 
w040  & 83.413316 & +10.0036900 &  No   & N?  & YY-?Y & Y   & NM?    & III  & --    & --   & 0975-01714590  & --  & --     \\ 
w041  & 83.429965 & +09.7279361 &  No   & N?  & YY-YY & Y   & NM?    & III  & --    & --   & 0975-01715801  & --  & --     \\ 
w042c & 83.463710 & +09.8034280 &  Yes  & N   & N--N- & --  & NM     & --   & --    & --   & --             & --  & --     \\ 
w042e & 83.464820 & +09.8034270 &  No   & N   & ---NY & --  & NM     & --   & --    & --   & --             & --  & --     \\ 
w042w & 83.462520 & +09.8044530 &  pos  & N   & N--NY & --  & NM     & --   & --    & --   & --             & --  & --     \\ 
w043  & 83.466746 & +09.8840859 &  ?    & N   & YY--Y & Y   & NM?    & III  & --    & --   & --             & --  & --     \\ 
w044  & 83.391590 & +09.8397680 &  ?    & --  & NN    & N   & NM     & II   & --    & --   & --             & --  & --     \\ 
w045  & 83.303580 & +09.9661210 &  ?    & --  & ----  & --  & noD    & --   & --    & --   & --             & --  & --     \\ 
w046  & 83.265781 & +09.9784980 &  ?    & Y?  & ---   & --  & NM?    & --   & --    & --   & --             & --  & --     \\ 
w047  & 83.298360 & +09.9841260 &  ?    & N   & ----  & --  & NM?    & --   & --    & --   & --             & --  & --     \\ 
w048  & 83.602812 & +09.9491913 &  No   & N   & N--NY & --  & NM     & --   & --    & --   & --             & --  & --     \\ 
w049  & 83.643163 & +10.0553090 &  Yes  & N   & N--N- & --  & NM     & --   & --    & --   & --             & --  & --     \\ 
w050  & 83.370230 & +09.8523960 &  ?    & --  & ----  & --  & noD    & --   & --    & --   & --             & --  & --     \\ 
w051  & 83.415110 & +09.8510330 &  ?    & --  & ----  & --  & noD    & --   & --    & --   & --             & --  & --     \\ 
w052  & 83.346180 & +09.9470260 &  ?    & --  & ----  & --  & noD    & --   & --    & --   & --             & --  & --     \\ 
\hline 
\end{tabular} 
\end{center} 
$\,$ \\ %
$^1$ ``Yes'' stands for objects located in the area of the CCD ($I-J$ versus $J-[3.6]$) where quasars are located, after \cite{Bouy09.1}.  See Fig. 5.\\
$^2$ Membership based on the HR diagram, see Fig. 6.\\
$^3$ Membership based on several color-magnitude and color-color diagrams, see Fig. 7.\\
$^4$ Final membership assigment: Probable member = Y, Possible member = Y?, Possible non-member = NM?, Probable non-member=NM,  noData=--\\ %
$^5$ Disk Class based on the Spitzer/IRAC color-color diagram, after \cite{Lada87.1} and \cite{Allen04}.\\
$^6$ Membership and spectra types after \cite{Sacco08.1}.\\
\end{table*}



\clearpage

\setcounter{table}{9}
%
%
\begin{table*}
\tiny 
\begin{center}
\caption{Properties for our probable and possible X-ray candidate members of Collinder 69, after the selection process. }  
\label{tab:finalmembers} 
\newcolumntype{d}[1]{D{.}{.}{#1}}
\begin{tabular}{l l                       r             r                  l          r             r            r       r        l      l        c       } \\ 
\hline                                           
Source         &   Other            &    Teff        & Lum(bol)        &Fraction&     Mass     &  Log L$_X$& Age     & Log     &Disk  &Disk    & IRAC    \\ 
                                        \cline{3-3}                               \cline{6-6}
          &   names                 &Logg=4.0 5Myr   & Logg=4.0        & Flux   & Logg=4  5Myr &           & Logg=4  &L$_X$/Lbol&class&slope   & slope      \\ 
C69\#          &                    &(K)             &(Lsun)           &        & (Msun)       &  (erg/s)  &  (Myr)  &         &      & s      &            \\ 
\hline                                                                                              

e001           &   LOri003   DM046  & 4000 K    4320  & 0.818 0.009 & 0.519 &  0.699    1.092 &     31.0  &  1.8 Si &  -2.50   & III  & No     &  -2.75     \\ 
e003$^{New}$    &   LOri---   --     & 5000 K    4909  & 2.356 0.032 & 0.580 &  1.622    1.661 &     30.7  & 10.0 Si &  -3.26   & III  & No     &  -2.77     \\ 
e004           &   LOri008   DM051  & 3750 K    4219  & 0.670 0.006 & 0.496 &  0.499    0.970 &     30.6  &  1.0 Si &  -2.81   & III  & No     &  -2.62     \\ 
e006           &   LOri017   DM060  & 4000 K    3967  & 0.418 0.005 & 0.520 &  0.726    0.696 &     30.3  &  2.4 Si &  -2.91   & III  & No     &  -2.77     \\ 
e007c          &   LOri014   DM058  & 4000 K    4074  & 0.508 0.007 & 0.519 &  0.707    0.813 &     30.4  &  2.0 Si &  -2.89   & III  & No     &  -2.72     \\ 
e008           &   LOri---   DM057  & 4750 K    4495  & 1.149 0.014 & 0.561 &  1.315    1.294 &     30.4  &  9.9 Si &  -3.24   & III  & No     &  -2.78     \\ 
e009$^{New}$    &   LOri---   --     & 4750 K    5229  & 4.283 0.018 & 0.171 &  1.900    1.869 &     31.1  &  4.9 Si &  -3.12   & III  & Thin   &  -2.27     \\ 
e010c          &   LOri---   DM045  & 4750 K    4718  & 1.686 0.021 & 0.581 &  1.501    1.505 &     30.4  &  7.3 Si &  -3.41   & III  & No     &  -2.75     \\ 
e011c          &   LOri025   DM059  & 4000 K    4000  & 0.445 0.011 & 0.519 &  0.719    0.732 &     30.3  &  2.2 Si &  -2.93   & III  & No     &  -2.73     \\ 
''  e$^{Y?}$    &                    & --   --         & onlyRI  --  & --    &  --            &      --    &         &   --     & --   & ----   &  --        \\ 
e012$^{New}$   &   LOri---   --     & 3750 K    5237  & 4.340 0.038 & 0.374 &  0.500    1.873 &     30.4  &  0.5 Si &  -3.82   & III  & Thin   &  -2.46     \\ 
e013$^{Y?,New}$&   LOri---   --      & 5500 K  11998   & 82.11 0.325 & 0.132 &  4.114    3.035 &     30.1  & 5.0 Si  & -5.40    & (III)& ----   &  --        \\ 
e014          &   LOri022   DM044  & 3750 K    3936  & 0.387 0.004 & 0.499 &  0.499    0.665 &     30.0  &  1.6 Si &  -3.17   & III  & No     &  -2.51     \\ 
e016          &   LOri018   DM056  & 3750 K    4005  & 0.450 0.005 & 0.494 &  0.499    0.738 &     29.9  &  1.4 Si &  -3.34   & III  & No     &  -2.66     \\ 
e017          &   LOri028   --     & 3700 NG   3757  & 0.245 0.003 & 0.475 &  0.699    0.498 &     29.9  &  7.8 NG &  -3.07   & III  & No     &  -2.68     \\ 
e018          &   LOri---   DM065  & 3750 K    3840  & 0.302 0.004 & 0.496 &  0.499    0.573 &     30.1  &  1.9 Si &  -2.96   & III  & No     &  -2.77     \\ 
e019          &   LOri---   DM070  & 3750 K    3757  & 0.245 0.003 & 0.498 &  0.499    0.498 &     30.0  &  1.9 Si &  -2.97   & III  & No     &  -2.76     \\ 
e020$^{New}$   &   LOri---   --     & 5000 K    5297  & 4.865 0.062 & 0.562 &  2.109    1.905 &     29.8  &  7.0 Si &  -4.47   & III  & No     &  -2.65     \\ 
e022          &   LOri032   DM055  & 3700 NG   3803  & 0.277 0.003 & 0.471 &  0.701    0.539 &     29.9  &  6.3 NG &  -3.13   & III  & No     &  -2.67     \\ 
e023          &   LOri004   --     & 4500 K    4435  & 1.027 0.004 & 0.191 &  1.283    1.227 &     30.0  &  5.9 Si &  -3.60   & III  & No     &  -2.66     \\ 
e025          &   LOri---   DM061  & 3750 K    3673  & 0.203 0.003 & 0.530 &  0.499    0.437 &     29.7  &  1.9 Si &  -3.19   & III  & No     &  -2.76     \\ 
e027c          &   LOri---   DM062  & 3750 K    3644  & 0.189 0.002 & 0.504 &  0.499    0.416 &     29.7  &  2.0 Si &  -3.16   & III  & No     &  -2.72     \\ 
e028c          &   LOri023   DM050  & 3750 K    3911  & 0.363 0.004 & 0.496 &  0.499    0.640 &     29.7  &  1.7 Si &  -3.44   & III  & No     &  -2.70     \\ 
e029c          &   LOri033   DM039  & 3750 K    3840  & 0.303 0.006 & 0.499 &  0.499    0.573 &     29.9  &  1.8 Si &  -3.17   & III  & No     &  -2.69     \\ 
e030          &   LOri---   DM053  & 4000 K    3952  & 0.403 0.005 & 0.517 &  0.729    0.681 &     30.0  &  2.5 Si &  -3.19   & III  & $^{TR}$&  -2.56     \\ 

e031          &   LOri---   DM067  & 3600 NG   3762  & 0.248 0.003 & 0.464 &  0.587    0.502 &     29.6  &  5.0 NG &  -3.38   & III  & No     &  -2.81     \\ 
e034c          &   LOri---   DM035  & 4000 K    3886  & 0.339 0.004 & 0.517 &  0.747    0.616 &     30.1  &  2.9 Si &  -3.01   & III  & No     &  -2.78     \\ 
e035c          &   LOri040   DM041  & 3500 NG   3764  & 0.249 0.003 & 0.455 &  0.465    0.503 &     29.7  &  3.1 NG &  -3.28   & III  & No     &  -2.65     \\ 
e036          &   LOri041   DM038  & 3200 NG   3750  & 0.242 0.003 & 0.440 &  --       0.493 &     29.8  &  --  -- &  -3.17   & III  & No     &  -2.67     \\ 
e037c          &   LOri064   --     & 3400 NG   3232  & 0.076 0.001 & 0.466 &  0.207    0.199 &     29.6  &  3.1 NG &  -2.87   & III  & No     &  -2.58     \\ 
e040$^{New}$   &   LOri---   --     & 3600 NG   3461  & 0.130 0.001 & 0.463 &  0.557    0.316 &     29.4  & 11.6 NG &  -3.30   & III  & No     &  -2.78     \\ 
e041$^{Y?}$    &  LOri---   DM071   & 3700 NG  3671   & 0.202 0.002 & 0.463 & 0.687    0.435  &     29.8  & 9.9 NG  & -3.09    & --   & $^{TR}$&  --        \\ 

e042$^{Y?,New}$ &  LOri---   --      & 7250 K  12323   &98.800 0.148 & 0.066 & 3.500    3.170  &     29.2  &18.3 Si  & -6.38    & (III)& ----   &  --      B \\ 
e054          &   LOri051   --     & 3500 NG   3509   & 0.145 0.001 & 0.444 &  0.449    0.341 &     29.1  &  6.3 NG &  -3.64   & III  & No     &  -2.74     \\ 
e055          &   LOri054   --     & 3300 NG   3501   & 0.142 0.001 & 0.433 &  0.287    0.337 &     29.5  &  2.5 NG &  -3.24   & III  & No     &  -2.64     \\ 
e057$^{New}$   &   LOri---   --     & 4750 K    5213  & 4.163 0.052 & 0.582 &  1.899    1.860 &     29.1  &  4.9 Si &  -5.10   & III  & No     &  -2.74     \\ 
e058          &   LOri---   DM069  & 4000 K    3857  & 0.315 0.004 & 0.519 &  0.752    0.589 &     29.3  &  2.9 Si &  -3.78   & III  & No     &  -2.70     \\ 
e064$^{New}$   &   LOri---   --     & 3750 K    3561  & 0.160 0.001 & 0.485 &  0.498    0.368 &     29.6  &  2.0 Si &  -3.19   & III  & No     &  -2.64     \\ 
e072$^{New}$   &   LOri---   --     & 3600 NG   3359  & 0.107 0.001 & 0.467 &  0.560    0.266 &     29.1  & 15.8 NG &  -3.51   & III  & No     &  -2.73     \\ 
e078          &   LOri090   --     & 3400 NG   3065  & 0.040 0.001 & 0.412 &  0.312    0.128 &     29.1  & 20.2 NG &  -3.08   & III  & No     &  -2.58     \\ 
e079          &   LOri055   --     & 3700 NG   3496  & 0.141 0.001 & 0.429 &  0.648    0.335 &     29.6  & 15.8 NG &  -3.13   & III  & No     &  -2.72     \\ 
e081c          &   LOri047   DM047  & 3300 NG   3651  & 0.192 0.002 & 0.451 &  0.299    0.421 &     29.3  &  1.7 NG &  -3.57   & III  & No     &  -2.64     \\ 
e082          &   LOri039   DM049  & 3700 NG   3688  & 0.211 0.003 & 0.472 &  0.691    0.448 &     29.4  &  9.3 NG &  -3.51   & III  & No     &  -2.66     \\ 
e084          &   LOri---   DM040  & 3700 NG   3482  & 0.137 0.002 & 0.470 &  0.646    0.328 &     29.3  & 16.1 NG &  -3.42   & III  & No     &  -2.83     \\ 
e086$^{New}$   &   LOri---   --     & 3200 NG   3202  & 0.070 0.001 & 0.336 &  0.201    0.186 &     29.1  &  3.7 NG &  -3.33   & III  & No     &  -2.56     \\ 
e091          &   LOri098   --     & 3200 NG   3043  & 0.035 0.001 & 0.407 &  0.169    0.119 &     29.2  &  6.5 NG &  -2.93   & III  & No     &  -2.63     \\ 
e096c$^{Y?}$    &  LOri058   --      & 3300 NG  3314   & 0.097 0.001 & 0.450 & 0.214    0.243  &     29.0  & 2.4 NG  & -3.57    & III  & No     & -2.99      \\ 
e104c$^{New}$  &   LOri---   --     & 3700 NG   3389   & 0.113 0.001 & 0.472 &  0.633    0.280 &     29.0  & 21.4 NG &  -3.64   & III  & No     &  -2.65     \\ 
''  e          &                    & 1200 C    1841  & 0.0003 0.0001& 0.365 &  --       0.009 &     --    & --   -- &   --     & --   & ----   &   --       \\ 
e105$^{New}$   &   LOri---   --     & 3300 NG   3141  & 0.057 0.001 & 0.300 &  0.250    0.161 &     29.0  &  7.8 NG &  -3.34   & III  & No     &  -2.74     \\ 
w001c$^{New}$   &   LOri---   --     & 5250 K    5175  & 3.865 0.012 & 0.246 &  1.800    1.847 &     31.0  & 11.8 Si &  -3.12   & III  & No     &  -2.77     \\ 
w002c          &   LOri026   DM012  & 3700 NG   3960  & 0.411 0.005 & 0.467 &  0.749    0.689 &     30.9  &  3.9 NG &  -2.30   & III  & No     &  -2.70     \\ 
w003          &   LOri013   DM004  & 3600 NG   4146  & 0.580 0.006 & 0.465 &  0.651    0.887 &     30.3  &  1.5 NG &  -3.05   & III  & No     &  -2.69     \\ 
w004          &   LOri001   DM001  & 4000 K    4376  & 0.916 0.006 & 0.521 &  0.698    1.158 &     30.5  &  1.7 Si &  -3.05   & III  & No     &  -2.81     \\ 
w005c          &   LOri---   DM009  & 4500 K    4196  & 0.640 0.009 & 0.546 &  1.099    0.943 &     30.4  &  7.7 Si &  -2.99   & III  & No     &  -2.82     \\ 
w006          &   LOri---   DM014  & 4000 K    3995  & 0.441 0.006 & 0.516 &  0.720    0.726 &     30.7  &  2.2 Si &  -2.53   & III  & No     &  -2.74     \\ 
w007c$^{New}$   &   LOri---   --     & 3600 NG   3442  & 0.125 0.001 & 0.463 &  0.557    0.307 &     30.4  & 12.4 NG &  -2.28   & III  & No     &  -2.85     \\ 
w008c          &   LOri038   DM002  & 3750 K    3790  & 0.267 0.003 & 0.432 &  0.499    0.527 &     30.1  &  1.9 Si &  -2.91   & I/II & Thick  &  -1.27     \\ 
w011c          &   LOri---   DM007  & 4750 K    4761  & 1.824 0.025 & 0.595 &  1.544    1.543 &     30.3  &  7.0 Si &  -3.55   & III  & No     &  -2.73     \\ 
w012          &   LOri---   DM006  & 3700 NG   3833  & 0.298 0.003 & 0.460 &  0.709    0.566 &     29.8  &  6.1 NG &  -3.26   & II   & Thick  &  -1.39     \\ 
w013c          &   LOri037   DM011  & 3700 NG   3805  & 0.278 0.003 & 0.468 &  0.702    0.541 &     30.0  &  6.3 NG &  -3.03   & III  & No     &  -2.67     \\ 
w014c$^{New}$   &   LOri---   --     & 6750 K    6031  &14.233 0.031 & 0.090 &   --      2.088 &     29.8  &  --  -- &  -4.47   & III  & $^{TR}$ &  -2.71     \\ 
w015          &   LOri044   --     & 3750 K    3616  & 0.176 0.001 & 0.512 &  0.499    0.397 &     29.7  &  2.0 Si &  -3.13   & III  & No     &  -2.67     \\ 
w020$^{New}$   &   LOri---   --     & 7000 K    5896  &12.483 0.019 & 0.081 &  --       2.058 &     29.8  & --   -- &  -4.88   & III  & No     &  -2.65     \\ 
w023          &   LOri036   --     & 3700 NG   3773  & 0.255 0.002 & 0.423 &  0.699    0.511 &     29.9  &  7.2 NG &  -3.09   & III  & No     &  -2.67     \\ 
w024c          &   LOri052   --     & 3700 NG   3531  & 0.151 0.001 & 0.450 &  0.654    0.353 &     29.5  & 14.8 NG &  -3.26   & III  & No     &  -2.65     \\ 
w028$^{Y?,New}$&  LOri---   --      & 2500 NG  2698   & 0.007 0.001 & 0.382 &  -       0.036 &     29.8  & --   --  & -1.65    & II   & Thick  & -1.69      \\ 
w029$^{New}$   &   LOri---   --     & 3200 NG   3587  & 0.167 0.002 & 0.431 &  0.221    0.382 &     29.6  & 01.3 NG &  -3.21   & II   & Thin   &  -1.97     \\ 
\hline                                                                      
\end{tabular} 
\end{center} 
$^{Y?}$ Possible member.\\
$^{New}$ New probable or possible member identified in this work.\\  
$^{TR}$ Transition disk based on the excess at 24 micron.\\
(III).- Class III based on SED with incomplete IRAC data, but detected flux at 24 micron.\\
Models: K=Kurucz, NG=NextGen (Baraffe et al. 1998), C=COND2000 (Chabrier et al. 2000), Si=Siess (\cite{Siess00.1}).\\
Effective temperatures and masses were derived using VOSA (Bayo et al. 2008, assuming Log g=4.0), and using  5 Myr isochrones (Siess et al. 2000 and Chabrier et al. 2000).\\
Ages are estimated from a HR diagram,  using VOSA (Bayo et al. 2008, assuming Log g=4.0) and several isochrones.\\
Spectral types and equivalent withs from: S=Sacco et al. 2008, D=D\&M 1999, B= Bayo 2009 (PhD thesis).\\
\end{table*}


\clearpage

\setcounter{table}{10}
%
%
\begin{landscape}
\begin{table}
\caption{Stellar parameters and X-ray upper limits for probable and possible members of Collinder\,69 (taken from D\&M1999, ByN2004 and Morales (2008)) that are undetected in X-rays.
(easter pointing --Col69\,E)}
\label{tab:stelpar_ulE} 
\begin{tabular}{lrrrcrrrcclrr}
Designation  & RA          & DEC          & Catalog&XMM- & Rate            &$\log{L_{\rm x}}$&$\log{(\frac{L_{\rm x}}{L_{\rm bol}})}$& IRAC &   Disk?      & ${T_{\rm eff}}$  & Lum(bol)      & $Mass$ \\
             &             &              &        &field&                 &                &                                     &class &  type       &Model Logg=4 5Myr&[$L_\odot$]    &[$M_\odot$]\\ 
             &             &              & (1)    &     &[$10^{-3}$\,cts/s]& [erg/s]        &                                    &  (2) &      (3)     &   (4)         &    (5)         &   (6)  \\ 
\hline 
\hline 
C69EI1-00790 & 5:36:55.330 & +09:46:48.00 &1 sq.deg&  E  &   $ <  1.00$    &   $ <  29.2$   &         $ <  -3.40$                &    II &        Thick &  NG  3700  3344& 0.1041 0.0015 &  0.259 \\ 
C69EI1-02229 & 5:36:48.880 & +09:55:42.40 &1 sq.deg&  E  &   $ <  0.83$    &   $ <  29.1$   &         $ <  -3.32$                &    II &        Thick &  Ku  4250  3197& 0.0687 0.0006 &  0.184 \\ 
C69EI1-03042 & 5:36:43.560 & +09:55:13.80 &1 sq.deg&  E  &   $ <  0.61$    &   $ <  29.0$   &         $ <  -2.65$                &    II &        Thick &  NG  3500  2850& 0.0117 0.0003 &  0.056 \\ 
C69EI1-04761 & 5:36:32.170 & +09:53:16.70 &1 sq.deg&  E  &   $ <  0.58$    &   $ <  29.0$   &         $ <  -4.64$                &   III &         Thin &  Ku  5750  4495& 1.1488 0.0120 &  1.294 \\ 
C69EI1-07620 & 5:36:11.230 & +09:46:26.40 &1 sq.deg&  E  &   $ <  0.89$    &   $ <  29.2$   &         $ <  -3.39$                &    II &         Thin &  NG  2900  3336& 0.1024 0.0011 &  0.254 \\ 
C69EI1-13002 & 5:35:37.210 & +09:56:51.70 &1 sq.deg&  E  &   $ <  1.05$    &   $ <  29.3$   &         $ <  -3.20$                &    II &        Thick &  NG  3200  3258& 0.0830 0.0008 &  0.213 \\ 
C69EI1-14424 & 5:35:28.450 & +10:02:27.50 &1 sq.deg&  E  &   $ <  1.89$    &   $ <  29.5$   &         $ <  -3.26$                &    II &        Thick &  NG  3700  3525& 0.1493 0.0018 &  0.350 \\ 
C69EI1-14609 & 5:35:25.340 & +09:54:47.70 &1 sq.deg&  E  &   $ <  1.38$    &   $ <  29.4$   &         $ <  -2.92$                &    II &        Thick &  NG  3300  3134& 0.0550 0.0004 &  0.157 \\ 
DM048        & 5:35:55.860 & +09:56:21.70 & DM1999 &  E  &   $ <  0.72$    &   $ <  29.1$   &         $ <  -3.73$                &    II &        Thick &  NG  3700  3623& 0.1780 0.0010 &  0.400 \\ 
LOri002      &  5 36 10.36 & +10  8 54.90 &CFHT1999&  E  &   $ <  1.70$    &   $ <  29.5$   &         $ <  -4.08$                &   III &     Diskless &  Ku  3750  4420& 0.9961 0.0101 &  1.210 \\ 
LOri063      &  5 35 19.16 & +09 54 41.90 &CFHT1999&  E  &   $ <  2.10$    &   $ <  29.6$   &         $ <  -2.91$                &  I/II &        Thick &  NG  3600  3260& 0.0835 0.0009 &  0.214 \\ 
LOri065      &  5 35 17.95 & +09 56 58.40 &CFHT1999&  E  &   $ <  2.11$    &   $ <  29.6$   &         $ <  -2.87$                &   III &   Transition &  NG  3600  3237& 0.0774 0.0008 &  0.201 \\ 
LOri067      &  5 36 26.37 & +09 45 46.80 &CFHT1999&  E  &   $ <  1.08$    &   $ <  29.3$   &         $ <  -3.08$                &   III &     Diskless &  NG  3700  3170& 0.0628 0.0007 &  0.173 \\ 
LOri070      &  5 36  0.03 & +10  5 49.00 &CFHT1999&  E  &   $ <  0.96$    &   $ <  29.2$   &         $ <  -3.17$                &   III &     Diskless &  NG  3600  3160& 0.0606 0.0007 &  0.168 \\ 
LOri074      &  5 36  0.57 & +09 42 38.15 &CFHT1999&  E  &   $ <  1.55$    &   $ <  29.4$   &         $ <  -3.09$                &   III &     Diskless &  NG  3400  3250& 0.0808 0.0007 &  0.208 \\ 
LOri078      &  5 36 16.43 & +09 50 16.42 &CFHT1999&  E  &   $ <  0.65$    &   $ <  29.0$   &         $ <  -3.28$                &   III &     Diskless &  NG  3600  3112& 0.0502 0.0010 &  0.148 \\ 
LOri080      &  5 35 30.04 & +09 59 25.84 &CFHT1999&  E  &   $ <  1.35$    &   $ <  29.4$   &         $ <  -3.01$                &    II &         Thin &  NG  3100  3188& 0.0667 0.0006 &  0.180 \\ 
LOri082      &  5 36  0.80 & +09 52 57.40 &CFHT1999&  E  &   $ <  0.70$    &   $ <  29.1$   &         $ <  -3.18$                &    II &     Diskless &  NG  3400  3109& 0.0495 0.0006 &  0.147 \\ 
LOri083      &  5 35 43.44 & +09 54 26.28 &CFHT1999&  E  &   $ <  0.94$    &   $ <  29.2$   &         $ <  -3.06$                &   III &     Diskless &  NG  3400  3100& 0.0475 0.0006 &  0.143 \\ 
LOri085      &  5 35 21.54 & +09 53 29.00 &CFHT1999&  E  &   $ <  1.85$    &   $ <  29.5$   &         $ <  -2.77$                &    II &        Thick &  NG  3300  3106& 0.0488 0.0005 &  0.145 \\ 
LOri092      &  5 35 50.95 & +09 51  3.86 &CFHT1999&  E  &   $ <  0.88$    &   $ <  29.2$   &         $ <  -2.98$                &    II &     Diskless &  NG  3400  3064& 0.0397 0.0005 &  0.128 \\ 
LOri095      &  5 35 24.21 & +09 55 14.94 &CFHT1999&  E  &   $ <  1.41$    &   $ <  29.4$   &         $ <  -2.74$                &   III &     Diskless &  NG  3300  3048& 0.0363 0.0005 &  0.121 \\ 
LOri102      &  5 35 22.04 & +09 52 52.19 &CFHT1999&  E  &   $ <  2.84$    &   $ <  29.7$   &         $ <  -2.38$                &   III &     Diskless &  NG  3200  3026& 0.0315 0.0006 &  0.112 \\ 
LOri106      &  5 35 28.80 & +09 54  9.92 &CFHT1999&  E  &   $ <  1.30$    &   $ <  29.3$   &         $ <  -2.74$                &    II &        Thick &  NG  3100  3013& 0.0286 0.0005 &  0.106 \\ 
LOri107      &  5 35 55.17 & +09 52 20.35 &CFHT1999&  E  &   $ <  1.04$    &   $ <  29.2$   &         $ <  -2.88$                &   III &     Diskless &  NG  2900  3025& 0.0312 0.0005 &  0.111 \\ 
LOri110      &  5 35 32.63 & +09 52 48.87 &CFHT1999&  E  &   $ <  1.32$    &   $ <  29.3$   &         $ <  -2.63$                &    II &         Thin &  NG  3300  2985& 0.0224 0.0005 &  0.094 \\ 
LOri114      &  5 36 18.10 & +09 52 25.53 &CFHT1999&  E  &   $ <  0.57$    &   $ <  29.0$   &         $ <  -2.92$                &    II &         Thin &  NG  3000  2982& 0.0219 0.0004 &  0.093 \\ 
LOri118      &  5 35 24.42 & +09 53 51.73 &CFHT1999&  E  &   $ <  1.42$    &   $ <  29.4$   &         $ <  -2.45$                &    II &        Thick &  NG  3000  2955& 0.0183 0.0003 &  0.083 \\ 
LOri126      &  5 35 39.88 & +09 53 23.64 &CFHT1999&  E  &   $ <  1.24$    &   $ <  29.3$   &         $ <  -2.39$                &    II &        Thick &  NG  3000  2880& 0.0128 5.0E-5 &  0.061 \\ 
LOri129      &  5 36  9.84 & +09 42 37.44 &CFHT1999&  E  &   $ <  1.30$    &   $ <  29.3$   &         $ <  -2.49$                &    II &        Thick &  NG  2900  2929& 0.0160 0.0004 &  0.075 \\ 
LOri131      &  5 36  7.02 & +09 52 51.65 &CFHT1999&  E  &   $ <  0.62$    &   $ <  29.0$   &         $ <  -2.72$                &    II &         Thin &  NG  2900  2895& 0.0136 0.0003 &  0.065 \\ 
LOri134      &  5 35 22.88 & +09 55  6.65 &CFHT1999&  E  &   $ <  1.48$    &   $ <  29.4$   &         $ <  -2.27$                &   III &         Thin &  NG  2900  2863& 0.0121 0.0003 &  0.058 \\ 
LOri139      &  5 35 44.34 & +10  5 54.11 &CFHT1999&  E  &   $ <  1.42$    &   $ <  29.4$   &         $ <  -2.09$                &    II &        Thick &  NG  3000  2729& 0.0080 3.1E-5 &  0.038 \\ 
LOri148      &  5 36 29.00 & +09 43 21.45 &CFHT1999&  E  &   $ <  1.26$    &   $ <  29.3$   &         $ <  -1.97$                &    -- &         Thin &  Du  2400  2588& 0.0049 2.6E-5 &  0.028 \\ 
LOri153      &  5 36 18.20 & +09 57 40.97 &CFHT1999&  E  &   $ <  0.52$    &   $ <  28.9$   &         $ <  -2.20$                &    -- &           -- &  NG  2900  2508& 0.0033 2.2E-5 &  0.023 \\ 
LOri155      &  5 36 25.07 & +10  1 54.32 &CFHT1999&  E  &   $ <  0.59$    &   $ <  29.0$   &         $ <  -2.13$                &   III &         Thin &  NG  2700  2518& 0.0035 2.7E-5 &  0.024 \\ 
LOri161      &  5 35 54.10 & +09 43 36.11 &CFHT1999&  E  &   $ <  1.44$    &   $ <  29.4$   &         $ <  -1.44$                &    -- &           -- &  NG  2700  2389& 0.0018 1.6E-5 &  0.018 \\ 
LOri163      &  5 35 18.36 & +09 56 52.78 &CFHT1999&  E  &   $ <  2.06$    &   $ <  29.5$   &         $ <  -1.26$                &    -- &           -- &  Du  2200  2340& 0.0015 3.6E-5 &  0.017 \\ 
\hline 
\end{tabular} 
\,\\ %
(1) First listed in Morales-Calder\'on(2008) --1 sq.deg with Spitzer, Dolan \& Mathieu (1999) --DM1999- or Barrado y Navascu\'es et al. (2004) --CFHT1999.\\
(2) Based on Spitzer/IRAC color-color diagram.\\
(3) IRAC slope, after Lada et al. (2007).\\
(4) Temperatures from VOSA fitting with logg=4.0 dex or from a 5 Myr isochrone.\\
(5) Bolometric luminosity from VOSA and logg=4.0 dex.\\
(6) Mass from a 5 My isochrone.\\
\end{table}
\end{landscape}

\clearpage

 \setcounter{table}{11}
%
%
\begin{landscape}
\begin{table}
\caption{Stellar parameters and X-ray upper limits for probable and possible members of Collinder\,69 (taken from D\&M1999, ByN2004 and Morales (2008)) that are undetected in X-rays.
(western pointing --Col69\,W)}
\label{tab:stelpar_ulW}
\begin{tabular}{l r            r              r       c      r                     r                  r                               c       c            l                   r             r}
Designation  & RA          & DEC          & Catalog&XMM- & Rate            &$\log{L_{\rm x}}$&$\log{(\frac{L_{\rm x}}{L_{\rm bol}})}$& IRAC &   Disk?      & ${T_{\rm eff}}$  & Lum(bol)      & $Mass$ \\ 
             &             &              &        &field&                 &                &                                     &class &  type       &Model Logg=4 5Myr&[$L_\odot$]    &[$M_\odot$]\\ 
             &             &              & (1)    &     &[$10^{-3}$\,cts/s]& [erg/s]        &                                    &  (2) &      (3)     &   (4)         &    (5)         &   (6)  \\ 
\hline
\hline
C69WI1-09288 & 5:34:03.910 & +09:52:12.30 &1 sq.deg&  W  &   $ <  0.91$    &   $ <  29.2$   &         $ <  -3.63$                &    II &         Thin &  NG  3000  3615& 0.1757 0.0016 &  0.396 \\ 
C69WI1-14892 & 5:33:21.960 & +10:02:02.00 &1 sq.deg&  W  &   $ <  1.31$    &   $ <  29.3$   &         $ <  -3.12$                &    II &        Thick &  Ku  4250  3196& 0.0685 0.0005 &  0.184 \\ 
LOri005      &  5 33 53.65 & +09 43  7.97 &CFHT1999&  W  &   $ <  1.94$    &   $ <  29.5$   &         $ <  -4.03$                &   III &     Diskless &  Ku  4000  4360& 0.8880 0.0089 &  1.140 \\ 
LOri007      &  5 34 29.55 & +09 48 58.70 &CFHT1999&  W  &   $ <  1.72$    &   $ <  29.5$   &         $ <  -3.88$                &   III &     Diskless &  Ku  4000  4188& 0.6309 0.0082 &  0.935 \\ 
LOri011      &  5 34 44.66 & +09 53 57.50 &CFHT1999&  W  &   $ <  3.64$    &   $ <  29.8$   &         $ <  -3.62$                &   III &     Diskless &  Ku  3750  4227& 0.6807 0.0075 &  0.980 \\ 
LOri015      &  5 34 21.84 & +10  4 14.50 &CFHT1999&  W  &   $ <  1.33$    &   $ <  29.4$   &         $ <  -3.97$                &   III &     Diskless &  Ku  4250  4168& 0.6054 0.0061 &  0.911 \\ 
LOri046      &  5 34 26.08 & +09 51 49.70 &CFHT1999&  W  &   $ <  1.30$    &   $ <  29.3$   &         $ <  -3.48$                &   III &     Diskless &  Ku  3750  3556& 0.1583 0.0017 &  0.366 \\ 
LOri053      &  5 34 36.71 & +09 52 58.10 &CFHT1999&  W  &   $ <  2.50$    &   $ <  29.6$   &         $ <  -3.12$                &   III &     Diskless &  NG  3700  3482& 0.1365 0.0016 &  0.327 \\ 
LOri071      &  5 34 15.78 & +10  6 54.44 &CFHT1999&  W  &   $ <  1.79$    &   $ <  29.5$   &         $ <  -2.96$                &   III &     Diskless &  NG  3300  3224& 0.0746 0.0008 &  0.195 \\ 
LOri081      &  5 33 56.64 & +10  6 14.70 &CFHT1999&  W  &   $ <  1.12$    &   $ <  29.3$   &         $ <  -2.91$                &    II &        Thick &  NG  3600  3075& 0.0422 0.0006 &  0.133 \\ 
LOri086      &  5 34 11.56 & +09 49 15.61 &CFHT1999&  W  &   $ <  1.28$    &   $ <  29.3$   &         $ <  -2.89$                &   III &     Diskless &  NG  3600  3066& 0.0402 0.0006 &  0.129 \\ 
LOri087      &  5 34 33.72 & +09 55 33.44 &CFHT1999&  W  &   $ <  1.72$    &   $ <  29.5$   &         $ <  -2.77$                &   III &         Thin &  NG  3400  3106& 0.0488 0.0007 &  0.145 \\ 
LOri091      &  5 34 35.82 & +09 54 25.99 &CFHT1999&  W  &   $ <  1.66$    &   $ <  29.5$   &         $ <  -2.76$                &    II &         Thin &  NG  3100  3098& 0.0472 0.0006 &  0.142 \\ 
LOri093      &  5 34 41.19 & +09 50 16.34 &CFHT1999&  W  &   $ <  5.57$    &   $ <  30.0$   &         $ <  -2.18$                &   III &     Diskless &  NG  3400  3062& 0.0393 0.0006 &  0.127 \\ 
LOri094      &  5 34 43.18 & +10  1 59.92 &CFHT1999&  W  &   $^b$          &   $^b$         &         $^b$                        &   III &     Diskless &  NG  3200  3067& 0.0403 0.0005 &  0.129 \\ 
LOri105      &  5 34 17.57 & +09 52 30.14 &CFHT1999&  W  &   $ <  1.23$    &   $ <  29.3$   &         $ <  -2.68$                &   III &     Diskless &  NG  3200  2996& 0.0248 0.0004 &  0.099 \\ 
LOri109      &  5 34  8.54 & +09 50 43.57 &CFHT1999&  W  &   $ <  1.10$    &   $ <  29.3$   &         $ <  -2.63$                &   III &     Diskless &  NG  3200  2985& 0.0224 9.1E-5 &  0.094 \\ 
LOri113      &  5 33 47.92 & +10  1 39.62 &CFHT1999&  W  &   $ <  0.90$    &   $ <  29.2$   &         $ <  -2.67$                &    II &        Thick &  NG  3300  2967& 0.0194 7.8E-5 &  0.087 \\ 
LOri120      &  5 34 46.20 & +09 55 36.90 &CFHT1999&  W  &   $ <  3.67$    &   $ <  29.8$   &         $ <  -2.00$                &    II &        Thick &  NG  3000  2936& 0.0166 0.0004 &  0.077 \\ 
LOri122      &  5 34 35.43 & +09 51 18.71 &CFHT1999&  W  &   $ <  1.77$    &   $ <  29.5$   &         $ <  -2.27$                &    II &         Thin &  NG  3000  2921& 0.0153 0.0004 &  0.072 \\ 
LOri124$^a$  &  5 34 14.24 & +09 48 26.97 &CFHT1999&  W  &   $ <  1.37$    &   $ <  29.4$   &         $ <  -2.28$                &   III &     Diskless &  NG  3200  2877& 0.0126 0.0004 &  0.060 \\ 
LOri132      &  5 34 29.18 & +09 47  7.70 &CFHT1999&  W  &   $ <  1.87$    &   $ <  29.5$   &         $ <  -2.14$                &    II &         Thin &  NG  2900  2841& 0.0114 0.0004 &  0.054 \\ 
LOri136      &  5 34 38.31 & +09 58 13.10 &CFHT1999&  W  &   $ <  2.56$    &   $ <  29.6$   &         $ <  -2.10$                &   III &         Thin &  NG  2900  2886& 0.0131 0.0004 &  0.063 \\ 
LOri138      &  5 33 43.44 & +09 45 22.81 &CFHT1999&  W  &   $ <  1.67$    &   $ <  29.4$   &         $ <  -2.20$                &    -- &         Thin &  NG  2900  2811& 0.0104 0.0003 &  0.049 \\ 
LOri140      &  5 34 19.29 & +09 48 28.02 &CFHT1999&  W  &   $ <  1.62$    &   $ <  29.4$   &         $ <  -2.18$                &    II &        Thick &  NG  2800  2792& 0.0098 0.0003 &  0.046 \\ 
LOri142      &  5 34 17.00 & +10  6 16.42 &CFHT1999&  W  &   $ <  1.71$    &   $ <  29.5$   &         $ <  -1.93$                &    -- &           -- &  NG  3000  2685& 0.0070 2.8E-5 &  0.035 \\ 
LOri154      &  5 34 19.78 & +09 54 20.84 &CFHT1999&  W  &   $ <  1.13$    &   $ <  29.3$   &         $ <  -1.92$                &    -- &           -- &  Du  2200  2558& 0.0043 0.0003 &  0.026 \\ 
LOri156      &  5 34 36.28 & +09 55 32.18 &CFHT1999&  W  &   $ <  2.06$    &   $ <  29.5$   &         $ <  -1.59$                &   III &        Thick &  Du  2200  2503& 0.0032 2.4E-5 &  0.023 \\ 
LOri160      &  5 34 11.27 & +09 45 10.72 &CFHT1999&  W  &   $ <  1.88$    &   $ <  29.5$   &         $ <  -1.29$                &    -- &           -- &  NG  2800  2356& 0.0016 1.6E-5 &  0.017 \\ 
LOri166      &  5 34  0.35 & +09 54 22.45 &CFHT1999&  W  &   $ <  1.14$    &   $ <  29.3$   &         $ <  -1.24$                &    -- &           -- &  Du  2300  2207& 0.0009 9.0E-6 &  0.014 \\ 
\hline 
\end{tabular} 
$\,$ \\ %
(1) First listed in Morales-Calder\'on(2008) --1 sq.deg with Spitzer, Dolan \& Mathieu (1999) --DM1999- or Barrado y Navascu\'es et al. (2004) --CFHT1999.\\
(2) Based on Spitzer/IRAC color-color diagram.\\
(3) IRAC slope, after Lada et al. (2007).\\
(4) Temperatures from VOSA fitting with logg=4.0 dex or from a 5 Myr isochrone.\\
(5) Bolometric luminosity from VOSA and logg=4.0 dex.\\
(6) Mass from a 5 My isochrone.\\
$^a$ LOri124=LOri125\\
$^b$ LOri094 was located very close to the XMM-Newton edge. It is a non-detection, but no reliable upper limit was derived.\\
\end{table}
\end{landscape}

\clearpage

\setcounter{table}{12}
%
%
\begin{table*}
\begin{center}
\caption{Median X-ray luminosity and disk fraction for sub-samples of Collinder\,69 representing different mass ranges.
`$N_{\rm II}$', `$N_{\rm III,D}$' and `$N_{\rm III,DL}$' are the number of Class\,II and ~III --with thin/transition disks, or diskless-- sources for the disk fraction.
}
\label{tab:diskfraction}
\begin{tabular}{l  r              r            r             c           r            c     r             r               r                c              }           
\hline                                                                                                                
             &   \multicolumn{4}{c}{X-ray detections}                              &    & \multicolumn{4}{c}{X-ray upper Limits}                      \\ 
                   \cline{2-6}                                                                   \cline{8-11}                                              
Mass range   &$N_{\rm II}$ &$N_{\rm III,D}$  &$N_{\rm III,DL}$&Disk fract.&$<L_{\rm x}>$ &$\,$&$N_{\rm II}$ &$N_{\rm III,D}$ &$N_{\rm III,DL}$ & Disk fract.   \\
$[M_\odot]$   &            &               &               & [\%]     &  [erg/s]    &    &            &               &               & [\%]    \\
\hline                                                                                                                                                 
$0.3 - 0.5$ &   1         &  1$^1$          & 16          & 11.1       &   29.6     &    &  3         &    0         &  2             & 21.7 \\
$0.5 - 0.9$ &   2         &  1              & 18          & 14.3       &   30.0     &    &  0         &    0         &  0             & 14.3 \\
$0.9 - 1.2$ &   0         &  0              &  4          &  0.0       &   30.6     &    &  0         &    0         &  4             &  0.0 \\
$1.2 - 1.5$ &   0         &  0              &  2          &  0.0       &   30.2     &    &  0         &    1         &  1             & 25.0 \\
\hline
\end{tabular}
\end{center}
$\,$\\
$^1$ Including C69-X-e041, with no IRAC data (unknown II/III, but clear excess at 24 micron).\\
\end{table*}

\end{document}